\documentclass[11pt,a4paper]{article}
\usepackage[utf8]{inputenc}
\usepackage[numbers,sort&compress]{natbib}
\usepackage[margin=1in]{geometry}
\usepackage{amsmath,amssymb,amsthm}
\usepackage{graphicx}
\usepackage[colorlinks=true,linkcolor=blue,citecolor=blue,urlcolor=blue]{hyperref}
\usepackage{bm}
\usepackage[T1]{fontenc}
\usepackage{lmodern}
\usepackage{booktabs}

\newenvironment{namedthm}[3]{%
  \par\medskip\noindent\textbf{#1~#2}%
  \if\relax\detokenize{#3}\relax\else\ (\emph{#3})\fi%
  \textbf{.}\hspace{0.4em}\ignorespaces}{\par\medskip}

\allowdisplaybreaks
\numberwithin{equation}{section}
\title{The spectral picture of self-similar collapse in the Constantin--Lax--Majda equation}
\author{Jie Xu\\[2pt]
\small Department of Mechanical and Industrial Engineering,\\
\small University of Illinois Chicago, Chicago, IL, USA\\
\small \texttt{jiexu@uic.edu}\\
\small ORCID: \href{https://orcid.org/0000-0001-5765-7431}{0000-0001-5765-7431}}
\date{}

\hypersetup{
  pdftitle={The spectral picture of self-similar collapse in the Constantin-Lax-Majda equation},
  pdfauthor={Jie Xu},
  pdfkeywords={Constantin-Lax-Majda equation; self-similar blow-up; essential spectrum; spectral gap; Hilbert transform; operator realization},
  pdfsubject={Spectral analysis of the CLM self-similar collapse profile}
}

\begin{document}
\maketitle

\begin{abstract}
We give a spectral description of the self-similar collapse profile of the Constantin-Lax-Majda (CLM) equation, the $a=0$ anchor of the generalized family $w_t + a\,u\,w_x = u_x\,w$, $u_x = Hw$. Linearizing about the exact profile $\Omega(y) = -y/(y^2+\tfrac14)$ and realizing $L_0$ as a closed operator on the origin-$H^2$ space, whose regularity at the origin matches the collapse point, we prove three things at $a=0$. Its essential spectrum meets the closed half-plane $\{\mathrm{Re}\,\lambda \ge -\tfrac12\}$ in the single vertical line $\{\mathrm{Re}\,\lambda = -\tfrac12\}$: the line is placed by a log-widening Weyl sequence, and an explicit Hardy-Mellin resolvent bound constructively empties the rest of the half-plane apart from $0$ and $1$. Its full point spectrum over $\mathbb{C}$, on the odd realization, is exactly $\{0,1\}$, the scaling and time-shift symmetry modes, with no embedded eigenvalues; removing these by the standard modulation leaves a spectral gap of $\tfrac12$ on $X$. The linear semigroup and its exact decay rate $e^{-\tau/2}$ are computed in closed form, but on a weighted space of the conjugated variable reached from $X$ by a bounded transfer map; we keep the two separate, since $L_0$ is non-normal and a spectral gap does not by itself give a decay rate in the $X$ norm. A realization dichotomy identifies the smear that discretizations without an origin condition place inside the strip as the faithful spectrum of the maximal $L^2$ realization, which origin-$H^2$ removes. For $a>0$ we prove a conditional two-line inclusion for each admissible smooth focusing profile, recompute the branch $c_l(a)$ of Lushnikov, Silantyev, and Siegel as a cross-check (matching their critical advection $a_c$ to $0.04\%$), and record the formal scaling-relevance exponent $s^*(a) = 1/c_l(a)$, below which fractional dissipation is asymptotically subdominant in self-similar variables for fixed sufficiently regular data. The contribution is the realization-dependent spectral picture of the collapse profile itself.
\end{abstract}


\section{Introduction}

The regularity of the three-dimensional incompressible Euler and Navier-Stokes equations turns on whether the vortex-stretching term can drive a finite-time singularity against the regularizing effect of advection (and, in the viscous case, of dissipation). Because the full problem remains out of reach, a hierarchy of one-dimensional caricatures has been used since the 1980s to isolate this competition. The base model is the Constantin-Lax-Majda (CLM) equation \cite{ConstantinLaxMajda1985}, which retains vortex stretching but discards transport and admits an explicit finite-time singularity. De Gregorio \cite{DeGregorio1990} reinstated the advection term, and Okamoto, Sakajo, and Wunsch \cite{OkamotoSakajoWunsch2008} interpolated between the two with a scalar advection parameter $a$, giving the generalized Constantin-Lax-Majda (gCLM) family that we study here.

We write the gCLM equation as
\begin{equation}
w_t + a\,u\,w_x = u_x\,w, \qquad u_x = H w,
\end{equation}
where $H$ denotes the Hilbert transform (Fourier multiplier $-i\,\mathrm{sgn}\,k$), $a=0$ is CLM, and $a=1$ is De Gregorio. On the real line the advection is regularizing for $a>0$ (the sign convention of \cite{OkamotoSakajoWunsch2008}), and high-precision computation locates a critical advection $a_c = 0.6890665\ldots$ \cite{LushnikovSilantyevSiegel2021}: for $a\in(0,a_c)$ solutions undergo self-similar collapse (the spatial extent shrinks, focusing exponent $c_l>0$), while for $a_c < a \le 1$ blow-up occurs with an expanding extent ($c_l<0$).

The last few years have seen a surge of rigorous and computer-assisted results on inviscid self-similar blow-up in this family and its relatives. Chen, Hou, and Huang \cite{ChenHouHuang2021} proved finite-time blow-up for the inviscid De Gregorio and gCLM models, the smooth-data De Gregorio case by a computer-assisted (interval-arithmetic) argument and the small-advection gCLM and Hölder-data cases analytically; Huang, Qin, Wang, and Wei \cite{HuangQinWangWei2024} and Huang, Tong, and Wang \cite{HuangTongWang2026} constructed inviscid self-similar profiles (interiorly smooth one-scale profiles for all $a \le 1$ \cite{HuangQinWangWei2024}; additional two-scale singular scenarios (numerical) from smooth data with derivative degeneracy for $a \le 0$ \cite{HuangTongWang2026}); Chen and Hou \cite{ChenHou2022partI,ChenHou2023partII} pushed the computer-assisted-proof methodology to the two-dimensional Boussinesq and three-dimensional axisymmetric Euler settings; and a parallel machine-learning thread has computed self-similar profiles by neural networks, for the axisymmetric-Euler scenario \cite{WangLaiGomezSerranoBuckmaster2023} and, in the DeepMind unstable-singularities work, selected unstable profiles to near machine precision \cite{DeepMindUnstable2025,WangLegerLaiBuckmaster2025}. Most recently, Hou, Wang, and Yang \cite{HouWangYang2025} brought this computer-assisted-proof architecture to the viscous three-dimensional Navier-Stokes equations themselves, proving nonuniqueness of Leray-Hopf weak solutions (not finite-time singularity): they construct a self-similar Leray-Hopf solution, split the linearization about it into a coercive part and a compact perturbation, approximate the compact part to controlled error by a finite-rank operator, and verify by a computer-assisted argument on the finite-rank image the invertibility that produces an unstable eigenpair, hence a second solution. The Evans-determinant diagnostic of Section 3.2 is a numerical analogue of this finite-rank-certification architecture (its three recorded gaps are listed there), applied to the collapse linearization of the gCLM model rather than to the viscous three-dimensional equations; the closed-form theorems of Section 3 are proved instead by Mellin diagonalization and an explicit resolvent, since for the collapse linearization the nonlocal term is genuinely non-compact (Section 3.1). Lushnikov, Silantyev, and Siegel \cite{LushnikovSilantyevSiegel2021} and Ambrose, Lushnikov, Siegel, and Silantyev \cite{AmbroseLushnikovSiegelSilantyev2024} developed the complex-pole-dynamics machinery that pins down $a_c$ and the real-line versus periodic dichotomy.

The literature on this family is rich, and self-similar blow-up together with its stability has been established by several methods. The spectrum of the linearization at the collapse profile is a different object, and it is the subject of this paper. Inviscid stable self-similar blow-up for this family of nonlocal transport models was proved by Elgindi, Ghoul, and Masmoudi \cite{ElgindiGhoulMasmoudi2021} using modulation, near a known family of self-similar solutions, and Elgindi and Jeong \cite{ElgindiJeong2020} obtained finite-time singularity formation for the De Gregorio model in a low-regularity (Hölder) class and for the Okamoto-Sakajo-Wunsch models from smooth data in a new range of parameter values. Dissipative blow-up is also known: Schochet \cite{Schochet1986} solved the viscous CLM ($a=0$) explicitly and showed it blows up even at the full Laplacian $s=2$; Sakajo \cite{Sakajo2003global,Sakajo2003blowup} treated the $a=0$ CLM with a generalized viscosity term of arbitrary derivative order, proving, on the periodic line, blow-up at sufficiently small viscosity regardless of the order \cite{Sakajo2003blowup} and global solutions at large viscosity \cite{Sakajo2003global}; J. Chen \cite{Chen2020} proved real-line, dissipative, self-similar blow-up for $a$ near $\tfrac12$ at $s=2$ by establishing nonlinear stability of an approximate self-similar profile (a stability-plus-persistence result in exactly the regime where we cross-check $s^*$); and the exact pole-dynamics solutions of Silantyev, Lushnikov, Siegel, and Ambrose \cite{SilantyevLushnikovSiegelAmbrose2025} give dissipative gCLM blow-up at isolated integer dissipation orders and isolated advection values on the periodic line. What these do not provide, and what we contribute, is a treatment of the collapse-profile family as a spectral object: the essential-spectrum structure and linear-stability gap of the self-similar linearization at $a=0$, and the scaling-relevance map $s^*(a) = 1/c_l(a)$. The sharp critical dissipation curve separating blow-up from global regularity is a different object, and remains unknown. (Throughout, $s$ is the order in $\Lambda^s$ with $\Lambda = (-\partial_{xx})^{1/2}$, so $s=2$ is the ordinary Laplacian.) How this contribution differs from each of these nearest neighbors, including what the overlap with \cite{SilantyevLushnikovSiegelAmbrose2025} amounts to at the two advection values where $s^*(a)$ is cross-checked, is set out in Section 8.

Our first main result is the spectral picture of the linearization at $a=0$, proved in full. Realized on the origin-$H^2$ space, the essential spectrum of $L_0$ meets the closed half-plane $\{\mathrm{Re}\ge-\tfrac12\}$ in the single vertical line $\{\mathrm{Re}=-\tfrac12\}$ (Theorem 1, Section 3.1), and its full point spectrum over $\mathbb{C}$, on the odd realization, is exactly $\{0,1\}$, the scaling and time-shift symmetry modes, with no embedded eigenvalues (Theorem 2 with the completeness Lemma 1, Section 3.3, extended to all of $\mathbb{C}$ by Theorem 3 of Section 5). Hence the CLM collapse profile is linearly stable at $a=0$ in the spectral sense, with a gap of $\tfrac12$ on $X$; the linear semigroup and its exact decay rate $e^{-\tau/2}$ are explicit on the conjugated weighted space, reached from $X$ by a bounded transfer map (Appendix A). A realization dichotomy (Proposition 2) identifies the essential-spectrum smear that discretizations without an origin condition place in the strip as the faithful spectrum of the maximal $L^2$ realization. The novelty lies in the completeness of the point spectrum, the realization dichotomy, and the constructive resolvent. The conjugation $v = (y+\tfrac{i}{2})^2u$ turns the semigroup into a profile-independent pure dilation, and it is that reduction which makes the exact decay rate explicit in closed form. The scaling and time-shift symmetries generically generate distinguished modes of a self-similar linearization (as in the gKdV self-similar linearization of \cite{Chapman2024}), and in the present normalization these sit at the eigenvalues $\{0,1\}$ by direct substitution, so the content of Theorem 2 is their completeness, that nothing else appears, rather than the values; and the realization-dependence of the essential spectrum is classical cone calculus \cite{Kondratev1967,CoriascoSchroheSeiler2007} (as is, for a linearized inviscid fluid operator, an essential spectrum that is a vertical band \cite{ShvydkoyLatushkin2003}), the new elements being the explicit closed-form resolution for the CLM operator and the reinterpretation of the numerical smear.

The second main result concerns $a>0$: the endpoint-inclusion result and the collapse-profile branch. The full high-precision branch $c_l(a)$ on $(0,a_c)$, together with the critical advection $a_c = 0.6890665\ldots$, was computed by Lushnikov, Silantyev, and Siegel \cite{LushnikovSilantyevSiegel2021}. We recompute $c_l(a)$ independently by a lower-order Newton continuation on a compactified real-line grid (Newton residual $10^{-13}$-$10^{-10}$, solution accuracy $\sim 3\times10^{-4}$ at the $a=0$ floor), as an implementation cross-check and to supply the operator coefficients $(\Omega,c_l)$ that Proposition 1 and the exponent $s^*$ below need at $a>0$; it reproduces the branch endpoint against the published $a_c$ to $0.04\%$ (Section 2). The $a=0$ results above do not use it: there the profile and $c_l(0)=1$ are exact. Existence and smoothness of such profiles are established by Huang, Qin, Wang, and Wei \cite{HuangQinWangWei2024}, with the $a=1$ spectral construction in \cite{HuangTongWei2023}; these are candidate profiles for the admissibility hypotheses; whether they coincide with the numerically continued branch, and whether they satisfy (H3)-(H4), is left open (Section 4.1). For $a>0$ the two-line essential-spectrum inclusion is proven for each admissible smooth focusing profile (Proposition 1, Section 3.1, a conditional inclusion under stated hypotheses); beyond $a=0$ the discrete exclusion rests on numerical evidence at the sampled advections (Section 3.2), with the small-$a$ perturbative exclusion summarized as a program in Section 8.

The third is the formal scaling-relevance exponent $s^*(a) = 1/c_l(a)$ (Section 6), read off the branch: the value below which fractional dissipation of order $s$ enters the self-similar variables with a coefficient that vanishes as $\tau\to\infty$, so it is asymptotically subdominant for fixed regular data, with two sub/supercritical cross-checks against Schochet \cite{Schochet1986} and J. Chen \cite{Chen2020}. It is a formal scaling diagnostic, a scaling-relevance exponent distinct from a proved persistence threshold or the sharp blow-up/regularity curve.

\section{\texorpdfstring{The collapse profile and the exponent $c_l(a)$}{The collapse profile and the exponent cl(a)}}

We seek an odd, self-similar collapse solution of the inviscid gCLM equation,
\begin{equation}
w(x,t) = (T-t)^{-1}\,\Omega(y), \qquad y = \frac{x}{(T-t)^{c_l}},
\end{equation}
where $T$ is the collapse time and $c_l>0$ the focusing exponent. Because $H$ is homogeneous of degree $0$ under dilation, substituting the ansatz reduces the equation to the profile equation
\begin{equation}
\Omega + (c_l\,y + a\,U)\,\Omega' = \Omega\,H\Omega, \qquad U' = H\Omega,
\end{equation}
with $\Omega$ odd and decaying, and $c_l = c_l(a)\in(0,1]$ (with $c_l(0)=1$) appearing as a nonlinear eigenvalue selected together with $\Omega$.

At $a=0$ the profile is exact and serves as our anchor:
\begin{equation}
\Omega(y) = -\frac{y}{y^2 + \tfrac14}, \qquad H\Omega(y) = \frac{\tfrac12}{y^2 + \tfrac14}, \qquad c_l = 1.
\end{equation}
This is the classical single-scale collapse; the $a=0$ model also admits distinct multi-scale asymptotically self-similar blowups from other initial data (Huang, Qin, and Wang \cite{HuangQinWang2024multiscale}), which lie outside the spectral description of this paper, whose object is the linearization at the exact single-scale profile.

The full high-precision branch $c_l(a)$ on $(0,a_c)$ was already computed by Lushnikov, Silantyev, and Siegel \cite{LushnikovSilantyevSiegel2021} (complex pole dynamics), and we take theirs as the reference; the recomputation here is an independent lower-order cross-check that also supplies the operator coefficients $(\Omega, c_l)$ needed by Section 3. We solve the profile equation numerically: the real line is compactified by $y = c\tan(\theta/2)$, $\theta\in(-\pi,\pi)$, which maps the endpoints $y=\pm\infty$ to $\theta = \pm\pi$; the Hilbert transform is transplanted exactly onto the $\theta$-circle, $(Hf)(y) = H_c\,g(\theta) - K$, where $H_c$ is the periodic (circular) Hilbert transform of $g(\theta) = f(y(\theta))$ and $K$ a constant fixed by the decay of $f$. We then run Newton continuation in $a$, starting from the exact CLM anchor $(\Omega,c_l) = (-y/(y^2+\tfrac14),\,1)$ and stepping $a$ upward in increments $\Delta a = 0.04$. The scheme drives the Newton residual to $\lVert F\rVert \sim 10^{-13}$-$10^{-10}$ depending on $a$ and $N$ ($F$ the residual of the discretized profile equation, i.e. how well the equation is solved on the grid, not the accuracy of the solution itself); the two-grid difference is $|c_l(N{=}1024) - c_l(2048)| \lesssim 5\times 10^{-4}$ (about three significant figures), a self-consistency measure rather than a certified continuum error bound (it can understate the true error against the high-precision branch of \cite{LushnikovSilantyevSiegel2021}, which at nonzero $a$ agrees only to about two or three significant figures; we do not tabulate a node-by-node comparison, and only the endpoint $a_c$ is checked against \cite{LushnikovSilantyevSiegel2021} quantitatively), while the $a=0$ solution accuracy is $\sim 3\times 10^{-4}$, set by the first-order accuracy of the transplanted Hilbert transform.

The recomputation is cross-checked two independent ways. First, at $a=0$ it recovers the exact CLM profile with $c_l(0) = 1.00024$ (agreeing with the exact $c_l=1$ to $\sim 2\times 10^{-4}$, the accuracy limited by the first-order Hilbert-transform floor, not by the Newton residual). Second, the computed $c_l(a)$ decreases monotonically from $1$ at $a=0$ and crosses zero at $a^* \approx 0.6888$ (linear interpolation between the computed branch values $c_l(0.68) = 0.0185$ and $c_l(0.72) = -0.0659$), to be compared with the published $a_c = 0.6890665$ of \cite{LushnikovSilantyevSiegel2021}. The two agree to $0.04\%$ relative error (Figure 1), which is the one quantitative check on the recomputed branch. This reproduced map $c_l(a)$ then supplies the inputs used below: the far-field-line distance $1-c_l/2$ and the exponent $s^*(a) = 1/c_l(a)$ are read off it.

Table 1 collects the recomputed branch and the derived quantities used throughout (the far-field-line value $1 - c_l/2$, the exponent $s^*(a) = 1/c_l(a)$; Section 6). The Newton branch is solved on a $\Delta a = 0.04$ grid; the tabulated $c_l$ at $a = 0.1, 0.3, 0.5, 0.65$, which fall between solved points, are degree-6 polynomial interpolations of the solved branch, and the remaining rows are Newton solves. The two-grid difference is $|c_l(1024) - c_l(2048)| \lesssim 5\times 10^{-4}$ (about three significant figures, a self-consistency measure not a continuum bound; the $a=0$ profile accuracy is $\sim 3\times 10^{-4}$, set by the first-order Hilbert-transform floor). The tabulated $1-c_l/2$ is the far-field line: it is the essential gap ($\tfrac12$) at $a=0$, while for $a\in(0,\tfrac12)$ the origin line of Proposition 1 lies to its right by the protrusion $-g(a)\in[0,0.0362]$, which vanishes as $a\to 0$ and is largest near $a\approx0.264$. Since Proposition 1 gives only an inclusion, the two lines bound the essential spectrum from within: the essential gap in $X$ is smaller than the far-field value by at least $-g(a)$, with equality precisely under the exactness question of Section 4.5 (Section 3.1). On the branch both lines stay strictly left of $\mathrm{Re} = -\tfrac12$ for $a>0$, so this correction never closes the gap.

\begin{table}[htbp]\centering
\caption{The recomputed collapse branch (reference branch from \cite{LushnikovSilantyevSiegel2021}): focusing exponent $c_l(a)$ ($N = 2048$; two-grid difference $\lesssim 5\times10^{-4}$; rows at $a = 0.1, 0.3, 0.5, 0.65$ are degree-6 interpolations of the $\Delta a = 0.04$ Newton branch), the far-field-line distance $1 - c_l/2$, the exponent $s^*(a) = 1/c_l(a)$, and the anchor cross-checks; $c_l(\tfrac12) = \tfrac13$ is exact (\cite{Chen2020}; also \cite[Thm. 2]{LushnikovSilantyevSiegel2021}).}
\begin{tabular}{rrrrp{0.40\textwidth}}\toprule
$a$ & $c_l(a)$ & far-field $1 - c_l/2$ & $s^*(a) = 1/c_l$ & cross-check \\ \midrule
0.0 & 1.0000 & 0.500 & 1.000 & exact CLM; Schochet \cite{Schochet1986}: full-Laplacian blow-up ($s=2>s^*$, supercritical) \\
0.1 & 0.8730 & 0.564 & 1.145 &  \\
0.2 & 0.7474 & 0.626 & 1.338 &  \\
0.3 & 0.6178 & 0.691 & 1.619 &  \\
0.4 & 0.4809 & 0.760 & 2.079 & ($s=2$ boundary at $\approx 0.39$, where $c_l = 1/2$) \\
0.5 & 0.3333 & 0.833 & 3.000 & J. Chen \cite{Chen2020}: blow-up at $s=2<3$ (subcritical) \\
0.6 & 0.1691 & 0.915 & 5.914 &  \\
0.65 & 0.0775 & 0.961 & 12.90 & branch endpoint $a^* \approx 0.6888$ vs $a_c = 0.6890665$ \cite{LushnikovSilantyevSiegel2021} \\
\bottomrule\end{tabular}\end{table}

\begin{figure}[htbp]\centering
\includegraphics[width=0.75\textwidth]{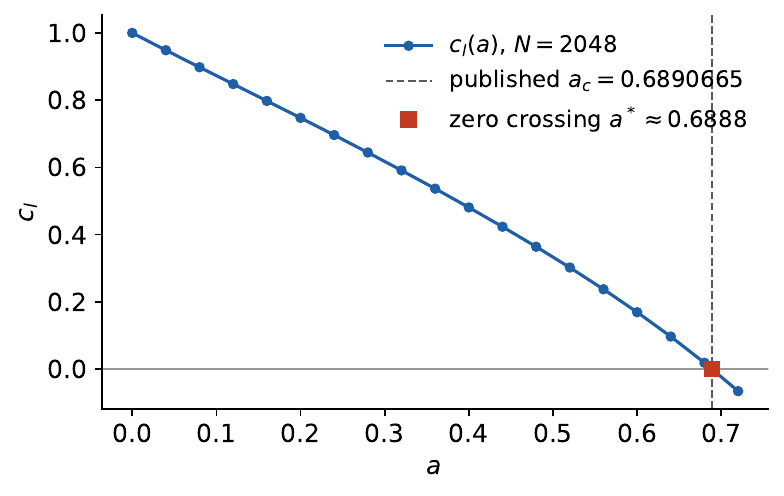}
\caption{The recomputed branch $c_l(a)$ from the compactified-grid Newton continuation at $N=2048$, plotted at the advections where the Newton solve succeeded, decreasing monotonically from the exact CLM anchor $c_l(0)=1$ to its zero crossing at $a^* \approx 0.6888$, which reproduces the published critical advection $a_c = 0.6890665$ of Lushnikov, Silantyev, and Siegel (dashed line) to $0.04\%$; the reference branch is theirs, this is an independent lower-order recomputation. The far-field-line distance $1 - c_l/2$ of Section 3 (the essential gap at $a=0$; for $a>0$ the far-field contribution to it) grows monotonically along this branch.}
\end{figure}

\section{Linear stability and the spectral gap}

Fix $a\in[0,a_c)$ and its profile $(\Omega, c_l)$. Linearizing the self-similar dynamics about $\Omega$ gives the operator $L_a = -\,d\,\mathrm{Res}/d\Omega$, where $\mathrm{Res}[W] := W + (c_l y + aU)W' - W\,HW$ is the residual of the profile equation, whose zero is the collapse profile (the rescaled evolution is $\partial_\tau W = -\mathrm{Res}[W]$; Section 6.1). It acts on odd perturbations $\varphi$ that are smooth at the origin ($\varphi\sim y$) and decaying at infinity. Explicitly, on the odd half-line $(0,\infty)$ with the measure $dy$,
\begin{equation}
L_a\varphi = -\varphi - (c_l\,y + a\,U)\,\varphi' - a\,V(\varphi)\,\Omega' + \varphi\,H\Omega + \Omega\,H\varphi,
\end{equation}
where $U = \int_0^y H\Omega$ and $V(\varphi) = \int_0^y H\varphi$.

We realize $L_a$ as a closed, densely defined operator on the origin-$H^2$ space $X$ of Section 3.1 (constructed in Section 4.1). Separate the recentered dilation part $A\varphi = -\varphi - (c_l y + aU)\varphi'$, with graph domain $D(A) = \{\varphi\in X : (c_l y + aU)\varphi'\in X\}$ and the odd, compactly supported smooth functions (whose linear coefficient need not vanish, e.g. $y$ times an even bump, which has $a_1\neq0$ and lies in $D(A)$ by compact support; dense in the odd part of $H^2$) as a core, from the profile part $B_a\varphi = -a\,V(\varphi)\,\Omega' + \varphi\,H\Omega + \Omega\,H\varphi$. Then $B_a\in\mathcal B(X)$ (Section 4.1: $H$ preserves both $H^2$ norm components while exchanging the odd and even parts, and the profile coefficients lie in $W^{2,\infty}$; at $a=0$ where $B_0\varphi = \varphi\,H\Omega + \Omega\,H\varphi$ this is immediate, and for $a>0$ the antiderivative term is $X$-bounded as a consequence of the admissibility conditions, derived in Section 4.1). Hence $L_a = A + B_a$ is closed on the graph domain $D(L_a) := D(A)$, densely defined, and the profile $\Omega$ itself lies in $D(L_a)$.

We read the essential spectrum in the Browder sense throughout, $\sigma_{\mathrm{ess}}(L) := \sigma(L)\setminus\sigma_{\mathrm{disc}}(L)$, where $\sigma_{\mathrm{disc}}$ collects the isolated eigenvalues of finite algebraic multiplicity (finite-rank Riesz projection). On the essential line we prove the stronger statement that $L_0-\lambda$ fails to be upper-semi-Fredholm there, via the singular Weyl sequences of Section 4, so the line lies in the Fredholm essential spectrum as well; any singular-Weyl-sequence inclusion produces a subset of the Browder essential spectrum. Stability is the statement that $L_a$ has no spectrum with $\mathrm{Re}\ge 0$ apart from the two symmetry modes at $0$ and $+1$ below, and a gap separating the rest of the spectrum from the imaginary axis.

\subsection{\texorpdfstring{The essential spectrum at $a=0$ (Theorem 1)}{The essential spectrum at a=0 (Theorem 1)}}

Realize $L_a$ on the origin-$H^2$ space
\begin{equation}
X = \{\varphi : \varphi \text{ odd},\ \varphi,\varphi'' \in L^2(0,\infty),\ \varphi(y) = a_1 y + o(y) \text{ as } y\to 0 \ (a_1\in\mathbb{C})\},
\end{equation}
which, extended oddly to the line, is exactly the odd part of $H^2(\mathbb{R})$ with the Hilbertian norm $\lVert\varphi\rVert_X^2 = \lVert\varphi\rVert_{L^2}^2 + \lVert\varphi''\rVert_{L^2}^2$; in particular $\varphi,\varphi'\to 0$ at infinity automatically, and $\varphi\in C^1$ with $\varphi(0)=0$. This is the realization whose origin regularity matches that of the collapse point (the origin second-derivative condition is the $L^2$ form of condition (D3) of Lemma 1). It is a modeling choice, not derived here from a dynamical well-posedness principle for the rescaled nonlinear flow, and the content of Proposition 2 below is precisely that the essential spectrum depends on this choice.

\begin{namedthm}{Theorem}{1}{essential spectrum at $a=0$}
On $X$, the essential spectrum of the CLM linearization $L_0$ meets the closed half-plane $\{\mathrm{Re}\,\lambda \ge -\tfrac12\}$ in the single vertical line
\begin{equation}
\sigma_{\mathrm{ess}}(L_0|_X) \cap \{\mathrm{Re}\,\lambda \ge -\tfrac12\} = \{\mathrm{Re}\,\lambda = -\tfrac12\}.
\end{equation}
\end{namedthm}

A Mellin analysis of the far-field dilation places the line in $\sigma_{\mathrm{ess}}$, and a constructive $X$-resolvent bound, an explicit analytic majorant $M_{\mathrm{an}}(z)\sim\alpha^{-3/2}$ for $R_0(z)=(L_0-z)^{-1}$ with $\alpha=\mathrm{Re}\,z+\tfrac12$, empties the open half-plane $\{\mathrm{Re}\,z>-\tfrac12\}$ apart from the two eigenvalues $0, 1$ (Section 4, Sections 4.2 to 4.4). The region $\mathrm{Re}\,\lambda<-\tfrac12$ plays no part in the stability conclusion and is not analyzed here.

Away from $a=0$ the same far-field line is joined by a second, origin line whose position is fixed by the second-derivative weight of $X$; the two coincide at $a=0$, which is why Theorem 1 sees a single line there. We record the general-$a$ inclusion, since it is the object the numerics of Section 3.2 track.

\begin{namedthm}{Proposition}{1}{endpoint essential-spectrum inclusion for an admissible focusing profile}
Fix $a\in(0,a_c)$ and a collapse profile $(\Omega,c_l)$ satisfying the admissibility conditions $\mathrm{Adm}(a)$ of Definition 4.2 (Section 4.1), which fix the regularity of $\Omega$ up to and including the origin, the origin expansions of $\Omega$, $H\Omega$ and $U$ together with their derivative-level remainders, far-field tail bounds, and the comparability of the transport weight $c_l y + aU$ with $y$. Then the essential spectrum of $L_a$ on $X$ contains the two vertical lines
\begin{equation}
\begin{gathered}
\{\mathrm{Re}\,\lambda = -1 + c_l/2\} \ \cup\ \{\mathrm{Re}\,\lambda = \lambda_X(a)\}, \\[4pt]
\lambda_X(a) = -\tfrac{\tilde c}{2}, \qquad \tilde c = c_l + a\,H\Omega(0), \qquad H\Omega(0) = \tfrac{1+c_l}{1-a},
\end{gathered}
\end{equation}
the closed form $\lambda_X = -\tilde c/2$ following from the identity $H\Omega(0)-\tilde c = 1$ (and $H\Omega(0) = (1+c_l)/(1-a)$ from differentiating the profile equation at the origin, using $\Omega'(0)\neq 0$).
\end{namedthm}

These are precisely the hypotheses the estimates below consume, and they yield $B_a\in\mathcal B(X)$, hence $L_a = A+B_a$ closed on $D(A)$; the smooth one-scale profiles of Huang, Qin, Wang, and Wei \cite{HuangQinWangWei2024} are a candidate supply of such profiles (Section 4.1). The far-field line is realized by a log-widening Weyl sequence; the origin line by a Weyl sequence concentrated at the exponent $y^{3/2+i\omega}$ that the second-derivative weight of $X$ makes marginal (Section 4.5). Writing $g(a) = 2c_l - H\Omega(0)(1-\tfrac32 a)$, the two lines coincide at $a=0$ and, exactly, at $a=\tfrac12$: the exact analytical collapse solution at $a=\tfrac12$ (J. Chen \cite{Chen2020}; found independently, in slightly generalized form, as \cite[Thm. 2]{LushnikovSilantyevSiegel2021}) has self-similar variable $x/(t_c-t)^{1/3}$, so $c_l(\tfrac12)=\tfrac13$ exactly and $g(\tfrac12) = (3c_l-1)/2 = 0$; the branch reproduces this within the grid tolerance ($c_l(\tfrac12) = 0.333$). For $a\in(0,\tfrac12)$ the origin line protrudes into the strip, on the recomputed branch of Section 2 by at most $0.037$ (Figure 2).

Whether $\sigma_{\mathrm{ess}}$ equals this union for $a>0$, and whether a stronger third-derivative realization ($\varphi'''\in L^2$ near the origin) leaves the single far-field line uniformly in $a$, are exactness questions (Section 4.5). They are not used below: the linear stability of this section, the point-spectrum theorems of Section 3.3, and the semigroup of Appendix A all live at $a=0$, where the two lines coincide and Theorem 1 is exact. The origin line is a feature of the $H^2$ realization that a stronger $H^3$ realization is expected to remove (the $H^3$ weight shifts the included origin exponent; exactness for the stronger realization is the same open question), so at $a>0$ the two-line picture is realization-relative; Proposition 2 below makes the mechanism precise.

On the plain maximal $L^2$ realization the picture is different: there every point of the strip between the two indicial lines lies in the essential spectrum, and the single line at $a=0$ is recovered only in the origin-$H^2$ topology (Proposition 2 below).

The proof (full detail in Section 4) combines a Mellin diagonalization of the two endpoint models with singular Weyl sequences at both endpoints: the log-widening far-field sequence at $y=\infty$ for the far-field line, and its transplant to $y=0$ at the $X$-marginal exponent $y^{3/2+i\omega}$ for the origin line. The two nontrivial steps, carried out with explicit estimates in Section 4, are the following: the nonlocal terms $\Omega\,H\varphi_n$ and $a\,V(\varphi_n)\,\Omega'$ vanish on the endpoint Weyl sequences (at infinity with explicit Hilbert-Schmidt bounds, splitting the kernels at $\sqrt{R_n}$ and using the profile tail rates $\Omega = O(y^{-1/c_l})$, $\Omega' = O(y^{-1-1/c_l})$; at the origin because their coefficient or order gains a full power of $y$), and the exclusion of everything else, the $L^2$ indicial line $\mathcal{L}_0$ at the origin (whose modes $y^{-1/2-i\omega}$ violate the domain requirement $\varphi\sim y$) and the bulk, which at $a=0$ is proven by the constructive resolvent bound of Section 4.4, and for $a>0$ is precisely the exactness question left open (Section 4.5).

\begin{figure}[htbp]\centering
\includegraphics[width=0.92\textwidth]{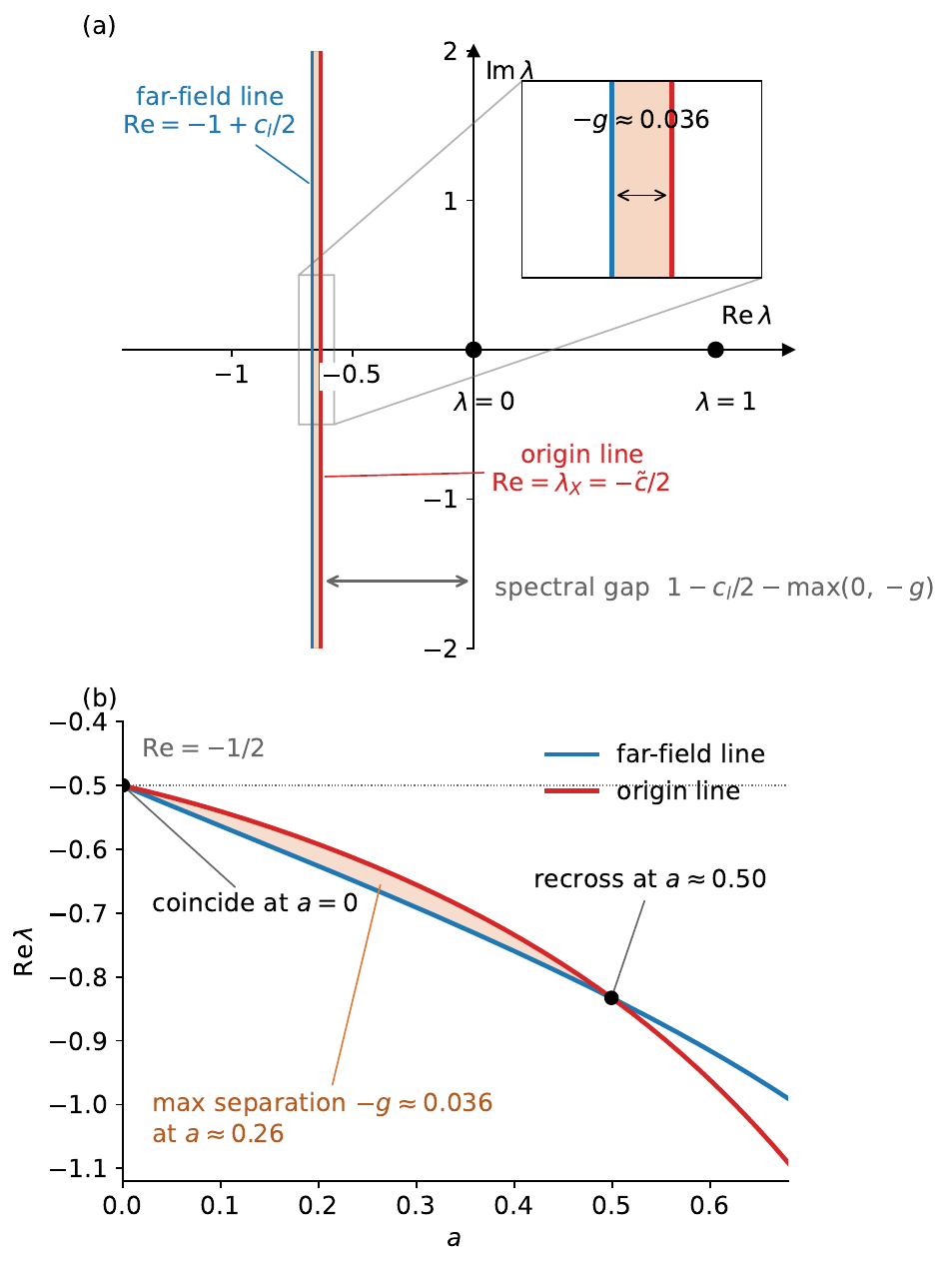}
\caption{The two essential-spectrum lines of Proposition 1 (for an admissible profile). Panel (a): the spectral $\lambda$-plane at $a=0.264$ (maximal protrusion of the origin line into the strip): far-field line $\mathrm{Re} = -1+c_l/2$ (blue), origin line $\mathrm{Re} = \lambda_X = -\tilde c/2$ (red), shaded strip, and the symmetry eigenvalues $\lambda = 0,1$; the inset resolves the separation $-g \approx 0.036$. Panel (b): both lines along the recomputed branch as functions of $(a, c_l(a))$: they coincide at $a=0$, the origin line lies at most $0.037$ right of the far-field line on $(0,\tfrac12)$ (minimum $-0.0362$ at $a\approx 0.264$), the two recross exactly at $a=\tfrac12$ ($c_l(\tfrac12)=\tfrac13$; \cite{Chen2020}, also \cite[Thm. 2]{LushnikovSilantyevSiegel2021}), and separate again for $a>\tfrac12$. Both lines stay strictly left of $\mathrm{Re}=-\tfrac12$ for $a>0$.}
\end{figure}

Ruling out singular sequences in the bulk is not a routine Fredholm fact for this nonlocal $L_a$, and compactness-based arguments genuinely fail: the Hilbert term $\Omega\,H\varphi$ is not relatively compact (a coefficient vanishing at both endpoints times an order-zero singular integral is not compact). What holds instead, and suffices, is that the nonlocal terms are subordinate at both endpoints, the coefficient of $\Omega\,H\varphi$ is $\Omega$ (not $H\Omega$), which vanishes linearly at the origin and decays at infinity, and $V$ is an antiderivative (order $-1$), so after the log substitution their translates vanish strongly at both ends and they drop out of both endpoint models; a two-ended Fredholm argument would then identify the essential spectrum with the union of the two endpoint model lines, an exactness question (Section 4.5). The numerical corroboration is diagnostic only: a log-Mellin discretization that imposes no origin regularity renders the maximal-$L^2$ realization (Proposition 2), so it cannot by itself certify either endpoint line, whose separation is small ($\le 0.037$ on the branch, and $0$ at the benchmark point $a=\tfrac12$). At $a=0$ the two lines coincide and Theorem 1 fixes the essential edge at $-\tfrac12$, so the essential contribution to the spectral gap is $\tfrac12$; for $a>0$ both lines are strictly in the left half-plane (for $c_l<2$), the right essential edge is the rightmost of the two lines of Proposition 1, and the resulting gap is corroborated by numerical evidence at the sampled advections (Section 3.2), contingent on the $a>0$ essential-spectrum exactness (Section 4.5) together with exclusion of discrete spectrum to the right of the two lines.

\emph{Map of realizations.} One realization is the one we work in and the others are named alternatives. The chosen realization is $X$ (origin-$H^2$): Theorem 1 and Proposition 1 live there, and its essential spectrum for $a\in(0,\tfrac12)$ contains two lines. The maximal $L^2$ realization (Section 4.6) is the \emph{weaker} alternative, in which the whole strip is spectrum. A third-derivative realization is a \emph{stronger} alternative in which the included origin line moves left and a single far-field line is expected to survive for all $a$; we do not pursue it here. Each statement names the realization it lives in.

\begin{namedthm}{Proposition}{2}{realization dichotomy at $a=0$}
The essential spectrum of $L_0$ depends on the realization. On the maximal $L^2$ realization ($\sigma_{\mathrm{ess}}$ read in the Browder sense there; Section 4.6),
\begin{equation}
\sigma_{\mathrm{ess}}(L_0) \supseteq \{\lambda : -\tfrac12 \le \mathrm{Re}\,\lambda \le \tfrac32\},
\end{equation}
the full vertical strip between the two indicial lines; on the origin-$H^2$ realization $X$ of Theorem 1, the open strip contains no spectrum beyond the two symmetry eigenvalues $0$ and $1$, so $\sigma_{\mathrm{ess}}(L_0|_X) \cap \{\mathrm{Re}\,\lambda \ge -\tfrac12\}$ is the single line $\{\mathrm{Re}\,\lambda = -\tfrac12\}$. The strip is filled entirely by the explicit family
\begin{equation}
u_\lambda(y) = \frac{y^{1-\lambda}}{(y+\tfrac{i}{2})^2}, \qquad -\tfrac12 < \mathrm{Re}\,\lambda < \tfrac32,
\end{equation}
each member of which, for $\lambda \notin \{0,1\}$, is $L^2$ but fails $u_\lambda'' \in L^2$ at the origin (it is the sharpness family of Lemma 1; at $\lambda \in \{0,1\}$ the family degenerates to the two smooth symmetry modes); the odd two-branch combinations of this family are genuine eigenfunctions of the maximal realization at every strip point (Section 4.6), and imposing the single second-derivative condition that defines $X$ removes the whole non-symmetry family and collapses the strip to the line. The proof is Section 4.6 (the strip-filling half) together with Proposition 4.6 (the strip-emptying half).
\end{namedthm}

This dichotomy is the mathematical content of the informal statement that the collapse point is a regular point of the perturbation, and the operator-norm statements of Sections 3, 4 and 5 are read in the topology of $X$ (the semigroup estimates of Appendix A use the scale-covariant weighted space $Y_\theta$). It also settles a discrepancy between the two families of discretization used here: the persistent essential smear that the grids without an origin condition place inside the strip is the faithful spectrum of the maximal (weak) realization, while discretizations that enforce origin regularity render the origin-$H^2$ line (Section 3.2), so the two families of grids were never inconsistent; they were computing different realizations. This inverts the usual spectral-pollution reading, on which persistent in-gap numerical spectrum is spurious, lying outside the spectrum of the intended target operator (in standard projection computations for self-adjoint operators) \cite{LevitinShargorodsky2004,DaviesPlum2004} and is suppressed by resolvent-based schemes \cite{ColbrookHorningTownsend2021}: here the in-strip spectrum is genuine, the essential spectrum of the maximal realization. The strip-emptying half is constructive: an exact Hardy-Mellin lemma gives $\lVert T_z\rVert = 1/(\mathrm{Re}\,z+\tfrac12)$ for the Hardy component, the $L^2$ divergence along $u_\lambda$ is isolated in a single origin-Taylor coefficient that $X$-balls control, and the assembled resolvent majorant is $M_{\mathrm{an}}\sim\alpha^{-3/2}$ (Section 4.4).

\subsection{The two symmetry modes and the numerical spectrum}

Everything in this paper is set in the \emph{odd sector}: the gCLM flow preserves oddness, the collapse profile $\Omega$ is odd, and the realization $X$ consists of odd functions. That restriction is what fixes the number of symmetry modes. A self-similar blow-up of this type carries \emph{three} continuous symmetries, not two: scaling, time-shift/amplitude, and spatial translation. The first two produce odd modes and appear below. The third produces the \emph{even} mode $\varphi = \Omega'$, which is an exact eigenfunction of $L_a$ for every admissible profile, at eigenvalue
\begin{equation}
\tilde c = c_l + a\,H\Omega(0) = \frac{c_l+a}{1-a} ,
\end{equation}
the same combination that fixes the origin transport slope and the origin line $\lambda_X = -\tilde c/2$. (Differentiating the profile equation and using $V(\Omega') = H\Omega - H\Omega(0)$ gives $L_a\Omega' = \tilde c\,\Omega'$ directly; the pinned gauge term $-a\,V(\varphi)\,\Omega'$ is what moves the eigenvalue off $c_l$.) It reduces to $c_l$ only at $a=0$, where one checks $L_0\Omega' = \Omega'$ exactly, with $\Omega' = (y^2-\tfrac14)/(y^2+\tfrac14)^2$ even and $O(y^{-2})$, so it lies in $L^2(\mathbb{R})$ but not in $X$. Centering the collapse at the origin, which is what the odd restriction does, removes this mode. At $a=0$ the eigenvalue $0$ likewise has the even partner $y^2/(y^2+\tfrac14)^2$. So on the full $L^2(\mathbb{R})$ the eigenvalues $0$ and $1$ each carry an even partner, while on the odd realization $X$ each eigenspace is one-dimensional, which is the situation the theorems below describe.

Within the odd sector, two exact eigenpairs exist for \emph{all} $a$, reflecting the two surviving symmetries: scaling contributes $\varphi = y\,\Omega'$ at eigenvalue $0$, and time-shift/amplitude contributes $\varphi = \Omega + c_l\,y\,\Omega'$ at eigenvalue $+1$. Both are exact, at $a=0$ and for every admissible profile at $a>0$ (Definition 4.2), by direct substitution. Admissibility gives $\Omega'\in L^1(\mathbb{R})$ with $\int_{\mathbb{R}}\Omega' = 0$, since $\Omega$ vanishes at $\pm\infty$ ((H1) and (H3)), so the commutator identity $H(y f) = y\,Hf - \pi^{-1}\!\int_{\mathbb{R}} f$ applied to $f = \Omega'$ gives $H(y\,\Omega') = y\,(H\Omega)'$, and integrating that by parts gives $V(y\,\Omega') = y\,H\Omega - U$. Substituting these into $L_a$ and eliminating $\Omega\,(H\Omega)'$ by the $y$-derivative of the profile equation $\Omega + (c_l y + aU)\,\Omega' = \Omega\,H\Omega$, every term cancels and $L_a(y\,\Omega') = 0$; substituting $V(\Omega) = U$ and eliminating $\Omega\,H\Omega$ by the profile equation itself gives $L_a\Omega = \Omega + c_l\,y\,\Omega'$, hence $L_a(\Omega + c_l\,y\,\Omega') = \Omega + c_l\,y\,\Omega'$. As independent checks, exact-symbolic substitution confirms both identities at $a=0$, and on the grid the nearest eigenvalues match $0$ and $+1$ to within $\sim 10^{-3}$, with eigenvector overlap $1.0000$ against the analytic modes. At $a=0$ these two eigenvalues exhaust the spectrum with $\mathrm{Re}\ge 0$ (Theorems 1 and 2); for $a>0$ the same statement is corroborated numerically at the sampled advections below. They are modded out by the standard modulation (choosing the collapse time and spatial scale) and do not represent instability.

Numerically, at the sampled advections $a = 0.3, 0.4, 0.5, 0.6, 0.65$ (with $N = 1024\ldots 8192$; Section 7), the compactified-grid Jacobian shows no eigenvalue with $\mathrm{Re}>0$ beyond the two symmetry modes, and no resolution-stable discrete eigenvalue in the open strip $(-1+c_l/2,\,0)$: the strip candidates that appear drift with $N$ rather than converging (for $a=0.5$ over $N = 1024/2048/4096$ the same behaviour appears under any of the natural selections -- rightmost real candidate, lowest-$\lvert\mathrm{Im}\rvert$ pair, or rightmost overall -- none of which settles as $N$ grows), and the apparent positive-real modes that proliferate as $a\to a_c$ are concentrated near the Nyquist frequency, i.e. grid pollution rather than instabilities. A log-Mellin discretization that imposes no origin regularity is consistent with this but does not independently certify the far-field edge: by the realization dichotomy (Proposition 2) it renders the maximal-$L^2$ realization, whose spectrum fills the strip up to the origin indicial line. We therefore read these as numerical evidence at the sampled parameters, supporting stability there; the discrete exclusion for $a>0$ remains open; the exclusion is a theorem only at $a=0$ (Section 3.3).

A fourth, structurally different piece of evidence is an Evans-determinant continuation in $a$. Splitting $L_a$ against its far field and forming the associated determinant on a weighted space in which the essential line is displaced left of the strip, the number of eigenvalues inside the strip window becomes an integer winding count $n_{\mathrm{disc}}(a)$ along a rectangular contour. This strand is exploratory: the computation records a per-point phase-increment and integer-defect diagnostic and anchors the $a=0$ value to the proven empty strip of Theorem 2, but it does not enforce these diagnostics as hard gates, and its run outputs are not retained. On the parameter values we ran it returns $n_{\mathrm{disc}}(a) = 0$ for $a \in [0, 0.65]$; we report this as exploratory evidence only. (For context, the spectral study of linearized operators in this family begins with Jia, Stewart, and Sverak \cite{JiaStewartSverak2019} and, subsequently, Lei, Liu, and Ren \cite{LeiLiuRen2020}, who analyzed the linearization of De Gregorio at its ground-state equilibria on the circle and proved stability of the ground states (Lei, Liu, and Ren also obtaining a conserved-quantity global well-posedness for one-sign data); Guo and Jiu \cite{GuoJiu2025} extend this line to the first excited state $-\sin 2\theta$. Both concern equilibria of De Gregorio on the torus, a different object from the real-line collapse profiles studied here, and the periodic methods do not transfer: on the real line the self-similar generator $L_a$ carries a dilation-generated continuous essential spectrum and admits no Fourier-series reduction, and it is strongly non-normal, so a torus-type energy or spectral argument cannot reach the spectral gap, the same obstruction identified as the coercivity limitation in Section 8.) Unlike eigenvalue-drift diagnostics, this is a topological count rather than a convergence judgement; it is still numerical evidence, not proof, with three explicitly recorded gaps (a uniform large-imaginary-part bound, trace-ideal membership of the kernel, and quadrature-error bounds in the trace norm); upgrading any of these would move this strand toward a computer-assisted proof.

\subsection{\texorpdfstring{Discrete-spectrum exclusion at $a=0$ (Theorem 2)}{Discrete-spectrum exclusion at a=0 (Theorem 2)}}

At $a=0$ the discrete exclusion is proven in closed form (full derivations in Section 5). The mechanism is that the exact CLM profile makes the nonlocal operator collapse to a scalar one.

\begin{namedthm}{Theorem}{2}{CLM discrete exclusion}
For the CLM linearization $L_0$ on the realization $X$ (the operator above with $a=0$, $\Omega = -y/(y^2+\tfrac14)$, $c_l=1$),
\begin{equation}
\sigma_{\mathrm{disc}}(L_0|_X)\cap\{\mathrm{Re} > -\tfrac12\} = \{0, +1\},
\end{equation}
so the open strip $(-\tfrac12, 0)$ contains no discrete eigenvalue (on $X$, the only spectrum with $\mathrm{Re} > -\tfrac12$ is the two symmetry modes at $0$ and $+1$).
\end{namedthm}

The argument is a Hardy reduction. On the upper-half-plane Hardy space $H^2_+$ (where $H = -i$), the exact profile gives the single-simple-pole identity
\begin{equation}
H\Omega - i\,\Omega = \frac{i}{y + \tfrac{i}{2}},
\end{equation}
so $L_0$ restricts to the scalar first-order operator
\begin{equation}
L_0^+ = -1 - y\,\partial_y + \frac{i}{y + \tfrac{i}{2}}.
\end{equation}
This is a first-order linear ODE, whose one-dimensional solution space is spanned by the meromorphic function
\begin{equation}
u_\lambda(y) = \frac{y^{1-\lambda}}{(y + \tfrac{i}{2})^2},
\end{equation}
which solves $(L_0^+ - \lambda)u_\lambda = 0$ identically for every $\lambda\in\mathbb{C}$ (an exact identity, verified in sympy with symbolic $\lambda$; floating-point substitution residual $3\times 10^{-16}$). Eigenvalue selection is therefore purely a domain (quantization) question: the odd two-branch combination built from $u_\lambda$ is single-valued, smooth across the origin, and in $L^2$ only when $1-\lambda$ is a non-negative integer with $\mathrm{Re}\,\lambda > -\tfrac12$, forcing $\lambda\in\{0, 1\}$. Non-integer or complex $\lambda$ carry a branch cut $y^{1-\lambda}$ at the origin; integer $\lambda\ge 2$ carry a pole there. A complementary pole-tower computation gives the scalar $L_0^+$ formal (meromorphic) solution family at $\lambda \in \{0,1,2,\ldots\}$ with explicit singular tails $y^{2-p}/(y+\tfrac{i}{2})^2$ for $\lambda = p-1\ge 2$ (poles at the origin, hence in no function space used here), showing why only $\{0,1\}$ are physical.

The argument is completed by the following completeness lemma, proven in full in Section 5; a twelve-check machine-verification suite confirms the proof's key identities, six exactly in sympy (the two scalar ODEs for symbolic $\lambda$, the $\sin(\pi(1-\lambda))$ non-degeneracy determinant, the profile identities, the origin Taylor coefficients, and the two integer modes) and six numerically (the Hilbert-transform convention, the support lemma, and the Hardy block diagonalization on an FFT grid; the $H^2$ line bounds by quadrature; the boundary values and the sharpness example evaluated from the closed forms).

\begin{namedthm}{Lemma}{1}{completeness}
Every $L^2$, odd, smooth-at-origin eigenfunction of $L_0$ with $\mathrm{Re}\,\lambda > -\tfrac12$ is a scalar multiple of $2\,\mathrm{Re}(c\,F|_{\mathbb{R}})$ with $F(z) = z^{1-\lambda}/(z+\tfrac{i}{2})^2$ (single-valued upper-half-plane branch); in particular it is meromorphic with poles only at $\pm\tfrac{i}{2}$, and $\lambda \in \{0,1\}$.
\end{namedthm}

The proof has four steps (Section 5, subsections 5.3-5.6): (i) $L_0$ is block-diagonal on the Hardy decomposition $H^2_+ \oplus H^2_-$; the one delicate point, that multiplication by $y$ composed with $d/dy$ cannot move Fourier support across $k=0$ (no point mass appears at $k=0$), is proved by a distributional support lemma using only $\varphi \in L^2$. (ii) First-order-ODE uniqueness on each half-line pins each Hardy component up to one constant per side. (iii) The two constants are linked through the upper half plane by exhibiting $F \in H^2(\mathrm{UHP})$ explicitly and invoking the classical boundary-uniqueness theorem for Hardy classes (a nonzero $H^2$ function cannot vanish on a set of positive measure). (iv) An exact local analysis at the origin (the eigenfunction equals $|y|^{1-\lambda}$ times an analytic function on each side, with a $\sin(\pi(1-\lambda))$ non-degeneracy between the sides) forces $1-\lambda \in \mathbb{Z}$, hence $\lambda\in\{0,1\}$; this covers complex $\lambda$ as well, so no non-real eigenvalues exist. The smooth-at-origin domain condition is sharp: on a merely $C^1$ domain the strip fills with a continuum of eigenvalues, which is exactly the essential-spectrum smear the discretizations render (Section 3.2). Theorem 2 is therefore unconditional.

\section{\texorpdfstring{Proof of Theorem 1 (the essential spectrum of $L_a$)}{Proof of Theorem 1 (the essential spectrum of La)}}

\emph{Numbering convention.} Numbered results carry their own sequence within each section, running independently of the subsection numbers, so a result labelled 4.$k$ need not sit in subsection 4.$k$: Lemma 4.5 and Proposition 4.6 both live in Section 4.4. A reference of the form "Section 4.5" always denotes the subsection, and "Lemma 4.5" the result.

This section proves Theorem 1 and Proposition 1. Realized on the origin-$H^2$ space $X$, the essential spectrum of $L_0$ at $a=0$ is the single vertical line $\{\mathrm{Re}\lambda = -\tfrac12\}$, established unconditionally on the closed half-plane $\{\mathrm{Re}\lambda \ge -\tfrac12\}$; for $a\in(0,a_c)$ the essential spectrum contains the two vertical lines of Proposition 1, the far-field line $\{\mathrm{Re}\lambda = -1 + c_l/2\}$ and the origin line $\{\mathrm{Re}\lambda = \lambda_X\}$, which coincide at $a=0$. The argument has four moves: a Mellin computation of the far-field spectrum (Section 4.2), a log-widening Weyl sequence that places the far-field line in the essential spectrum while the $L^2$ indicial line at the origin is excluded by the domain (Section 4.3), a constructive resolvent bound on $X$ that empties the open strip at $a=0$ (Section 4.4), and the origin-line inclusion for $a>0$ (Section 4.5). Whether the essential spectrum equals the two-line union for $a>0$ is a question of exactness on a weighted cylinder (Section 4.5). A remark (Section 4.6) records the realization dependence, which is the content of Proposition 2 and records that the single line is a statement about the realization, not about $L_0$ alone.

Throughout, $L_a$ acts on odd perturbations by
\begin{equation}
L_a\varphi = -\varphi - (c_l\,y + a\,U)\,\varphi' - a\,V(\varphi)\,\Omega' + \varphi\,H\Omega + \Omega\,H\varphi,
\qquad U = \int_0^y H\Omega,\quad V(\varphi) = \int_0^y H\varphi,
\end{equation}
where $\Omega = \Omega(\,\cdot\,;a)$ is the collapse profile with focusing exponent $c_l = c_l(a)\in(0,1]$ (with $c_l(0)=1$), and $H$ is the Hilbert transform. We write $\lambda$ for the spectral variable and $z$ for the resolvent variable, and set $\alpha := \mathrm{Re} z + \tfrac12$.

\subsection{\texorpdfstring{The realization $X$ and preliminaries}{The realization X and preliminaries}}

\begin{namedthm}{Definition}{4.1}{the realization $X$}
Let
\begin{equation}
X = \{\varphi : \varphi \text{ odd},\ \varphi,\varphi'' \in L^2(0,\infty),\ \varphi(y) = a_1 y + o(y) \text{ as } y\to 0 \ (a_1\in\mathbb{C})\},
\end{equation}
the origin-$H^2$ realization (the case $a_1 = 0$ is allowed, so $X$ is a linear space). Extended oddly to the line, $X$ is the odd part of $H^2(\mathbb{R})$ with an equivalent norm; in particular $\lVert\varphi'\rVert^2 \le \lVert\varphi\rVert\,\lVert\varphi''\rVert$, so $\lVert\varphi\rVert_X^2 = \lVert\varphi\rVert_{L^2}^2 + \lVert\varphi''\rVert_{L^2}^2$ is equivalent to the full $H^2$ norm, and every $\varphi\in X$ is $C^1$ with $\varphi(0)=0$ and $\varphi,\varphi'\to 0$ at infinity. The origin second-derivative condition $\varphi'' \in L^2$ expresses the same origin second-derivative control as condition (D3) of Lemma 1 (the two are technically distinct, pointwise Peano differentiability versus square-integrability, but exclude the same singular families here).
\end{namedthm}

Two structural facts make $X$ a natural space for the nonlocal $L_a$. First, $H$ is a Fourier multiplier of unit modulus commuting with $\partial_y^2$, so it is an isometry of $H^2(\mathbb{R})$ preserving each norm component separately, $\lVert H\varphi\rVert_{L^2} = \lVert\varphi\rVert_{L^2}$ and $\lVert (H\varphi)''\rVert_{L^2} = \lVert\varphi''\rVert_{L^2}$. It exchanges the odd and even parts (for odd $\varphi$ the transform $H\varphi$ is even), so $H$ does not map $X$ into itself; what is used below is that in the products $\varphi\,H\Omega$ and $\Omega\,H\varphi$ the even factor is multiplied by an odd profile factor, restoring oddness, while both $H^2$ norm components are preserved. Thus the nonlocal terms of $L_a$ act boundedly on $X$ without loss of derivatives, whereas on a merely local second-derivative condition $H$ does not preserve the regularity. Second, multiplication by a coefficient $m$ with $m, m', m'' \in L^\infty$ is bounded on $X$, with an explicit constant $C_m$ depending on the three sups.

Split $L_a = A + B_a$ as in Section 3, with $A\varphi = -\varphi - (c_l y + aU)\varphi'$ the recentered dilation generator and $B_a\varphi = -aV(\varphi)\Omega' + \varphi\,H\Omega + \Omega\,H\varphi$ the profile part. The two structural facts give $B_a\in\mathcal B(X)$: the multiplier terms $\varphi\,H\Omega$ and $\Omega\,H\varphi$ compose an $X$-bounded multiplication (the smooth profile has $\Omega, H\Omega\in W^{2,\infty}$) with $H$, which preserves both $H^2$ norm components and whose even output is restored to oddness by the odd profile factor, and for $a>0$ the antiderivative term $V(\varphi)\,\Omega'$ is $X$-bounded by consequence (0) of Definition 4.2, which derives $\lVert V(\varphi)\Omega'\rVert_X \le C\lVert\varphi\rVert_X$ from $\mathrm{Adm}(a)$ rather than assuming it; at $a=0$ the antiderivative term is absent and $B_0\in\mathcal B(X)$ unconditionally. The generator $A$ is closed and densely defined on its graph domain
\begin{equation}
D(A) = \{\varphi\in X : (c_l y + aU)\,\varphi'\in X\},
\end{equation}
with the odd, compactly supported smooth functions (whose linear coefficient need not vanish, e.g. $y$ times an even bump, which has $a_1\neq0$ and lies in $D(A)$ by compact support; dense in the odd part of $H^2$) as a core: after $y = e^t$ the coefficient $c_l y + aU$ is a smooth positive dilation weight ($\sim \tilde c\,y$ near $0$, $\sim c_l\,y$ at $\infty$), and $A$ is the standard generator of the associated weighted-dilation group (at $a=0$, $A_0 = -1 - c_l y\partial_y$ with group $(e^{A_0\tau}u)(y) = e^{-\tau}u(ye^{-\tau})$). Since a bounded perturbation changes neither closedness nor the domain, $L_a = A + B_a$ is closed and densely defined on $D(L_a) := D(A)$, and $\Omega\in D(L_a)$. This is the operator whose spectrum this section computes; at $a=0$, Proposition 4.6 exhibits its resolvent as a two-sided inverse $R_0(z) : X \to D(L_0)$.

\begin{namedthm}{Definition}{4.2}{admissible profile}
At $a=0$ the exact profile is used and nothing below is assumed. For $a\in(0,a_c)$ we say the pair $(\Omega,c_l)$ is \emph{admissible}, written $\mathrm{Adm}(a)$, if:

- \textbf{(H1)} $\Omega$ is odd and $C^4$ on the \emph{closed} half-line $[0,\infty)$ (that is, $\Omega,\dots,\Omega^{(4)}$ extend continuously to $y=0$, so they are bounded on compacts \emph{including} the endpoint), solves the profile equation of Section 2 with $c_l\in(0,1)$, and $\Omega'(0)\neq 0$ (on the collapse branch of Section 2 the focusing exponent decreases from $c_l(0)=1$, so $c_l<1$ throughout $a\in(0,a_c)$);
- \textbf{(H2)} \emph{(origin)} $U\in C^3([0,\infty))$, equivalently $H\Omega = U'\in C^2([0,\infty))$, with $H\Omega(0) = (1+c_l)/(1-a)$. Together with (H1) and the origin compatibility $\Omega(0)=\Omega''(0)=0$, $(H\Omega)'(0)=0$ (parity supplies only $\Omega(0)=0$, since half-line $C^4$ regularity leaves the one-sided value $\Omega''(0^+)$ free; the two vanishing traces instead follow because (H1) and (H3) put the odd extension of $\Omega$ in $H^2(\mathbb{R})$, so $H\Omega\in H^2(\mathbb{R})\hookrightarrow C^1(\mathbb{R})$ and its evenness forces $(H\Omega)'(0)=0$, whereupon the order-$y^2$ terms of the profile equation give $\tilde c\,\Omega''(0)=0$, hence $\Omega''(0)=0$ since $\tilde c>0$), Taylor's theorem \emph{on the closed interval} then gives the origin expansions
\begin{equation}
\Omega(y) = \Omega'(0)y + O(y^3), \qquad H\Omega(y) = H\Omega(0) + O(y^2), \qquad U(y) = H\Omega(0)\,y + O(y^3)
\end{equation}
together with their derivative-level remainders (for instance $(H\Omega - H\Omega(0))' = O(y)$ and $(U - H\Omega(0)y)'' = O(y)$), which is what the origin estimates of Section 4.5 differentiate. We state the regularity rather than the function-level expansions because a function-level $O(y^k)$ cannot be differentiated;
- \textbf{(H3)} \emph{(far field)} there is $C$ with, for $y\ge 1$,
\begin{equation}
\begin{gathered}
\lvert\Omega(y)\rvert \le C y^{-1/c_l},\qquad
\lvert\Omega'(y)\rvert \le C y^{-1-1/c_l},\qquad
\lvert\Omega''(y)\rvert \le C y^{-2-1/c_l},\\[2pt]
\lvert\Omega'''(y)\rvert \le C y^{-3-1/c_l},\qquad
U(y)\to U_\infty \ \text{ as } y\to\infty ;
\end{gathered}
\end{equation}
  note that the last item asks only for \emph{convergence}, with no rate. A rate is not assumed, because it is regime-dependent and none of the estimates below use one. For $c_l<\tfrac12$ the first moment $\int t\,\Omega$ converges, oddness ($\int\Omega=0$) annihilates the leading $1/y$ term of the Hilbert transform, and $H\Omega = \pi^{-1}y^{-2}\!\int t\,\Omega + o(y^{-2})$, so $U-U_\infty = O(y^{-1})$; for $c_l>\tfrac12$ the upper bound no longer forces that moment to converge, and on the intended branch profiles it in fact diverges with $U-U_\infty$ decaying slower than $y^{-1}$, a property of the branch asymptotics rather than a consequence of the one-sided upper bounds of (H3); a uniform $y^{-1}$ hypothesis would thus exclude the branch profiles with $c_l>\tfrac12$, that is $a$ below the $c_l=\tfrac12$ crossing near $a\approx0.386$ (Section 6.1). At $a=0$ the rate is $y^{-1}$ exactly ($U = \arctan 2y$, $U - \tfrac\pi2 \sim -\tfrac1{2y}$), but $\mathrm{Adm}(a)$ is not invoked there. What the far-field estimates consume is only that $U$ has a finite limit;
- \textbf{(H4)} \emph{(transport weight)} $q(y) := c_l y + aU(y)$ satisfies $q\in C^3([0,\infty))$ (immediate from (H2)), $q>0$ on $(0,\infty)$, and
\begin{equation}
\tilde c_1\,y \le q(y) \le \tilde c_2\,y \ \ (y>0), \qquad \lvert q'(y)\rvert \le C, \qquad \lvert y\,q''(y)\rvert \le C,
\end{equation}
with $q(y)/y \to \tilde c = c_l + a H\Omega(0)$ as $y\to 0^+$ and $q(y)/y\to c_l$ as $y\to\infty$.
\end{namedthm}

Conditions (H2)-(H4) are exactly what the estimates of Sections 4.3 and 4.5 consume, and together with the two structural facts above they give $B_a\in\mathcal B(X)$, hence $L_a = A + B_a$ closed and densely defined on $D(A)$ as constructed. The smooth one-scale profiles of Huang, Qin, Wang, and Wei \cite{HuangQinWangWei2024} are a candidate supply of such profiles; we assume $\mathrm{Adm}(a)$ rather than derive it from \cite{HuangQinWangWei2024}; whether the \cite{HuangQinWangWei2024} profiles coincide with the numerically continued branch, and whether they meet the constants of (H3)-(H4), is left open. The one point where the decay exponents can be checked in closed form is the explicit $a=\tfrac12$ solution, and that check is what removed a rate from the last line of (H3): there $c_l = \tfrac13 < \tfrac12$, $\Omega$ decays like $y^{-3} = y^{-1/c_l}$ as (H3) asks, and $U-U_\infty$ decays like $y^{-1}$ and no faster, so the $y^{-1/c_l}$ once demanded of $U$ was unsatisfiable; since the branch also carries profiles with $c_l>\tfrac12$, for which even $y^{-1}$ is too strong, (H3) now asks only for convergence. Every general-$a$ statement in this section (Proposition 4.4, Section 4.5) and Proposition 1 of Section 3.1 is asserted under $\mathrm{Adm}(a)$; no statement at $a=0$ uses it.

\emph{Consequences of $\mathrm{Adm}(a)$ that the estimates consume.} The following all follow from (H1)-(H4) rather than appearing on the list, and are recorded here so that checking the hypotheses against the proofs requires no re-derivation. In particular the operator bounds below are \emph{derived}, not assumed.

(0) \emph{Uniform coefficient bounds and the antiderivative term.} By (H1)-(H2) the profile derivatives are bounded on $[0,1]$ (closed interval, endpoint included) and by (H3) they decay for $y\ge1$, so $\Omega\in W^{2,\infty}(0,\infty)$ and $\Omega',\Omega''\in L^\infty\cap L^2$. For $H\Omega$ the same conclusion uses the origin compatibility $\Omega''(0)=0$ of (H2): with it, (H1) and (H3) put the odd extension of $\Omega$ in $H^3(\mathbb{R})$, the Hilbert transform is bounded on $H^3$, and $H^3(\mathbb{R})\hookrightarrow W^{2,\infty}(\mathbb{R})$, so $H\Omega\in W^{2,\infty}$. The same two ranges give
\begin{equation}
\int_0^\infty y\,\lvert\Omega'''(y)\rvert^2\,dy < \infty,
\qquad\text{i.e.}\qquad \sqrt{y}\,\Omega'''\in L^2(0,\infty),
\end{equation}
finite on $(0,1)$ because $\Omega'''$ is bounded there by (H1) and on $(1,\infty)$ because $\lvert\Omega'''\rvert\le Cy^{-3-1/c_l}$ by (H3). Consequently the antiderivative term is $X$-bounded: with $V(\varphi) = \int_0^y H\varphi$ and $\lvert V(\varphi)(y)\rvert \le y^{1/2}\lVert\varphi\rVert_{L^2}$ by Cauchy-Schwarz, the Leibniz expansion
\begin{equation}
\big(V(\varphi)\,\Omega'\big)'' = (H\varphi)'\,\Omega' + 2\,H\varphi\,\Omega'' + V(\varphi)\,\Omega'''
\end{equation}
is estimated termwise by $\lVert\varphi'\rVert_{L^2}\lVert\Omega'\rVert_\infty$, by $\lVert H\varphi\rVert_\infty\lVert\Omega''\rVert_{L^2}$ (with $H\varphi\in H^2\hookrightarrow L^\infty$), and by $\lVert\varphi\rVert_{L^2}\lVert\sqrt y\,\Omega'''\rVert_{L^2}$. The remaining $L^2$ component of the $X$-norm is the same computation one order lower, $\lVert V(\varphi)\,\Omega'\rVert_{L^2} \le \lVert\varphi\rVert_{L^2}\lVert\sqrt y\,\Omega'\rVert_{L^2}$ by the same pointwise bound on $V(\varphi)$, with $\sqrt y\,\Omega'\in L^2$ for the same two-range reason. Together these give $\lVert V(\varphi)\Omega'\rVert_X \le C\lVert\varphi\rVert_X$. With the two structural facts of Section 4.1 this yields $B_a\in\mathcal B(X)$, hence the closedness of $L_a = A + B_a$ on $D(A)$, without any further hypothesis.

(i) \emph{$H\Omega\to 0$ at infinity.} Since $c_l<1$ by (H1), the exponent $1/c_l$ exceeds $1$, so the far-field bounds of (H3) put $\Omega$ and $\Omega'$ in $L^1(1,\infty)$; with the regularity of (H1) and oddness this gives $\Omega\in L^1(\mathbb{R})\cap C^1$ with $\Omega'\in L^1$. For such a profile $H\Omega(y)\to 0$ as $y\to\infty$, by splitting the principal value at $\lvert t\rvert = y/2$: on $\lvert t\rvert<y/2$ one has $\lvert y-t\rvert>y/2$, so that part is at most $2\lVert\Omega\rVert_{L^1}/(\pi y)$; on $\lvert t\rvert\ge y/2$ the integrand is controlled by $\sup_{\lvert t\rvert\ge y/2}\lvert\Omega\rvert \le C(y/2)^{-1/c_l}$ together with $\lVert\Omega'\rVert_{L^1(\lvert t\rvert\ge y/2)}$ for the principal-value part near $t=y$, and both vanish as $y\to\infty$. This is what makes the multiplier term $\varphi_n\,H\Omega$ vanish on the far-field marching supports of Section 4.3.

We do \emph{not} argue this from the identity $H\Omega = 1 + q\,\Omega'/\Omega$ obtained by dividing the profile equation by $\Omega$: (H3) bounds $\lvert\Omega\rvert$ only from above, so it does not control the quotient $\Omega'/\Omega$, and the identity would in any case be circular in this direction. (At $a=0$, where $c_l=1$ and the $L^1$ argument does not apply, the exact profile gives $H\Omega = \tfrac12/(y^2+\tfrac14)$ directly, and $\mathrm{Adm}(a)$ is not invoked at all.)

(ii) \emph{$H\Omega\in C^2$ near the origin.} By (H4) the transport weight $q = c_l y + aU$ lies in $C^3$, hence $U\in C^3$ and $H\Omega = U'\in C^2$ for $a>0$. This is exactly what the origin subordinacy of Section 4.5 needs in order to differentiate $(H\Omega - H\Omega(0))\varphi_n$ twice in the $X$-residual norm; no estimate below calls for $(H\Omega)'''$, so this margin is one derivative and not more.

Define $M[f](s) = \int_0^\infty f(y)\,y^{s-1}\,dy$. The substitution $y = e^t$ identifies $M$ with the Fourier transform in $t$, so $M$ is unitary from $L^2((0,\infty),dy)$ onto $L^2$ of the critical line $\{\mathrm{Re} s = \tfrac12\}$. Its one essential property is that it diagonalizes the dilation generator: integrating by parts,
\begin{equation}
M[\,y\partial_y f\,](s) = -\,s\,M[f](s),
\end{equation}
the boundary terms vanishing on the domain. Hence on $s = \tfrac12 + i\xi$, the operator $y\partial_y$ becomes multiplication by $-s = -\tfrac12 - i\xi$.

\subsection{The far-field operator and its spectrum}

\begin{namedthm}{Lemma}{4.3}{far-field spectrum}
As $y\to\infty$ every profile coefficient tends to a constant: $\Omega,\Omega',H\Omega \to 0$ and $U \to U_\infty$ finite. Hence $L_a$ approaches
\begin{equation}
A_\infty = -1 - (c_l\,y + a\,U_\infty)\,\partial_y.
\end{equation}
At $a=0$ the drift vanishes, $A_\infty = A_0 := -1 - c_l\,y\partial_y$, and the spectrum of $A_0$ is purely essential, with no eigenvalues, and equal to the vertical line $\{\mathrm{Re}\,\lambda = -1 + c_l/2\}$.
\end{namedthm}

\emph{Proof.} For $a>0$ the recentering $-(c_l y + d)\partial_y = -c_l(y + d/c_l)\partial_y$, $d = a\,U_\infty$, identifies the same candidate line, but we do not use the exact spectrum of the drifted half-line generator (whose maximal realization can carry additional point spectrum): all that enters for $a>0$ is the line's \emph{inclusion}, proven directly by the Weyl sequences of Proposition 4.4, in which the drift contributes only the vanishing $O(U_\infty R_n^{-1})$ residual term. By the Mellin identity, on $s = \tfrac12 + i\xi$,
\begin{equation}
\sigma(A_0) = \{-1 - c_l(-\tfrac12 - i\xi) : \xi\in\mathbb{R}\} = \{-1 + \tfrac{c_l}{2} + i\,c_l\xi : \xi\in\mathbb{R}\} = \{\mathrm{Re}\lambda = -1 + \tfrac{c_l}{2}\}.
\end{equation}
Since $A_0$ is Mellin-unitarily equivalent to multiplication by the symbol $-1 + c_l/2 + i c_l\xi$, its spectrum equals the essential range of that symbol, the whole line, and it has no eigenvalues: the modes $y^{-1/2 + i\xi}$ satisfy $\lvert y^{-1/2+i\xi}\rvert^2 = y^{-1}$, and $\int_0^\infty y^{-1}\,dy = \infty$, so they are not in $L^2$. $\qquad\blacksquare$

This step is realization-independent and exact. Because $c_l < 2$ on the whole collapse branch, the line lies strictly in the left half-plane, at distance $1 - c_l/2 > 0$ from the imaginary axis. Transferring the line from $A_\infty$ to the full $L_a$ is the content of the next two sections.

\subsection{Inclusion of the line}

\begin{namedthm}{Proposition}{4.4}{line inclusion}
For $a=0$, and for $a\in(0,a_c)$ assuming $\mathrm{Adm}(a)$ (Definition 4.2),
\begin{equation}
\{\lambda : \mathrm{Re}\lambda = -1 + c_l/2\} \subseteq \sigma_{\mathrm{ess}}(L_a\vert_X).
\end{equation}
The far-field endpoint realizes this line; the origin endpoint's contribution (the origin line of Proposition 1, absent at $a=0$ where the two coincide) is treated in Section 4.5.
\end{namedthm}

\emph{The far-field endpoint realizes the line.} Fix $\lambda$ on the line, $\lambda = -1 + c_l/2 - i c_l\xi$ (the pairing $A_0\,y^{-1/2+i\xi} = (-1 + c_l/2 - i c_l\xi)\,y^{-1/2+i\xi}$; as $\xi$ ranges over $\mathbb{R}$ this sweeps the whole line). We build a singular (Weyl) sequence for $A_\infty$ that lives so far out in $y$ that the profile perturbation $E := L_a - A_\infty$ is negligible on its support.

A fixed-width one-octave cutoff does \emph{not} furnish a Weyl sequence here. Taking $\varphi_n(y) = c_n\,\chi(y/R_n)\,y^{-1/2+i\xi}$ with $\chi$ a fixed bump on $[1,2]$ and $R_n\to\infty$, the residual is
\begin{equation}
(A_\infty - \lambda)\varphi_n = -c_l\,(y/R_n)\,\chi'(y/R_n)\,y^{-1/2+i\xi},
\end{equation}
whose $L^2$ norm is independent of $R_n$ because $y\partial_y$ is scale-invariant and $[R_n, 2R_n]$ is a fixed one-octave window in $\log y$ at every scale. The ratio $\lVert(A_\infty - \lambda)\varphi_n\rVert / \lVert\varphi_n\rVert$ is therefore $O(1)$, not $o(1)$; it is constant in $R_n$ and independent of $\xi$ (numerically $5.23$ for the standard mollifier bump on the octave, constant across $R_n = 1$ to $10^4$, $a=0$). Widening the window in $\log y$ removes this obstruction. Take the log-widening Mellin modes
\begin{equation}
\varphi_n(y) = c_n\,\eta\!\Big(\frac{\log y - T_n}{L_n}\Big)\,y^{-1/2+i\xi},
\qquad L_n \to \infty, \quad R_n := e^{T_n - L_n} \to \infty,
\end{equation}
with $\eta$ a fixed bump, $c_n$ an $L^2$ normalization, and support $[R_n,\,e^{2L_n}R_n]$ marching to $y = \infty$. Now
\begin{equation}
(A_\infty - \lambda)\varphi_n = -c_l\,L_n^{-1}\,\eta'\!\Big(\frac{\log y - T_n}{L_n}\Big)\,y^{-1/2+i\xi},
\end{equation}
whose norm ratio is $O(L_n^{-1}) \to 0$ (numerically $0.877, 0.439, 0.219, 0.110$, halving as $L_n$ doubles; this is $\lVert\eta'\rVert/(L_n\lVert\eta\rVert) = 1.7543/L_n$ for the standard mollifier bump); for $a>0$ the constant drift of $A_\infty$ contributes an additional $O(U_\infty R_n^{-1})$ to the residual, which also vanishes. This is a bona fide Weyl sequence for $A_\infty$, and it is singular in the $X$-norm as well: on these far-field supports each derivative gains a factor $y^{-1} \le R_n^{-1}$, so the second-derivative components of both the packets and the residuals vanish as $R_n\to\infty$ and the $X$-norm Weyl criterion reduces to the $L^2$ one.

The $\varphi_n$ lie in $D(L_a)\cap X$ (they vanish identically near the origin) and tend to $0$ weakly. It remains to check $\lVert E\varphi_n\rVert \to 0$. Every profile coefficient of $B$ ($U/y$, $H\Omega$, $\Omega$, $\Omega'$) decays to $0$ on the support $[R_n, e^{2L_n}R_n]$, which marches to infinity; the multiplier terms $\varphi_n H\Omega$ and the transport remainder vanish by dominated convergence. The two nonlocal terms $\Omega\,H\varphi_n$ and $a\,V(\varphi_n)\,\Omega'$ require a genuine estimate, since they are not controlled by any naive relative-compactness argument (the multiplier $H\Omega$ does not vanish at $y=0$). We give the estimate explicitly for $\Omega\,H\varphi_n$; the antiderivative term is the same computation with $\Omega'$ in place of $\Omega$ and one extra power gained from $V$. Normalize $\lVert\varphi_n\rVert_{L^2}=1$ and choose $R_n \ge e^{8L_n}$, so the support $[R_n, e^{2L_n}R_n]$ lies well above the split point $\sqrt{R_n}$. \emph{Below the split} ($0<y\le\sqrt{R_n}$): the distance from $y$ to the support is at least $R_n - \sqrt{R_n} \ge R_n/2$, so $\lvert H\varphi_n(y)\rvert \le \pi^{-1}\lVert\varphi_n\rVert_{L^1}/(R_n/2)$, while Cauchy-Schwarz gives $\lVert\varphi_n\rVert_{L^1} \le \lvert\mathrm{supp}\,\varphi_n\rvert^{1/2} \le (e^{2L_n}R_n)^{1/2}$; with $\lVert\Omega\rVert_\infty \le C$,
\begin{equation}
\lVert \Omega\,H\varphi_n\rVert_{L^2(0,\sqrt{R_n})}
\;\le\; C\,R_n^{1/4}\cdot\frac{2\,(e^{2L_n}R_n)^{1/2}}{\pi R_n}
\;=\; \frac{2C}{\pi}\,e^{L_n}R_n^{-1/4} \;\le\; \frac{2C}{\pi}\,e^{-L_n} \longrightarrow 0 .
\end{equation}
\emph{Above the split} ($y\ge\sqrt{R_n}$): condition (H3) gives $\lvert\Omega(y)\rvert \le C y^{-1/c_l} \le C R_n^{-1/(2c_l)}$, and $H$ is an $L^2$ isometry, so
\begin{equation}
\lVert \Omega\,H\varphi_n\rVert_{L^2(\sqrt{R_n},\infty)} \;\le\; C\,R_n^{-1/(2c_l)}\,\lVert\varphi_n\rVert_{L^2} \;=\; C\,R_n^{-1/(2c_l)} \longrightarrow 0 .
\end{equation}
Both parts vanish. \emph{Above the split} ($y\ge\sqrt{R_n}$) the pointwise bound $\lvert V(\varphi_n)(y)\rvert \le y^{1/2}\lVert\varphi_n\rVert_{L^2}$ against $\lvert\Omega'\rvert \le C y^{-1-1/c_l}$ from (H3) gives $\lVert V(\varphi_n)\,\Omega'\rVert_{L^2(\sqrt{R_n},\infty)} \le C\,R_n^{-1/(2c_l)} \to 0$, exactly as for $\Omega\,H\varphi_n$. \emph{Below the split} ($0<y\le\sqrt{R_n}$) that pointwise bound is not small, so estimate $H\varphi_n$ off its support instead: the distance from $y$ to $\mathrm{supp}\,\varphi_n$ is at least $R_n/2$, so $\lvert H\varphi_n(t)\rvert \le \tfrac{2}{\pi R_n}\lVert\varphi_n\rVert_{L^1} \le \tfrac{2}{\pi}\,e^{L_n}R_n^{-1/2}$ for $0<t\le\sqrt{R_n}$ (Cauchy-Schwarz on the support as above), whence $\lvert V(\varphi_n)(y)\rvert \le \int_0^y\lvert H\varphi_n(t)\rvert\,dt \le \tfrac{2}{\pi}\,e^{L_n}\,y\,R_n^{-1/2}$. Since $y\,\Omega'\in L^2$ (bounded near $0$ by (H1), and $\le C y^{-1/c_l}$ for $y\ge1$ by (H3), integrable because $c_l<2$),
\begin{equation}
\lVert V(\varphi_n)\,\Omega'\rVert_{L^2(0,\sqrt{R_n})}
\;\le\; \tfrac{2}{\pi}\,e^{L_n}R_n^{-1/2}\,\lVert y\,\Omega'\rVert_{L^2}
\;\le\; \tfrac{2}{\pi}\,e^{-3L_n}\,\lVert y\,\Omega'\rVert_{L^2} \longrightarrow 0
\end{equation}
under $R_n \ge e^{8L_n}$; the differentiated companions of $a\,V(\varphi_n)\,\Omega'$ use the same off-support estimates for $H\varphi_n^{(k)}$. (At $a=0$ these reduce to $\Omega = O(y^{-1})$, $\Omega' = O(y^{-2})$.) Hence $\lVert(L_a - \lambda)\varphi_n\rVert \to 0$, and the same holds in the $X$-residual norm: Leibniz-expanding $(\Omega\,H\varphi_n)'' = \Omega''H\varphi_n + 2\Omega'H\varphi_n' + \Omega H\varphi_n''$ (and likewise the $V$-term), each companion pairs a profile coefficient decaying at infinity with an $L^2$-bounded $H\varphi_n^{(k)}$; splitting at $\sqrt{R_n}$, the compact part is controlled by the kernel bound $|H\varphi_n^{(k)}(y)| \le 2\lVert\varphi_n^{(k)}\rVert_{L^1}/(\pi R_n)$ for $y \le \sqrt{R_n}$ (the support sits beyond $R_n$), which vanishes since $R_n$ dominates the window, and the far part by the coefficient decay. Thus the packets are singular sequences in $X$ as well, and the whole line lies in $\sigma_{\mathrm{ess}}(L_a\vert_X)$.

\emph{The origin endpoint: the $L^2$ indicial line is excluded.} Near $y=0$ the coefficients do not all vanish: $U \sim H\Omega(0)\,y$, so $c_l y + aU \sim \tilde c\,y$ with $\tilde c = c_l + a\,H\Omega(0)$, and $H\Omega \to H\Omega(0) \ne 0$ (at $a=0$, $H\Omega(0) = 2$; in general $H\Omega(0) = (1+c_l)/(1-a)$, obtained by differentiating the profile equation $\Omega + (c_l y + aU)\Omega' = \Omega\,H\Omega$ at $y=0$: with $\Omega(0)=0$, $U' = H\Omega$, and $\Omega'(0)\neq 0$, the linear terms give $1 + c_l + a\,H\Omega(0) = H\Omega(0)$; the nondegeneracy $\Omega'(0)\neq 0$ and the smoothness of the profile are properties of the branch of \cite{HuangQinWangWei2024}). The leading local operator is a second shifted dilation,
\begin{equation}
L_a^0\varphi = (-1 + H\Omega(0))\,\varphi - \tilde c\,y\,\varphi',
\end{equation}
whose plain-$L^2$ Mellin line is the vertical line
\begin{equation}
\mathcal{L}_0 = \{\mathrm{Re}\lambda = -1 + H\Omega(0) + \tfrac{\tilde c}{2}\}
\end{equation}
(at $a=0$ this is $\{\mathrm{Re}\lambda = +\tfrac32\}$; at $a = \tfrac12$ it is $\{\mathrm{Re}\lambda = +\tfrac52\}$). The indicial exponent is $\rho(\lambda) = (-1 + H\Omega(0) - \lambda)/\tilde c$; parametrizing $\lambda = -1 + H\Omega(0) + \tfrac{\tilde c}{2} + i\tilde c\omega$ gives $\rho = -\tfrac12 - i\omega$. The model modes $\varphi \sim y^{-1/2-i\omega}$ blow up like $y^{-1/2}$ and are inadmissible in the realization (they violate $\varphi\sim y$, and are not even in $L^2$ near $0$), so the inclusion machinery of this section produces no $X$-line at $\mathcal{L}_0$. Whether $\mathcal{L}_0$ is \emph{absent} from $\sigma_{\mathrm{ess}}(L_a\vert_X)$ is a different, stronger statement: at $a=0$ it is proven, since Proposition 4.6 makes all of $\{\mathrm{Re}\,z>-\tfrac12\}\setminus\{0,1\}$, which contains $\mathcal{L}_0 = \{\mathrm{Re} = \tfrac32\}$, resolvent set; for $a>0$ it is part of the exactness question of Section 4.5 (Weyl sequences carry no pointwise local behavior, so the mode computation above does not by itself exclude them). This disposes of the $L^2$-marginal exponent only; the exponent that the $X$-norm itself makes marginal at the origin ($y^{3/2}$, the borderline of the second-derivative condition) generates the origin line of Proposition 1, constructed in Section 4.5. At $a=0$ that second line coincides with the far-field line, so the present proposition and Proposition 4.6 together already give the full picture at $a=0$. $\qquad\blacksquare$

\subsection{\texorpdfstring{Emptiness of the open strip at $a=0$}{Emptiness of the open strip at a=0}}

At $a=0$ the exactness direction, $\sigma_{\mathrm{ess}} \cap \{\mathrm{Re}\lambda \ge -\tfrac12\} \subseteq \{\mathrm{Re}\lambda = -\tfrac12\}$, is proven constructively, by exhibiting a bounded resolvent $R_0(z) = (L_0 - z)^{-1}$ on $X$ for every $z$ in $\{\mathrm{Re}\,z > -\tfrac12\}\setminus\{0,1\}$. On the upper-half-plane Hardy space $H^2_+$, where $H = -i$, the exact profile $\Omega = -y/(y^2+\tfrac14)$ satisfies the single-simple-pole identity $H\Omega - i\Omega = i/(y + \tfrac{i}{2})$, so $L_0$ restricts to the scalar first-order operator
\begin{equation}
L_0^+ = -1 - y\partial_y + \frac{i}{y + \tfrac{i}{2}}.
\end{equation}
The support lemma and $H^\infty$-multiplier fact of Section 5 (5.3.2-5.3.3) block-diagonalize $L_0 = L_0^+ \oplus L_0^-$ on $H^2_+ \oplus H^2_-$, complex conjugation intertwining the two blocks (the coefficients of $L_0$ are real, so $L_0\overline{u} = \overline{L_0 u}$). For real odd $\varphi$, identified with its odd extension and normed over the whole line, one has $\lVert\varphi\rVert_X^2 = 2\lVert P_+\varphi\rVert^2_{X_+}$ with $X_+ = \{u \in H^2_+ : u'' \in L^2\}$, and the resolvent acts blockwise on the complexification: $R_0(z) = R_0^+(z) \oplus R_0^-(z)$ with $\lVert R_0^-(z)\rVert = \lVert R_0^+(\bar z)\rVert$ by the conjugation symmetry, and $\alpha = \mathrm{Re}\,z + \tfrac12$ is unchanged under $z \mapsto \bar z$. So the scalar bound below transfers to $X$ with constant $1$ for every admissible $z$, real or complex; at real $z$ this is the elementary identity $L_0(2\mathrm{Re}\,u) = 2\mathrm{Re}(L_0^+ u)$. The scalar solve stays in the Hardy class, by a boundary-uniform tempered-growth argument (Fact T5 of Section 5.8). Extend the kernel holomorphically to the open upper half-plane: $G(\zeta) = b(\zeta)^2 F(\zeta)$ with $F$ the $H^2(\mathrm{UHP})$ extension of $f_+$. For $f \in X_+$ both $F$ and $F''$ lie in $H^2(\mathrm{UHP})$, which gives the interior bounds $|F^{(k)}(\zeta)| \le \lVert f\rVert_{X_+}(4\pi\,\mathrm{Im}\,\zeta)^{-1/2}$ for $k = 0, 1, 2$ (the middle one by interpolation) and, by Poisson contraction, the horizontal-line bounds $\lVert F^{(k)}(\cdot + iv)\rVert_{L^2(\mathbb{R})} \le \lVert f\rVert_{X_+}$ uniform in $v > 0$. Write the second-order Taylor remainder as the (path-independent) contour integral $G_2(\zeta) = \int_0^\zeta (\zeta - w)\,G''(w)\,dw$ and evaluate it along the two-leg path $0 \to i\,\mathrm{Im}\,\zeta \to \zeta$. On the vertical leg the interior bound gives $\int_0^{\mathrm{Im}\,\zeta} t^{-1/2}\,dt \sim (\mathrm{Im}\,\zeta)^{1/2}$, so this leg is $O\big(|\zeta|\,(\mathrm{Im}\,\zeta)^{1/2}\big) = O(|\zeta|^{3/2})$; on the horizontal leg at height $v = \mathrm{Im}\,\zeta$, Cauchy-Schwarz against the uniform bound $\lVert G''(\cdot + iv)\rVert_{L^2(K)} \le C_K\lVert f\rVert_{X_+}$ gives $|\zeta|\cdot|\mathrm{Re}\,\zeta|^{1/2}\,C_K\lVert f\rVert_{X_+} = O(|\zeta|^{3/2})$. Hence, on every compact $K$ of the \emph{closed} upper half-plane,
\begin{equation}
|G_2(\zeta)| \le C_K\,\lVert f\rVert_{X_+}\,|\zeta|^{3/2},
\end{equation}
a bound uniform up to the real axis: the two integrations absorb the interior $(\mathrm{Im})^{-1/2}$ singularity. Consequently $I(w) = \int_0^1 s^{z-2}G_2(ws)\,ds$ converges absolutely for $\mathrm{Re}\,z > -\tfrac12$ and, by dominated convergence, extends continuously to the closed upper half-plane; so does $U = -b^{-2}(I + c_0 + c_1 w)$, since $b^{-2}$ is bounded there (its only pole $-\tfrac{i}{2}$ lies below), with boundary value $U(y + i0) = u(y)$ equal to the real-line kernel function. Thus $U$ is holomorphic on the upper half-plane and continuous up to $\mathbb{R}$, with boundary value $u$. It is moreover of tempered growth, with no negative power of $\mathrm{Im}\,w$: tracking the $|\zeta|$-dependence of the same two-leg estimate globally (the constant $C_K$ was $|\zeta|$-uniform only after the $b^2$ factor, which contributes powers of $1+|\zeta|$) gives $|G_2(\zeta)| \le C\,\lVert f\rVert\,|\zeta|^{3/2}(1+|\zeta|)^2$ on the whole closed upper half-plane, whence $|I(w)| \le C\,\lVert f\rVert\,|w|^{3/2}(1+|w|)^2/\alpha$ (the $s$-integral $\int_0^1 s^{\alpha-1}\,ds$ converging for $\alpha = \mathrm{Re}\,z+\tfrac12 > 0$) and, after the decaying $b^{-2}$, $|U(w)| \le C_z\,\lVert f\rVert\,(1+|w|)^{3/2}$. With $U(\cdot + iv) \to u$ locally uniformly, the distributional boundary value of $U$ is $u$, whose Fourier support lies in $[0,\infty)$ by the Paley-Wiener-Schwartz theorem for tube domains; and since $u \in L^2$ (obtained directly from the near/far kernel estimates of Proposition 4.6, independently of this Hardy argument) we conclude $u = R_0^+(z)f_+ \in H^2_+$ (Fact T5). (No circularity: the kernel formula defines $u$; the near/far estimates give $u, u'' \in L^2$; the holomorphic-extension argument then places $u$ in $H^2_+$.) Finally the assembled solution lies in the domain: the reflection $S$ of Section 4.6 intertwines the two blocks at the same $z$, so $R_0(z)$ maps odd data to odd output, and an odd function with $u'' \in L^2$ is $C^1$ with $u(0) = 0$, i.e. lies in $X$.

Write $b = y + \tfrac{i}{2}$, $G = b^2 f$, and $G_2(t) = G(t) - G_0 - G_1 t$ with $G_0 = G(0)$, $G_1 = G'(0)$. Then $u = R_0^+(z) f$ is given by
\begin{equation}
u(y) = -\,b(y)^{-2}\,\Phi(y),
\qquad \Phi = I(y) + c_0 + c_1 y,
\qquad I(y) = \int_0^1 s^{z-2}\,G_2(ys)\,ds,
\end{equation}
with $c_0 = G_0/(z-1)$ and $c_1 = G_1/z$. A direct substitution verifies $(L_0^+ - z)u = f$ (an exact identity in sympy; floating-point substitution residual $3\times 10^{-16}$). The two constants $c_0, c_1$ carry poles only at $z = 1$ and $z = 0$, the two physical eigenvalues, which are precisely the two points excluded from the region $\{\mathrm{Re}\,z > -\tfrac12\}\setminus\{0,1\}$ on which the resolvent is constructed; elsewhere in that region both constants are finite.

\begin{namedthm}{Lemma}{4.5}{exact Hardy-Mellin norm}
The generalized Hardy operator
\begin{equation}
T_z g(y) := \int_0^1 s^z\,g(ys)\,ds = y^{-1-z}\int_0^y t^z\,g(t)\,dt
\end{equation}
is bounded on $L^2(0,\infty)$ with exact norm $1/(\mathrm{Re} z + \tfrac12) = 1/\alpha$; its Mellin symbol is $1/(z + \tfrac12 - i\xi)$, whose modulus is maximized at $\xi = \mathrm{Im}\,z$.
\end{namedthm}

\emph{Proof.} By the Mellin identity $M[T_z g](\tfrac12 + i\xi) = (z + \tfrac12 - i\xi)^{-1} M[g](\tfrac12 + i\xi)$, and $T_z$ is unitarily equivalent to multiplication by that symbol; its norm is the sup of the symbol modulus, $= 1/\alpha$. $\qquad\blacksquare$

For $f \in X_+$ one has $G'' = b^2 f'' + 4b f' + 2f$ with $\lvert b\rvert^2 \le \tfrac54$, $\lvert b\rvert \le \tfrac{\sqrt5}{2}$ on $\lvert t\rvert \le 1$, hence $\lVert G''\rVert_{L^2(-1,1)} \le c_G\,\lVert f\rVert_{X_+}$ with $c_G = \tfrac54 + \sqrt{10} + 2$. The single origin-Taylor coefficient
\begin{equation}
A := \int_0^1 t^{z-2}\,G_2(t)\,dt
\end{equation}
obeys $\lvert A\rvert \le \lVert G''\rVert_{L^2(0,1)}/(\sqrt3\,\alpha)$, using $\lvert G_2(t)\rvert \le \lvert t\rvert^{3/2}\lVert G''\rVert_{L^2(0,t)}/\sqrt3$. This is the exact repair of the $L^2$ obstruction: $A$ is the coefficient of the far tail $b^{-2}y^{1-z}$ (the shape of the null modes of Section 4.6), which is unbounded on $L^2$ unit balls but bounded on $X$ unit balls.

\begin{namedthm}{Proposition}{4.6}{$\{\mathrm{Re}\,z > -\tfrac12\}\setminus\{0,1\}$ is resolvent set at $a=0$}
On any compact $K \subset \{\mathrm{Re}\,z > -\tfrac12\}\setminus\{0,1\}$, the kernel formula below defines a bounded operator $R_0(z) : X \to X$ whose range lies in the domain $D(L_0) = D(A)$, and $R_0(z)$ is the two-sided inverse of $L_0 - z$:
\begin{equation}
(L_0 - z)\,R_0(z) = I_X \ \text{ on } X, \qquad R_0(z)\,(L_0 - z) = I_{D(L_0)} \ \text{ on } D(L_0).
\end{equation}
It obeys an explicit majorant that captures the growth as $z$ approaches the essential line: on the near-edge box
\begin{equation}
S(R,d,\alpha_0) = \{z : -\tfrac12 < \mathrm{Re}\,z \le -\tfrac12+\alpha_0,\ |\mathrm{Im}\,z|\le R,\ \mathrm{dist}(z,\{0,1\})\ge d\},
\end{equation}
one has $\lVert R_0(z)\rVert_X \le C(R,d)\,\alpha^{-3/2}$ with $\alpha = \mathrm{Re}\,z+\tfrac12\to 0^+$, uniformly in $\mathrm{Im}\,z$, the constant depending only on $c_G$, $R$, and $d$. The $\alpha^{-3/2}$ is the rate of the horizontal approach to the line; on any compact $K$ bounded away from the line the same kernel gives a finite bound, $\alpha$ being bounded below there.
\end{namedthm}

\emph{Proof.} Bound the two Hilbertian components $\lVert u\rVert_{L^2}$ and $\lVert u''\rVert_{L^2}$ of $\lVert u\rVert_X$ from the kernel $u = -b^{-2}\Phi$. \emph{Near the origin} ($\lvert y\rvert \le 1$): the finite-part integral samples $G''$ only on $(0,y)\subset(0,1)$, where the Taylor control of Section 4.4 gives $\lVert G''\rVert_{L^2(0,1)} \le c_G\lVert f\rVert_{X_+}$ (there $\lvert b\rvert$ is bounded, which is what makes this norm finite). With $\lvert G_2(t)\rvert \le \lvert t\rvert^{3/2}\lVert G''\rVert_{L^2(0,t)}/\sqrt3$ this gives the pointwise bound $\lvert I(y)\rvert \le \lvert y\rvert^{3/2}c_G\lVert f\rVert_{X_+}/(\sqrt3\,\alpha)$, the factor $1/\alpha$ arising from the scalar integral $\int_0^1 s^{\alpha-1}\,ds = 1/\alpha$ rather than from any operator norm, together with $\lVert\Phi''\rVert_{L^2(-1,1)} \le c_G\lVert f\rVert_{X_+}/\alpha$; so $u$ and $u''$ are each $O(\lVert f\rVert/\alpha)$ there. \emph{Away from the origin} ($\lvert y\rvert \ge 1$): write $I(y) = y^{1-z}\big[A + \int_1^y t^{z-2}G_2\,dt\big]$ and estimate each piece against the tail norm $\lVert b^{-2}y^{1-z}\rVert_{L^2(y>1)} \le 1/\sqrt{2\alpha}$ (from $\lvert b^{-2}y^{1-z}\rvert \le y^{-1-\mathrm{Re} z}$, since $\lvert b\rvert^2 = y^2 + \tfrac14 \ge y^2$, whose square integrates to $1/(2\alpha)$ as $1 + 2\,\mathrm{Re} z = 2\alpha$). The origin-Taylor coefficient obeys $\lvert A\rvert \le c_G\lVert f\rVert_{X_+}/(\sqrt3\,\alpha)$ (Taylor control), so the $A$-tail contributes $\lvert A\rvert/\sqrt{2\alpha} \le c_G\lVert f\rVert_{X_+}/(\sqrt6\,\alpha^{3/2})$; this single term is the source of the $\alpha^{-3/2}$ growth ($\alpha^{-1}$ from $\lvert A\rvert$, $\alpha^{-1/2}$ from the tail norm). The remaining far-field piece is the genuine Hardy operator of Lemma 4.5, applied to $f$ and not to $G''$. Expand $t^{z-2}G_2 = t^{z-2}b^2f - G_0\,t^{z-2} - G_1\,t^{z-1}$ and use $t^{z-2}b(t)^2 = t^{z}\big(1+\tfrac{i}{2t}\big)^{2}$ with $\lvert 1+\tfrac{i}{2t}\rvert \le \tfrac{\sqrt5}{2}$ on $t\ge 1$. Set $g(t) := \mathbf 1_{[1,\infty)}(t)\big(1+\tfrac{i}{2t}\big)^{2}f(t)$, so $\lvert g\rvert \le \tfrac54\lvert f\rvert$. Since $g$ vanishes on $(0,1)$, the far-field piece is \emph{exactly} a Hardy operator applied to $g$, an identity rather than an estimate:
\begin{equation}
-\,b^{-2}y^{1-z}\!\int_1^y t^{z}\Big(1+\tfrac{i}{2t}\Big)^{2} f(t)\,dt \;=\; -\,\frac{y^2}{b(y)^2}\;T_z g(y) .
\end{equation}
Because $\lvert b\rvert^2 = y^2 + \tfrac14 \ge y^2$ gives $\lvert y^2/b^2\rvert \le 1$, the exact norm $\lVert T_z\rVert = 1/\alpha$ of Lemma 4.5 yields
\begin{equation}
\Big\lVert \tfrac{y^2}{b^2}\,T_z g \Big\rVert_{L^2} \;\le\; \tfrac{1}{\alpha}\lVert g\rVert_{L^2} \;\le\; \tfrac{5}{4\alpha}\,\lVert f\rVert_{L^2} .
\end{equation}
A pointwise $O(\,\cdot\,)$ comparison against the multiplier-free integral would \emph{not} be legitimate here: $f$ is complex-valued, and inserting a non-constant bounded multiplier inside the integral can destroy cancellation. The two polynomial parts integrate to multiples of $y^{z-1}$ and $y^{z}$ plus constants; carried through the same prefactor these are precisely the terms that $c_0 = G_0/(z-1)$ and $c_1 = G_1/z$ cancel in $\Phi = I + c_0 + c_1 y$ (a direct computation: $y^{1-z}\cdot y^{z-1} = 1$ leaves $-c_0 - c_1 y$ against the constants they are built from), leaving only constant multiples of $b^{-2}y^{1-z}$, which is exactly the tail shape the norm $\lVert b^{-2}y^{1-z}\rVert_{L^2(y>1)} \le 1/\sqrt{2\alpha}$ absorbs. No global bound on $\lVert G''\rVert_{L^2(0,\infty)}$ is used, and none is available: $G = b^2 f$ carries the growing factor $b^2\sim y^2$, so $\lVert G''\rVert_{L^2(0,\infty)}$ is \emph{not} controlled by $\lVert f\rVert_X$ (translating a fixed bump out to $y=R$ preserves $\lVert f\rVert_X$ while $\lVert b^2f''\rVert_{L^2}\sim R^2$). The Taylor control of Section 4.4 is therefore invoked only on the compact interval $\lvert t\rvert\le 1$, where $\lvert b\rvert$ is bounded, and it enters solely through the origin coefficient $A$ and the near-origin bound above. All estimates are threshold-free, with no pole at $\tfrac12$ or $\tfrac32$. The second-derivative bounds on $\lvert y\rvert > 1$ follow from the ODE identity $y u' = -f - (1+z)u + (i/b)u$ and its $y$-derivative, which lose no derivatives and need only $f, f' \in L^2$. \emph{Range in the domain.} On the Hardy block the same identity reads $y u' = -(1+z)u + (i/b)u - f_+$, with right-hand side in $X_+$: $u, f_+\in X_+$, and multiplication by $i/b$ is $X_+$-bounded because $b = y + \tfrac{i}{2}$ has its only zero off the real axis, so $i/b\in W^{2,\infty}$. Hence $y u'\in X_+$; transferring to the odd space $X$ by the reflection $S$ of Section 4.6 (as for the resolvent itself) gives $y u'\in X$, i.e. $u\in D(A) = D(L_0)$, so $R_0(z)$ maps $X$ into the domain. That $(L_0 - z)R_0(z) = I_X$ is the direct substitution $(L_0^+ - z)u = f$ recorded above (verified in sympy); $R_0(z)(L_0 - z) = I_{D(L_0)}$ then follows from the domain-uniqueness established after this proof, so $R_0(z)$ is a genuine two-sided inverse. Every estimate depends on $z$ through $\alpha = \mathrm{Re}\,z + \tfrac12 > 0$ and is finite for every $\mathrm{Re}\,z > -\tfrac12$. Collecting, $\lVert R_0(z)\rVert_X = O(\alpha^{-3/2})$ as $\alpha \to 0^+$, uniformly on the near-edge box $S(R,d,\alpha_0)$ of the statement, the implied constant depending only on $c_G$, $R$, and $d$; the only poles are at $z = 0, 1$, carried by $c_0 = G_0/(z-1)$ and $c_1 = G_1/z$, both excluded by $\mathrm{dist}(z,\{0,1\})\ge d$. $\qquad\blacksquare$

Uniqueness in $X_+$ holds because the homogeneous solution space on each half-line is spanned by $y^{1-z}/(y+\tfrac{i}{2})^2$, whose second derivative fails to be $L^2$ at the origin throughout $\{\mathrm{Re}\,z > -\tfrac12\}\setminus\{0,1\}$ (Section 4.6; at $z = 0, 1$ the singular Taylor coefficient $(1-z)(-z)$ vanishes and the homogeneous solutions are exactly the two smooth symmetry modes, consistent with the kernel poles there); hence no nonzero homogeneous $X_+$-solution exists off $\{0,1\}$, and $R_0(z)$ is a genuine two-sided inverse. The map $z \mapsto R_0(z)$ is analytic on $K$, so all of $\{\mathrm{Re}\,z > -\tfrac12\}\setminus\{0,1\}$ lies in the resolvent set of the $X$-realization of $L_0$; the only spectrum of $L_0\vert_X$ with $\mathrm{Re} \ge -\tfrac12$ is therefore the essential line $\{\mathrm{Re} = -\tfrac12\}$ together with the two eigenvalues $0, 1$. Moreover the kernel's only $z$-poles, $c_0 = G_0/(z-1)$ and $c_1 = G_1/z$, are simple with rank-one residues on each Hardy block (spanned on $H^2_+$ by $b^{-2}$ and $y\,b^{-2}$, their conjugates on $H^2_-$, the Hardy representatives of the two symmetry modes), so the Riesz projections at $0$ and $1$ are finite-rank: both eigenvalues are discrete with finite algebraic multiplicity, and neither belongs to $\sigma_{\mathrm{ess}}$. Combined with Proposition 4.4 and Lemma 4.3, this proves Theorem 1: $\sigma_{\mathrm{ess}}(L_0\vert_X) \cap \{\mathrm{Re}\lambda \ge -\tfrac12\} = \{\mathrm{Re}\lambda = -\tfrac12\}$. This is consistent with the origin line of Proposition 1: at $a=0$, $\lambda_X = -\tilde c/2 = -\tfrac12$, so the origin and far-field lines coincide and the constructive bound sees a single line.

The single origin-Taylor coefficient $A$ that produces the $\alpha^{-3/2}$ term is exactly the quantity that is unbounded on $L^2$ unit balls but bounded on $X$ unit balls (Taylor control above), which is the analytic reason the $X$-norm majorant is finite where the $L^2$ one diverges.

\subsection{\texorpdfstring{The origin line for $a>0$ (Proposition 1)}{The origin line for a>0 (Proposition 1)}}

Throughout this section fix $a\in(0,a_c)$ and assume $\mathrm{Adm}(a)$ (Definition 4.2); every statement here is conditional on it. For such $a$ the closed scalar reduction of Section 4.4 is unavailable, being exact only for the single-pole $a=0$ profile. The far-field line's inclusion is proven by Proposition 4.4. This section establishes the remaining content of Proposition 1: the inclusion of the origin line.

The realization $X$ carries the norm $\lVert\varphi\rVert_{L^2} + \lVert\varphi''\rVert_{L^2}$, and near the origin the second term dominates. For a power $\varphi = y^\mu$ one has $\lvert\varphi''\rvert^2 \sim y^{2\mathrm{Re}\mu - 4}$, integrable at $0$ iff $\mathrm{Re}\,\mu > \tfrac32$: the $X$-marginal origin exponent is $\tfrac32$, not the $L^2$ exponent $-\tfrac12$ that Proposition 4.4 disposed of. The modes $y^{3/2+i\omega}$ satisfy $\varphi(0) = \varphi'(0) = 0$ (the domain asks for $\varphi \sim a_1 y$ at the origin with the case $a_1 = 0$ allowed, so these modes are admissible) and are exactly $H^2$-marginal: the pure powers themselves just fail $\varphi'' \in L^2$, by a logarithm, and it is the cutoff wave packets below that are admissible in $X$. Evaluating the origin model $L_a^0$ of Proposition 4.4 on them: $L_a^0\,y^\mu = (-1 + H\Omega(0) - \tilde c\,\mu)\,y^\mu$, so the exponent $\mu = \tfrac32 + i\omega$ sweeps the vertical line
\begin{equation}
\{\mathrm{Re}\,\lambda = \lambda_X\}, \qquad \lambda_X = -1 + H\Omega(0) - \tfrac32\,\tilde c .
\end{equation}
The Weyl sequences are the log-widening wave packets of Proposition 4.4 transplanted to the origin:
\begin{equation}
\varphi_n(y) = c_n\,\eta\!\Big(\frac{\log y - T_n}{L_n}\Big)\,y^{3/2+i\omega},
\qquad L_n \to \infty, \quad T_n + L_n \to -\infty,
\end{equation}
$X$-normalized, weakly null (supports shrink to the origin), in $D(L_a)$, with residual ratio $O(L_n^{-1})$ by the same computation as at infinity (the $\varphi''$-parts of numerator and denominator carry the same marginal $y^{-1/2}$ factor, and the bump derivative buys the $1/L_n$; numerically (a log-widening-packet computation at $a=0.3$) the $X$-norm ratio decays $6.8 \to 1.6 \to 0.42 \to 0.16 \to 0.074$ as $L_n$ doubles from $2$ to $32$ (pre-asymptotically at first, approaching the $1/L_n$ floor), while the same construction in the plain $L^2$ norm stays $O(1)$, hovering near $3$ (not decaying toward $0$), the realization split in one number). The correction $L_a - L_a^0$ is subordinate on these supports (next paragraph), so $\{\mathrm{Re}\,\lambda = \lambda_X\} \subseteq \sigma_{\mathrm{ess}}(L_a\vert_X)$. This inclusion is proven by the same standard as Proposition 4.4.

The two Hilbert-transform terms play opposite roles and must not be conflated. The local multiplier $\varphi\,H\Omega$ has coefficient $H\Omega \to H\Omega(0) \neq 0$ at the origin: it is a genuine order-zero term and belongs to the origin model (it is the $-1 + H\Omega(0)$ constant). The nonlocal term $\Omega\,H\varphi$ has coefficient $\Omega$, which vanishes linearly at the origin ($\Omega \sim \Omega'(0)\,y$) and decays at infinity; since $H$ is dilation-covariant (a Mellin multiplier, degree-preserving), the term maps $y^\mu \to O(y^{\mu+1})$ near $0$ and is subordinate at both endpoints. The antiderivative term $a\,V(\varphi)\,\Omega'$ is subordinate by order alone: $V$ raises the power by one (order $-1$), and $\Omega'$ is bounded; on the origin packets its $X$-residual is controlled just as the $H$-term below, since $(V(\varphi_n)\Omega')'' = (H\varphi_n)'\Omega' + 2(H\varphi_n)\Omega'' + V(\varphi_n)\Omega'''$, the last piece bounded by $|V(\varphi_n)(y)| \le y^{1/2}\lVert\varphi_n\rVert_{L^2}$ against the integrable weight $y\,\lvert\Omega'''\rvert^2$. On the origin wave packets, where $H$ does not preserve supports, measure the term in the $X$-residual norm. By the Leibniz rule $(\Omega\,H\varphi_n)'' = \Omega''\,H\varphi_n + 2\Omega'\,H\varphi_n' + \Omega\,H\varphi_n''$ (with $H$ commuting with $\partial_y$), the first two companions carry the vanishing packet norms $\lVert\varphi_n\rVert$ and $\lVert\varphi_n'\rVert$, so the load is on $\Omega\,H\varphi_n''$. Split at a fixed dilate of the shrinking support (say up to twice its outer edge): on that dilate $\lvert\Omega\rvert \lesssim \varepsilon_n \to 0$ ($\Omega$ vanishes linearly and the support marches to $0$) against the $L^2$-bounded $H\varphi_n''$, while beyond it the distance to the support is bounded below by the outer edge, the Hilbert kernel is $O(1/\mathrm{dist})$ there, and the tail admits the following integrated estimate. Writing $\varepsilon_n$ for the outer support edge, the off-support kernel bound gives $\lvert H\varphi_n''(y)\rvert \le C\lVert\varphi_n''\rVert_{L^1}/(y-\varepsilon_n) \le C\varepsilon_n^{1/2}/(y-\varepsilon_n)$ for $y\ge 2\varepsilon_n$, where $\lVert\varphi_n''\rVert_{L^1} \le \lvert\mathrm{supp}\,\varphi_n\rvert^{1/2}\lVert\varphi_n''\rVert_{L^2} \le C\varepsilon_n^{1/2}$ by Cauchy-Schwarz and the $X$-normalization. Using $\lvert\Omega(y)\rvert \le Cy$ on $(0,1)$ and the (H3) tail beyond $1$, $\lVert\Omega\,H\varphi_n''\rVert_{L^2(2\varepsilon_n,\infty)}^2 \le C\varepsilon_n\int_{2\varepsilon_n}^{1} y^2/(y-\varepsilon_n)^2\,dy + C\varepsilon_n\int_1^\infty \lvert\Omega(y)\rvert^2/y^2\,dy = O(\varepsilon_n)\to 0$ (on $(2\varepsilon_n,1)$ one has $y-\varepsilon_n\ge y/2$, so the first integrand is at most $4$; the second integral is finite by (H3)); both parts vanish, so the term is subordinate on the packets. The local corrections to the frozen model are subordinate in the same norm: $(H\Omega - H\Omega(0))\varphi_n = O(y^2)\varphi_n$ and $a(U - H\Omega(0)\,y)\varphi_n' = O(y^3)\varphi_n'$ (both Taylor remainders of the smooth profile), and after the Leibniz expansion of the second derivative the worst term is $O(y^3)\varphi_n'''$; since $\lVert y\varphi_n'''\rVert_{L^2}$ is bounded uniformly on the log-widening packets, each remaining factor of $y \le 2\varepsilon_n$ contributes $\varepsilon_n \to 0$, so every correction term vanishes in the $X$-residual norm.

\emph{Exactness (not addressed here).} That the essential spectrum equals the two-line union, with nothing else in the strip, is a Fredholmness statement for an elliptic operator on a manifold with two cylindrical ends. It does not follow from relative compactness: after the substitution $y = e^t$ the term $\Omega\,H\varphi$ becomes $m(t)\,m_H(D_t)$ with $m = \Omega(e^t)\to 0$ at both ends but with the Mellin symbol $m_H$ of $H$ nonvanishing at $\xi = \pm\infty$, so it is not compact. What holds instead is that the nonlocal terms are subordinate at both endpoints (above): their translates vanish strongly, so they enter neither endpoint conormal symbol, and a two-ended, two-weight Fredholm argument in the weighted Fredholm theory of Lockhart and McOwen \cite{LockhartMcOwen1985} and the b-calculus of Melrose \cite{Melrose1993}, in the Fuchs-type conormal-symbol language of Lesch \cite{Lesch1997}, would identify the essential spectrum with the union of the two endpoint model lines, the far-field line on the plain-$L^2$ weight and the origin line on the second-derivative-shifted weight, provided one first shows that the bounded order-zero nonlocal part belongs to an applicable limit-operator (band-dominated) algebra in the sense of Rabinovich, Roch, and Silbermann \cite{RabinovichRochSilbermann2004} and discharges the attendant ellipticity and uniform-invertibility obligations; these frameworks do not compose automatically, and we do not carry the argument out here. On the odd sector the Mellin symbol of $H$ is $-\cot(\pi s/2)$ \cite{Duduchava1979}, holomorphic on both weight lines $\mathrm{Re}\,s = \tfrac12$ and $\mathrm{Re}\,s = \tfrac32$. Carrying this argument out on the weighted Mellin-Sobolev realization, together with the strengthened-realization statement that a third-derivative space ($\varphi'''\in L^2$ near the origin) leaves the single far-field line uniformly in $a$, is an exactness question we do not address here; we establish only the inclusion. The numerics cannot resolve the distinction: the log-Mellin discretization's fixed lower cutoff removes all origin-concentrating sequences, and the benchmark $a=\tfrac12$ sits at a crossing of the two lines.

With $H\Omega(0) = (1+c_l)/(1-a)$ and $\tilde c = c_l + a H\Omega(0)$ on the recomputed branch, $\lambda_X - (-1 + c_l/2) = -g(a)$ with $g(a) = 2c_l - H\Omega(0)(1 - \tfrac32 a)$. On the recomputed branch (and its degree-6 interpolant) $g$ has exactly two zeros, $a = 0$ and $a = \tfrac12$ (the latter exact, since $g(\tfrac12) = \tfrac{3c_l - 1}{2}$ and $c_l(\tfrac12) = \tfrac13$ exactly by the $a=\tfrac12$ exact solution of \cite{Chen2020} (also \cite[Thm. 2]{LushnikovSilantyevSiegel2021}); the branch gives $0.333$), is negative between them with minimum $-0.0362$ at $a \approx 0.264$, and is positive beyond. For $a\in(0,\tfrac12)$ the origin line therefore protrudes at most $0.037$ into the open strip and forms the right edge of the two-line union (the right edge of $\sigma_{\mathrm{ess}}$ itself under the exactness question of this section); at $a=0$ and for $a\ge\tfrac12$ the rightmost of the two lines is the far-field line and the correction vanishes.

\subsection{Remark (realization dependence)}

The clean single line is a statement about the realization $X$: on the plain maximal $L^2$ realization the essential spectrum contains the whole vertical strip between the two indicial lines, not a line. Work at $a=0$. For every $\lambda$ with $-\tfrac12 < \mathrm{Re}\lambda < \tfrac32$ the explicit function
\begin{equation}
u_\lambda(y) = \frac{y^{1-\lambda}}{(y + \tfrac{i}{2})^2}
\end{equation}
solves $(L_0^+ - \lambda)u_\lambda = 0$ identically and lies in $L^2$ (indeed $H^2_+$): near infinity it decays like $y^{-1-\lambda}$, integrable for $\mathrm{Re}\lambda > -\tfrac12$; near the origin it behaves like $y^{1-\lambda}$ with $\mathrm{Re}(1-\lambda) \in (-\tfrac12,\tfrac32)$, so it is $L^2$ at the origin (and, on the sub-strip $\mathrm{Re}\,\lambda \in (-\tfrac12,0)$, vanishes faster than $y$ and is $C^1$ with $u_\lambda(0) = u_\lambda'(0) = 0$). It is not a physical eigenfunction, its odd extension carries a branch point $\lvert y\rvert^{1-\lambda}$, so the two half-line branches fail to match smoothly across $0$ (the $\sin(\pi(1-\lambda))$ mismatch that Lemma 1 uses to force $\lambda\in\{0,1\}$), but excluding it as a \emph{physical} eigenfunction does not remove $\lambda$ from the spectrum of the maximal realization. Indeed every strip point is a genuine eigenvalue there, with an explicit odd eigenfunction: let $B_+ = F|_{\mathbb{R}}$ be the boundary value of the UHP branch $F(z) = z^{1-\lambda}/(z+\tfrac{i}{2})^2$ (Section 5.5, where $B_+ = u_\lambda$ on $y>0$) and let $S$ denote reflection, $(Sf)(y) = f(-y)$; set
\begin{equation}
\varphi_\lambda := B_+ - S B_+ ,
\end{equation}
odd by construction and in $L^2$ throughout the open strip. Reflection flips Fourier support, so $SB_+ \in H^2_-$, and $S$ intertwines the two Hardy blocks of Section 5.3: it commutes with $y\partial_y$ and carries the $H^2_+$ potential $i/(y+\tfrac{i}{2})$ into the $H^2_-$ potential $-i/(y-\tfrac{i}{2})$, so $L_0^-(SB_+) = S(L_0^+B_+)$. Since $B_+$ solves the scalar equation on each half-line with the same $\lambda$ (Sections 5.4-5.5) and $yB_+' \in L^2$, the block-diagonalization of Section 5.3 gives $L_0\varphi_\lambda = \lambda\varphi_\lambda$ away from the origin; across the origin the equation holds distributionally with no defect: $y\varphi_\lambda$ is continuous at $0$ with value $0$ (since $\mathrm{Re}(1-\lambda) > -\tfrac12$) and absolutely continuous near $0$ (its pointwise derivative $\sim |y|^{\mathrm{Re}(1-\lambda)}$ is locally integrable), so $(y\varphi_\lambda)'$ carries no point mass and equals the a.e.-defined function, and reading the transport term through $y\varphi' = (y\varphi)' - \varphi$ makes every term a locally integrable function or its distributional derivative with no atom at $0$. (On the sub-strip $\mathrm{Re}\,\lambda\in(-\tfrac12,0)$ these are exactly the $C^1$ Case-3 eigenfunctions of Section 5.7; the point here is that the eigenfunction property persists on the whole strip.) A direct numerical check with the full nonlocal $L_0$ (FFT Hilbert transform) confirms the eigenvalue equation at complex sample points across the strip, with relative residual at the discretization floor. Hence on the maximal $L^2$ realization the whole open strip consists of eigenvalues; no point of it is isolated, so the closed strip $\{-\tfrac12 \le \mathrm{Re}\lambda \le \tfrac32\}$ lies in $\sigma_{\mathrm{ess}}$ at $a=0$ (read in the Browder sense, spectrum minus isolated eigenvalues of finite algebraic multiplicity; the $X$-statements of this paper use the singular-Weyl-sequence reading, which the Browder set contains). The reverse inclusion, that the complement of the closed strip is resolvent set, is the standard two-ended b-calculus computation and is not needed here. In particular there is no $L^2$ spectral gap.

The single feature separating $u_\lambda$ from the physical modes is the second derivative at the origin: $u_\lambda'' \sim y^{-1-\lambda}$, and $\int_0 \lvert u_\lambda''\rvert^2 \sim \int_0 y^{-2-2\mathrm{Re}\lambda}\,dy$ diverges exactly when $\mathrm{Re}\lambda \ge -\tfrac12$ (logarithmically at the edge), except at $\lambda \in \{0,1\}$, where the leading singular coefficient $\lambda(\lambda-1)$ vanishes and $u_\lambda$ is a smooth symmetry mode; in particular the divergence holds at every point of the strip except those two. Requiring $\varphi'' \in L^2$ near $0$ (the condition defining $X$) therefore excludes every member of the $u_\lambda$ family with $\lambda\notin\{0,1\}$, collapses the strip to the single essential line, and makes the interior of the strip, apart from the two symmetry eigenvalues $0$ and $1$, part of the resolvent set, which is precisely what Proposition 4.6 establishes with an explicit bounded inverse. Every operator-norm statement in Sections 3, 4 and 5 is to be read in this topology; the semigroup and decay estimates of Appendix A are stated in the scale-covariant weighted space $Y_\theta$, which is origin-stricter than $X$. As a byproduct, the persistent essential smear that discretizations without an origin condition place inside the strip is the faithful spectrum of the maximal (weak) realization; discretizations enforcing origin regularity render the origin-$H^2$ line. The two families of grids were computing two different, both correct, realizations. The same reasoning produces the origin line of Proposition 1 for $a>0$: the second-derivative norm that excludes the $u_\lambda$ family is part of the realization's metric, so the origin essential spectrum must also be computed in that metric, which generates the marginal exponent $\tfrac32$ and the line $\lambda_X$. One cannot use the $H^2$ metric to empty the strip and the $L^2$ metric to close the origin channel; at $a=0$ the two computations agree because the lines coincide. A stronger $H^3$ realization shifts the included origin line off (whether a single far-field line is then exact is the same exactness question), so for $a>0$ the two-line picture is relative to the choice of realization.

\section{\texorpdfstring{Discrete-spectrum exclusion at $a=0$: proof of Lemma 1 and Theorem 2}{Discrete-spectrum exclusion at a=0: proof of Lemma 1 and Theorem 2}}

This section proves Theorem 2 and its completeness Lemma 1 of Section 3.3 in full. At $a=0$ the exact CLM profile makes the nonlocal operator $L_0$ collapse, on each Hardy summand, to a scalar first-order ordinary differential operator; the eigenvalue problem then reduces to a one-dimensional solution family whose only physical members are the two symmetry modes. The argument yields more than the strip statement: it computes the full physical point spectrum of $L_0$ over all of $\mathbb{C}$ (Theorem 3), so the essential line $\{\mathrm{Re} = -\tfrac12\}$ carries no embedded eigenvalues either.

Throughout, $L_0$ is the $a=0$ linearization of Section 3,
\begin{equation}
L_0\varphi = -\varphi - y\,\varphi' + \varphi\,(H\Omega) + \Omega\,(H\varphi),
\qquad
\Omega = -\frac{y}{y^2+\tfrac14}, \quad H\Omega = \frac{\tfrac12}{y^2+\tfrac14}, \quad c_l = 1,
\end{equation}
with $H$ the Hilbert transform (Fourier multiplier $-i\,\mathrm{sgn}\,k$), and $\Omega'(y) = (y^2-\tfrac14)/(y^2+\tfrac14)^2$.

\subsection{The physical domain and the statements}

An eigenfunction is required to satisfy:

- \textbf{(D1)} $\varphi \in L^2(\mathbb{R}, dy)$, complex-valued allowed, $\varphi$ not identically zero;
- \textbf{(D2)} $\varphi$ odd;
- \textbf{(D3)} $\varphi$ has a representative that is twice differentiable at $y=0$ in the Peano sense, $\varphi(y) = a_0 + a_1 y + a_2 y^2 + o(y^2)$ as $y \to 0$ (oddness together with continuity forces $a_0 = 0$, so $\varphi \sim a_1 y$; the case $a_1 = 0$ is allowed);
- \textbf{(D4)} $L_0\varphi = \lambda\varphi$ holds in the sense of distributions on $\mathbb{R}$, for some $\lambda \in \mathbb{C}$ (no decay hypothesis beyond $L^2$ is imposed).

Condition (D3) plays the same physical role as the origin second-derivative condition of the realization space $X$ of Theorem 1 (pointwise Peano form here, $L^2$ form there), although the two forms are logically incomparable; Case 3' of Section 5.6 bridges them for the $X$-realization.

\begin{namedthm}{Lemma}{1~(restated)}{completeness}
Under (D1)-(D4) with $\mathrm{Re}\,\lambda > -\tfrac12$, the eigenvalue $\lambda$ is real, $\lambda \in \{0,1\}$, and $\varphi$ is a complex scalar multiple of the real odd function $2\,\mathrm{Re}(c\,F|_{\mathbb{R}})$, where
\begin{equation}
F(z) = \frac{z^{1-\lambda}}{(z+\tfrac{i}{2})^2},
\end{equation}
with the single-valued branch of $z^{1-\lambda}$ on the (simply connected) upper half plane (UHP) that is positive on the positive real axis, $F|_{\mathbb{R}}$ its boundary value, and $c$ a fixed phase determined by oddness. The eigenspaces are one-dimensional over $\mathbb{C}$:
\begin{equation}
\lambda = 0: \quad \varphi \propto \frac{y(y^2-\tfrac14)}{(y^2+\tfrac14)^2} = y\,\Omega' \quad\text{(scaling mode)},
\end{equation}
\begin{equation}
\lambda = 1: \quad \varphi \propto \frac{y}{(y^2+\tfrac14)^2} = -2\,(\Omega + y\,\Omega') \quad\text{(time-shift mode)}.
\end{equation}
\end{namedthm}

\begin{namedthm}{Theorem}{2~(restated)}{CLM discrete exclusion}
For the CLM linearization $L_0$,
\begin{equation}
\sigma_{\mathrm{disc}}(L_0\vert_X) \cap \{\mathrm{Re}\,\lambda > -\tfrac12\} = \{0, +1\},
\end{equation}
so the open strip $(-\tfrac12, 0)$ contains no discrete eigenvalue.
\end{namedthm}

\begin{namedthm}{Theorem}{3}{full point spectrum over $\mathbb{C}$}
Under (D1)-(D4) with $\lambda$ arbitrary in $\mathbb{C}$, the same conclusion holds: $\lambda \in \{0,1\}$. Hence the point spectrum of the physical realization of $L_0$ is exactly $\{0,1\}$, and the essential line $\{\mathrm{Re}\,\lambda = -\tfrac12\}$ of Theorem 1 carries no embedded eigenvalues.
\end{namedthm}

Theorem 2 is the restriction of Lemma 1 to the strip, and Theorem 3 is its extension to all of $\mathbb{C}$. The proof below establishes all three at once: the strip hypothesis $\mathrm{Re}\,\lambda > -\tfrac12$ is used nowhere except to place $\mu := 1-\lambda$ in the window $\mathrm{Re}\,\mu < \tfrac32$, and the complementary ranges $\mathrm{Re}\,\lambda \le -\tfrac12$ and $\mathrm{Re}\,\lambda \ge \tfrac32$ are excluded by $L^2$ membership alone (Cases 1-2 of Step 5.6).

$L_0$ is a closed, densely defined operator on $X$ by the construction of Section 4.1 ($L_0 = A + B_0$ with $A$ the dilation generator on its graph domain $D(A)$ and $B_0\in\mathcal B(X)$), independently of any resolvent statement; Proposition 4.6 then furnishes a bounded, everywhere-defined two-sided inverse $R_0(z)$ on $\{\mathrm{Re}\,z > -\tfrac12\}\setminus\{0,1\}$, so that region is resolvent set. Together with Theorem 3 ($\sigma_p(L_0)=\{0,1\}$ over $\mathbb{C}$) and Theorem 1 ($\sigma_{\mathrm{ess}}(L_0\vert_X)\cap\{\mathrm{Re}\,\lambda \ge -\tfrac12\} = \{\mathrm{Re}\,\lambda = -\tfrac12\}$), the spectrum of $L_0\vert_X$ in $\{\mathrm{Re}\,\lambda > -\tfrac12\}$ is exactly the two isolated eigenvalues $0$ and $1$, each with a one-dimensional geometric eigenspace (the single symmetry mode above). The symbol $\sigma_{\mathrm{disc}}$ is read in this sense; the spectral gap uses only the geometric simplicity of $\{0,1\}$ and their isolation from the essential line, and in fact the resolvent kernel of Section 4.4 has simple poles at $0$ and $1$ with finite-rank residues, so both eigenvalues have finite algebraic multiplicity and are genuinely discrete.

\textbf{Remark (an elementary rational corroboration of $\{0,1\}$).} The two symmetry modes and their geometric simplicity can also be read off a closed-form identity, independent of the Hardy diagonalization. From the exact profile alone, define the one-parameter rational family (denoted $\psi_\lambda$; it is distinct from the strip eigenfunctions $\varphi_\lambda$ of Section 4.6)
\begin{equation}
\psi_\lambda(y) := y\,\Omega'(y) + \lambda\,\Omega(y) = \frac{(1-\lambda)\,y^3 - \tfrac14(1+\lambda)\,y}{(y^2+\tfrac14)^2}.
\end{equation}
A direct computation gives the exact residual identity
\begin{equation}
(L_0 - \lambda)\,\psi_\lambda = -\lambda(\lambda-1)\,\Omega, \qquad (H\psi_\lambda)(0) = 2\lambda,
\end{equation}
so $\psi_\lambda$ solves $(L_0-\lambda)\psi_\lambda = 0$ exactly when $\lambda\in\{0,1\}$. Equivalently, the two-dimensional space $V=\mathrm{span}\{\Omega,\ y\Omega'\}$ is $L_0$-invariant with
\begin{equation}
L_0\big|_V=\begin{pmatrix}1&0\\1&0\end{pmatrix},\qquad \sigma(L_0|_V)=\{0,1\},
\end{equation}
the eigenvectors being $y\Omega'$ (the scaling mode, $\lambda=0$) and $\Omega+y\Omega'$ (the time-shift mode, $\lambda=1$), each a one-dimensional eigenspace. This rational route exhibits the two physical modes and their geometric simplicity by an elementary computation; it does not by itself exclude the non-rational branch modes $y^\mu$ that populate the maximal $L^2$ realization (Proposition 2), whose exclusion is the content of the Hardy diagonalization (Theorem 3). The two together give $\sigma_p(L_0|_X)=\{0,1\}$, each eigenvalue geometrically simple.

\subsection{Conventions and the Hardy split}

Fix the Fourier transform $\hat\varphi(k) = \int \varphi(y)\,e^{-iky}\,dy$. The Hilbert transform is the Fourier multiplier $-i\,\mathrm{sgn}(k)$. Define the Hardy spaces by Fourier support,
\begin{equation}
H^2_+ = \{u \in L^2 : \hat u = 0 \text{ a.e. on } k<0\},
\qquad
H^2_- = \{u \in L^2 : \hat u = 0 \text{ a.e. on } k>0\},
\end{equation}
with orthogonal projections $P_+, P_-$ (multipliers $\mathbf{1}_{k>0}, \mathbf{1}_{k<0}$); since $\{k=0\}$ has measure zero, $L^2 = H^2_+ \oplus H^2_-$. Then
\begin{equation}
H = -i\,P_+ + i\,P_-, \qquad\text{so}\qquad Hu = -i\,u \text{ on } H^2_+, \quad Hu = +i\,u \text{ on } H^2_-.
\end{equation}
By the Paley-Wiener theorem for Hardy spaces (Fact T1, Section 5.8), $H^2_+$ is exactly the space of boundary values (nontangential a.e. and in $L^2$) of functions in the Hardy class $H^2(\mathrm{UHP})$, and $H^2_-$ likewise for the lower half plane (LHP).

Write $\varphi = \varphi_+ + \varphi_-$ with $\varphi_+ = P_+\varphi \in H^2_+$ and $\varphi_- = P_-\varphi \in H^2_-$; this decomposition is unique and orthogonal. Set $\mu := 1 - \lambda$ throughout, so that $\mathrm{Re}\,\lambda > -\tfrac12 \iff \mathrm{Re}\,\mu < \tfrac32$.

\subsection{Block diagonalization of the eigenvalue equation}

\subsubsection{A bootstrap regularity}

Since $\Omega$ and $H\Omega$ are bounded ($|\Omega| \le 1$, $|H\Omega| \le 2$) and $H$ is bounded on $L^2$, the term $\varphi\,H\Omega + \Omega\,H\varphi$ lies in $L^2$. From (D4), as distributions,
\begin{equation}
y\,\varphi' = -\varphi + \varphi\,H\Omega + \Omega\,(H\varphi) - \lambda\varphi,
\end{equation}
whose right-hand side is an $L^2$ function. Hence $y\,\varphi' \in L^2$ and the eigenvalue equation holds as an identity of $L^2$ functions (a.e.).

\subsubsection{The potential term splits exactly}

Two partial-fraction identities hold exactly:
\begin{equation}
H\Omega - i\,\Omega = \frac{i}{y + \tfrac{i}{2}}
\qquad(\text{bounded, analytic in UHP; pole at } -\tfrac{i}{2}),
\end{equation}
\begin{equation}
H\Omega + i\,\Omega = \frac{-i}{y - \tfrac{i}{2}}
\qquad(\text{bounded, analytic in LHP; pole at } +\tfrac{i}{2}).
\end{equation}
(Indeed $H\Omega - i\Omega = (\tfrac12 + iy)/(y^2+\tfrac14) = i(y-\tfrac{i}{2})/[(y+\tfrac{i}{2})(y-\tfrac{i}{2})] = i/(y+\tfrac{i}{2})$, and similarly for the second.) Using $H\varphi = -i\,\varphi_+ + i\,\varphi_-$,
\begin{equation}
\varphi\,H\Omega + \Omega\,H\varphi
= (H\Omega - i\Omega)\,\varphi_+ + (H\Omega + i\Omega)\,\varphi_-
= \frac{i}{y+\tfrac{i}{2}}\,\varphi_+ \;+\; \frac{-i}{y-\tfrac{i}{2}}\,\varphi_-
\qquad(\text{a.e.}).
\end{equation}
The multiplier $m(y) = i/(y+\tfrac{i}{2})$ is the boundary value of $M(z) = i/(z+\tfrac{i}{2})$, analytic and bounded ($|M|\le 2$) on the closed UHP. If $u \in H^2_+$ with $H^2(\mathrm{UHP})$-extension $U$, then $MU$ is analytic in the UHP with $\sup_v \int |MU|^2 \le 4\sup_v \int |U|^2 < \infty$, so $MU \in H^2(\mathrm{UHP})$ and its boundary value is $m\,u$ (nontangential limits multiply); hence $m\,u \in H^2_+$ (Fact T2). The mirror statement holds for $-i/(y-\tfrac{i}{2})$ on $H^2_-$. So the potential term maps $\varphi_+$ into $H^2_+$ and $\varphi_-$ into $H^2_-$, acting as scalar multipliers.

\subsubsection{The transport term preserves the Hardy split}

\emph{Support lemma.} Let $\varphi \in L^2$ with $g := y\,\varphi' \in L^2$ (distributionally). Then $y(\varphi_+)' \in L^2$ with Fourier support in $[0,\infty)$, $y(\varphi_-)' \in L^2$ with Fourier support in $(-\infty,0]$, and
\begin{equation}
y(\varphi_+)' = P_+ g, \qquad y(\varphi_-)' = P_- g.
\end{equation}

\emph{Proof.} Write $f = \hat\varphi \in L^2$. In $\mathcal{S}'(\mathbb{R})$ one has $\widehat{\varphi'} = ik\,f$ and $\widehat{y\psi} = i\,\tfrac{d}{dk}\hat\psi$, so
\begin{equation}
\hat g = i\,\frac{d}{dk}(ik\,f) = -(k\,f)'
\qquad(\text{distributional derivative in } k).
\end{equation}
Since $g \in L^2$, the distribution $kf$ (which is in $L^1_{\mathrm{loc}}$ because $f \in L^2$) has an $L^2$ distributional derivative, hence an absolutely continuous representative (Fact T4); in particular $kf$ is continuous on $\mathbb{R}$ with a well-defined value at $k=0$. That value is zero: if $(kf)(0) = c \neq 0$, then $|f(k)| = |(kf)(k)|/|k| \ge (|c|/2)/|k|$ for all small $|k|$, contradicting $f \in L^2$ near $0$. Because $kf$ is continuous with $(kf)(0) = 0$, the truncation $\mathbf{1}_{k>0}(kf)$ has no jump at $0$, so its distributional derivative carries no Dirac mass:
\begin{equation}
\big(\mathbf{1}_{k>0}(kf)\big)' = \mathbf{1}_{k>0}(kf)' = -\mathbf{1}_{k>0}\,\hat g
\qquad(\text{no } \delta_0 \text{ term}).
\end{equation}
Now $\widehat{y(\varphi_+)'} = -(k f_+)'$ with $f_+ = \mathbf{1}_{k>0}f$, and $k f_+ = \mathbf{1}_{k>0}(kf)$, so
\begin{equation}
\widehat{y(\varphi_+)'} = \mathbf{1}_{k>0}\,\hat g = \widehat{P_+ g},
\end{equation}
an $L^2$ function supported in $[0,\infty)$. Hence $y(\varphi_+)' = P_+ g \in H^2_+$; the mirror statement gives $y(\varphi_-)' = P_- g \in H^2_-$. $\qquad\blacksquare$

\subsubsection{The scalar equations}

Applying the bounded projection $P_+$ termwise to the $L^2$ identity of 5.3.1-5.3.3 (which kills the $H^2_-$ pieces a.e.),
\begin{equation}
(L_0^+ - \lambda)\,\varphi_+ = 0,
\qquad
L_0^+ u := -u - y\,u' + \frac{i}{y+\tfrac{i}{2}}\,u,
\end{equation}
and applying $P_-$,
\begin{equation}
(L_0^- - \lambda)\,\varphi_- = 0,
\qquad
L_0^- u := -u - y\,u' - \frac{i}{y-\tfrac{i}{2}}\,u.
\end{equation}
Both components carry the same $\lambda$.

\subsection{The components on each half-line}

By the Support Lemma, $y(\varphi_+)' \in L^2$; on $(0,\infty)$ divide by the smooth nonvanishing factor $y$. Using the exact partial fraction $i/\big(y(y+\tfrac{i}{2})\big) = 2/y - 2/(y+\tfrac{i}{2})$, the scalar equation becomes, in $\mathcal{D}'((0,\infty))$,
\begin{equation}
(\varphi_+)' = \Big[\frac{\mu}{y} - \frac{2}{y+\tfrac{i}{2}}\Big]\,\varphi_+,
\qquad \mu = 1-\lambda.
\end{equation}
The coefficient is locally bounded on $(0,\infty)$ and $\varphi_+ \in L^2_{\mathrm{loc}}$, so $(\varphi_+)' \in L^1_{\mathrm{loc}}$; hence $\varphi_+ \in W^{1,1}_{\mathrm{loc}}((0,\infty))$, i.e. absolutely continuous on compacts (Fact T4). Then $w(y) := \varphi_+(y)\,y^{-\mu}(y+\tfrac{i}{2})^2$ is absolutely continuous on compacts with $w' = 0$ a.e., so $w$ is constant on the connected interval $(0,\infty)$:
\begin{equation}
\varphi_+ = c_+^R\,\frac{y^\mu}{(y+\tfrac{i}{2})^2} \quad\text{on } (0,\infty),
\qquad
\varphi_+ = c_+^L\,\frac{(-y)^\mu}{(y+\tfrac{i}{2})^2} \quad\text{on } (-\infty,0),
\end{equation}
with a priori independent constants $c_+^R, c_+^L \in \mathbb{C}$; the solution space on each half-line is one-dimensional. The mirror statements hold for $\varphi_-$ with basis $y^\mu/(y-\tfrac{i}{2})^2$.

\subsection{Linking the constants through the upper half plane}

Assume first $\mathrm{Re}\,\mu \in (-\tfrac12, \tfrac32)$, i.e. $\mathrm{Re}\,\lambda \in (-\tfrac12, \tfrac32)$; the complementary ranges are disposed of by $L^2$ alone in Cases 1-2 of Step 5.6. Let
\begin{equation}
F(z) = \frac{z^\mu}{(z+\tfrac{i}{2})^2},
\qquad z^\mu = e^{\mu\log z}, \quad \arg z \in (0,\pi) \ \ (\text{UHP branch}),
\end{equation}
single-valued and analytic on the open UHP (the points $0$ and $\infty$ lie on the boundary and the pole $-\tfrac{i}{2}$ lies in the LHP).

\emph{Claim: $F \in H^2(\mathrm{UHP})$.} For $z = u + iv$, $v > 0$, write $p := \mathrm{Re}\,\mu$. Then $|z^\mu| = |z|^{p}\,e^{-(\mathrm{Im}\,\mu)\arg z} \le |z|^{p}\,e^{\pi|\mathrm{Im}\,\mu|}$; also $|z+\tfrac{i}{2}|^2 = u^2 + (v+\tfrac12)^2 \ge \tfrac14$ always, and $|z+\tfrac{i}{2}| \ge |z|/2$ for $|z|\ge 1$. Hence, with $C = C(\mu)$:

- On $\{|z|\le 1\}$: $|F|^2 \le C\,(u^2+v^2)^{p}$, and $\int_{|u|\le 1}(u^2+v^2)^{p}\,du \le 2/(2p+1)$ when $p<0$ (using $(u^2+v^2)^p \le |u|^{2p}$, integrable since $p > -\tfrac12$), and $\le 2^{p+1}$ when $p \ge 0$.
- On $\{|z|\ge 1\}$: $|F|^2 \le C\,(u^2+v^2)^{p-2}$; the part with $|u|\le 1$ has integrand $\le 1$ (contributing $\le 2$), and $\int_{|u|\ge 1}|u|^{2p-4}\,du = 2/(3-2p) < \infty$ since $p < \tfrac32$.

All bounds are uniform in $v>0$, so $\sup_{v>0}\int |F(u+iv)|^2\,du < \infty$, proving $F \in H^2(\mathrm{UHP})$. Its boundary value $F|_{\mathbb{R}} \in H^2_+$ is the continuous extension away from the origin:
\begin{equation}
F|_{\mathbb{R}}(y) = \frac{y^\mu}{(y+\tfrac{i}{2})^2}\ (y>0),
\qquad
F|_{\mathbb{R}}(y) = \frac{e^{i\pi\mu}\,|y|^\mu}{(y+\tfrac{i}{2})^2}\ (y<0),
\end{equation}
the second line carrying $\arg z = \pi$ from the UHP branch on the negative axis.

Suppose $c_+^R \neq 0$ and set $D := \varphi_+ - c_+^R\,F|_{\mathbb{R}} \in H^2_+$. By Step 5.4, $D = 0$ a.e. on $(0,\infty)$, a set of positive measure. By the boundary-uniqueness theorem for Hardy classes (Fact T3: a nonzero $H^2(\mathrm{UHP})$ function has nontangential boundary values nonzero a.e.), $D \equiv 0$. Hence
\begin{equation}
\varphi_+ = c_+^R\,F|_{\mathbb{R}} \text{ on all of } \mathbb{R},
\qquad c_+^L = c_+^R\,e^{i\pi\mu}.
\end{equation}
If instead $c_+^R = 0$, then $\varphi_+$ vanishes a.e. on $(0,\infty)$ and Fact T3 already forces $\varphi_+ \equiv 0$, hence $c_+^L = 0$ as well. In all cases $\varphi_+ = c_+\,B_+$ with $B_+ := F|_{\mathbb{R}}$ and $c_+ \in \mathbb{C}$ (possibly $0$).

The function $\psi := \overline{\varphi_-}$ lies in $H^2_+$ (conjugation flips Fourier support) and satisfies $L_0^+\psi = \bar\lambda\,\psi$ (conjugating $-i/(y-\tfrac{i}{2})$ gives $i/(y+\tfrac{i}{2})$, while $y$ and $d/dy$ are real). Applying the analysis above with exponent $\bar\mu$ gives $\psi = c\,B_+^{\bar\mu}$, hence
\begin{equation}
\varphi_- = c_-\,B_-,
\qquad
B_-(y) = \frac{y^\mu}{(y-\tfrac{i}{2})^2}\ (y>0),
\qquad
B_-(y) = \frac{e^{-i\pi\mu}|y|^\mu}{(y-\tfrac{i}{2})^2}\ (y<0),
\end{equation}
which is exactly the boundary value of the LHP branch ($\arg z \in (-\pi,0)$) of $z^\mu/(z-\tfrac{i}{2})^2$. Therefore
\begin{equation}
\varphi = c_+\,B_+ + c_-\,B_- \quad\text{on } \mathbb{R}\setminus\{0\},
\qquad (c_+,c_-) \neq (0,0).
\end{equation}
When $\varphi$ is real (in particular whenever $\lambda$ is real, on choosing a real eigenfunction) one has $\varphi_- = \overline{\varphi_+}$, so $c_- = \overline{c_+}$ and $\varphi = 2\,\mathrm{Re}(c_+ B_+)$, the form stated in Lemma 1.

\subsection{Quantization at the origin}

Near $y=0$ the eigenfunction is exactly a power times an analytic function, with no free regular remainder. Writing $A_+(y) = 1/(y+\tfrac{i}{2})^2$ and $A_-(y) = 1/(y-\tfrac{i}{2})^2$ (both analytic on $|y| < \tfrac12$, with $A_\pm(0) = -4$),
\begin{equation}
y>0: \quad \varphi(y) = y^\mu\,g_R(y), \qquad g_R = c_+ A_+ + c_- A_-,
\end{equation}
\begin{equation}
y<0: \quad \varphi(y) = |y|^\mu\,g_L(|y|), \qquad g_L(t) = c_+ e^{i\pi\mu}A_+(-t) + c_- e^{-i\pi\mu}A_-(-t),
\end{equation}
with leading coefficients
\begin{equation}
g_R(0) = -4\,(c_+ + c_-),
\qquad
g_L(0) = -4\,(c_+ e^{i\pi\mu} + c_- e^{-i\pi\mu}).
\end{equation}

If $g_R(0) = g_L(0) = 0$, then $(c_+, c_-)$ is annihilated by the matrix $\begin{pmatrix}1 & 1 \\ e^{i\pi\mu} & e^{-i\pi\mu}\end{pmatrix}$, whose determinant is $-2i\sin(\pi\mu)$. For $\mu \notin \mathbb{Z}$ this forces $c_+ = c_- = 0$, i.e. $\varphi \equiv 0$, a contradiction. So for non-integer $\mu$ at least one side of the origin carries the exact form $|y|^\mu g(|y|)$ with $g(0) \neq 0$.

In the case analysis below, $\mu = 1-\lambda$, and each half-line component is $c\,y^\mu/(y \mp \tfrac{i}{2})^2$ regardless of the window assumption.

\emph{Case 1: $\mathrm{Re}\,\mu \ge \tfrac32$ (equivalently $\mathrm{Re}\,\lambda \le -\tfrac12$).} On $(1,\infty)$, $|\varphi_+| = |c_+^R|\,y^{\mathrm{Re}\,\mu - 2}(1 + O(1/y))$, and since $\varphi_+ \sim c\,u_\lambda$ decays only as $y^{\mathrm{Re}\,\mu - 2} = y^{-1-\mathrm{Re}\,\lambda}$, the integral $\int_1^\infty y^{2\mathrm{Re}\,\mu - 4}\,dy$ diverges for $\mathrm{Re}\,\mu \ge \tfrac32$. Membership $\varphi_+ \in L^2$ forces $c_+^R = 0$, and likewise $c_+^L$ and both lower constants vanish, so $\varphi \equiv 0$. There are no eigenvalues with $\mathrm{Re}\,\lambda \le -\tfrac12$: no embedded point spectrum on or to the left of the essential line.

\emph{Case 2: $\mathrm{Re}\,\mu \le -\tfrac12$ (equivalently $\mathrm{Re}\,\lambda \ge \tfrac32$).} On $(0,\delta)$, $|\varphi_+| = |c_+^R|\,y^{\mathrm{Re}\,\mu}(1+O(y))$, and $y^{2\mathrm{Re}\,\mu}$ is not integrable at $0$ for $\mathrm{Re}\,\mu \le -\tfrac12$; $L^2$ forces all four constants to vanish, so $\varphi \equiv 0$.

\emph{Case 3: $\mathrm{Re}\,\mu \in (-\tfrac12,\tfrac32)$, $\mu \notin \mathbb{Z}$.} By non-degeneracy pick a side where $\varphi = |y|^\mu g(|y|)$ exactly, with $g$ analytic and $g(0) \neq 0$. Then (D3) fails:

- If $\mathrm{Re}\,\mu \in (-\tfrac12, 0]$: for $\mathrm{Re}\,\mu < 0$, $|\varphi| \sim 4|c|\,|y|^{\mathrm{Re}\,\mu} \to \infty$; for $\mathrm{Re}\,\mu = 0$ (so $\mathrm{Im}\,\mu \neq 0$, since $\mu \neq 0$), $|\varphi| \to |g(0)| \neq 0$ while oddness and continuity require $\varphi(0)=0$. Either way $\varphi$ is not continuous at $0$ with value $0$: (D3) is violated.
- If $\mathrm{Re}\,\mu \in (0,1]$: $\varphi(y)/y = (\pm)|y|^{\mu-1}g$; for $\mathrm{Re}\,\mu < 1$ the modulus diverges, and for $\mathrm{Re}\,\mu = 1$ with $\mathrm{Im}\,\mu \neq 0$ the factor $|y|^{i\,\mathrm{Im}\,\mu}$ oscillates with modulus $|g| \to 4|c| \neq 0$, so no limit exists. Thus $\varphi'(0)$ does not exist: (D3) is violated. ($\mu=1$ is an integer, excluded here.)
- If $\mathrm{Re}\,\mu \in (1, \tfrac32)$: $\varphi(0)=0$ and $\varphi'(0)=0$, so (D3) requires $\varphi(y)/y^2 \to a_2$ finite; but $|\varphi|/y^2 = |y|^{\mathrm{Re}\,\mu-2}|g| \to \infty$ since $\mathrm{Re}\,\mu - 2 < -\tfrac12 < 0$. (D3) is violated.

In each sub-case the obstruction is detected along every sequence $y \to 0$ inside the full-measure set where $\varphi$ agrees with the closed form: either the modulus diverges, or (when $\mathrm{Im}\,\mu \neq 0$) the modulus tends to a nonzero limit while the phase $|y|^{i\,\mathrm{Im}\,\mu} = e^{i\,\mathrm{Im}\,\mu\,\log|y|}$ sweeps the unit circle as $y\to 0$, so the pointwise limit demanded by (D3) does not exist. Hence Case 3 forces $\mu \in \mathbb{Z}$.

\emph{Case 3' (the $X$-form of the origin condition).} The same conclusion holds with (D3) replaced by the realization condition $\varphi'' \in L^2$ near the origin (the two conditions are logically incomparable, so this is not a special case): steps 5.3-5.5 use only (D1), (D2), (D4), and on the resulting closed form $|y|^\mu g(|y|)$ with $g(0) \neq 0$ the second derivative has leading term $\mu(\mu-1)g(0)\,|y|^{\mu-2}$, which fails $L^2$ near $0$ for every $\mathrm{Re}\,\mu < \tfrac32$ unless $\mu \in \{0,1\}$. Hence eigenfunctions of $L_0\vert_X$ obey the same quantization, and Theorem 2 applies verbatim to the $X$-realization.

\emph{Case 4: $\mu \in \mathbb{Z}$ within the window $\mathrm{Re}\,\mu \in (-\tfrac12,\tfrac32)$, i.e. $\mu \in \{0,1\}$ and $\lambda \in \{1,0\}$.} Now $z^\mu$ is entire, $B_\pm = y^\mu/(y\pm\tfrac{i}{2})^2$, and $\varphi = c_+ B_+ + c_- B_-$ is smooth on $\mathbb{R}$ and in $L^2$ for any constants. Oddness pins the combination:

- $\mu = 1$ ($\lambda = 0$): $\varphi(-y) = -\varphi(y) \iff c_- = c_+$, giving
\begin{equation}
\varphi = c_+\Big[\frac{y}{(y+\tfrac{i}{2})^2} + \frac{y}{(y-\tfrac{i}{2})^2}\Big]
= 2c_+\,\frac{y(y^2-\tfrac14)}{(y^2+\tfrac14)^2} = 2c_+\,y\,\Omega',
\end{equation}
one complex dimension: the scaling mode.
- $\mu = 0$ ($\lambda = 1$): $\varphi(-y) = -\varphi(y) \iff c_- = -c_+$, giving
\begin{equation}
\varphi = c_+\Big[\frac{1}{(y+\tfrac{i}{2})^2} - \frac{1}{(y-\tfrac{i}{2})^2}\Big]
= -2i\,c_+\,\frac{y}{(y^2+\tfrac14)^2} \propto \Omega + y\,\Omega',
\end{equation}
one complex dimension: the time-shift mode.

In both cases $\varphi$ is a complex multiple of a real odd eigenfunction of the form $2\,\mathrm{Re}(c\,B_+)$, with $c$ real for $\mu=1$ and $c$ imaginary for $\mu=0$.

Since non-real $\lambda$ gives non-integer $\mu$ (Case 3 applies), there are no non-real eigenvalues anywhere in $\mathbb{C}$. Cases 1-4 together prove Theorem 3, and hence Lemma 1 and Theorem 2. $\qquad\blacksquare$

\subsection{The smoothness hypothesis is sharp}

Dropping condition (D3) admits a continuum of eigenfunctions. The Case-3 functions are genuine odd, $L^2$, $C^1(\mathbb{R})$ distributional eigenfunctions of $L_0$ for every real $\lambda \in (-\tfrac12, 0)$ (and for complex $\lambda$ with $\mathrm{Re}\,\lambda \in (-\tfrac12,0)$): they solve the equation classically away from the origin and distributionally across it (a $C^1$ function carries no distributional surplus at a point). On a merely $C^1$ domain the open strip would therefore fill with a continuum of point spectrum. This is precisely the family $u_\lambda = y^{1-\lambda}/(y+\tfrac{i}{2})^2$ of Proposition 2 (Section 3.1), and it is exactly the essential-spectrum smear the discretizations render (Section 3.2). The requirement of two derivatives at the origin is the mathematical content of the informal statement that the collapse point is a regular point of the perturbation, and it is what distinguishes the realization $X$ in which the strip is empty. A representative computation (at $\lambda = -\tfrac14$, $\mu = \tfrac54$) confirms the sharpness: the odd real combination $2\,\mathrm{Re}(c\,B_+)$ is $C^1$ (its quotient $\varphi/y$ vanishes like $y^{1/4}$) but its second derivative diverges like $y^{\mu-2}$ at the origin, so it satisfies (D1), (D2), (D4) but fails (D3), exactly as the Case-3 analysis requires (verified numerically).

\subsection{Consequences and imported facts}

Theorem 2 is unconditional: $\sigma_{\mathrm{disc}}(L_0\vert_X) \cap \{\mathrm{Re} > -\tfrac12\} = \{0, +1\}$, realized only by the two symmetry modes, and the open strip $(-\tfrac12, 0)$ is empty. By Theorem 3 the full physical point spectrum of $L_0$ is $\{0,1\}$ over all of $\mathbb{C}$, so the essential line $\{\mathrm{Re}=-\tfrac12\}$ of Theorem 1 carries no embedded eigenvalues, and there is no discrete spectrum to its left. Combined with Theorem 1, the CLM collapse profile is linearly stable at $a=0$ in the spectral sense, with a gap $1 - c_l/2 = \tfrac12$ on $X$.

The following are standard Hardy-space and real-analysis results, cited but not reproved.

- \textbf{T1} (Paley-Wiener/Hardy). $H^2(\mathrm{UHP})$ boundary values are exactly the $L^2$ functions with Fourier support in $[0,\infty)$, with nontangential a.e. and $L^2$ convergence to the boundary. (Koosis \cite[Ch. VI]{Koosis1998}; Duren \cite[Ch. 11]{Duren1970}; Garnett \cite[Ch. II]{Garnett2007}.)
- \textbf{T2}. $H^\infty(\mathrm{UHP})$ multipliers preserve $H^2(\mathrm{UHP})$, and boundary values multiply via nontangential limits (the two-line argument is given in 5.3.2).
- \textbf{T3} (boundary uniqueness). A nonzero $H^p(\mathrm{UHP})$ function ($p\ge 1$) has boundary values nonzero a.e.; vanishing on a set of positive Lebesgue measure forces the zero function. (Koosis \cite[Ch. VI]{Koosis1998}; Garnett \cite[Cor. II.4.2]{Garnett2007}; the F. and M. Riesz theorem.)
- \textbf{T4}. A distribution on an interval with locally integrable distributional derivative is a.e. equal to an absolutely continuous function, with the a.e.-classical derivative; $W^{1,1}_{\mathrm{loc}} = \mathrm{AC}_{\mathrm{loc}}$. (Brezis \cite[Thm. 8.2, Rem. 8]{Brezis2011} for the distributional step; Rudin \cite[Thm. 7.18-7.20]{Rudin1987} for the classical AC characterization.)
- \textbf{T5} (Paley-Wiener-Schwartz for tube domains). A function holomorphic on the upper half-plane of tempered growth, $|U(w)| \le C(1+|w|)^N(\mathrm{Im}\,w)^{-M}$, has a boundary value in $\mathcal{S}'(\mathbb{R})$ with Fourier support in $[0,\infty)$; if the boundary value lies in $L^2$, the function belongs to $H^2(\mathrm{UHP})$. (Stein and Weiss \cite[Ch. III]{SteinWeiss1971}; Hörmander \cite[Thm. 7.4.2-7.4.3]{Hormander1990}.)

\section{\texorpdfstring{The scaling-critical dissipation exponent $s^*(a)$}{The scaling-critical dissipation exponent s*(a)}}

We now add fractional dissipation and ask when the collapse survives. Consider
\begin{equation}
w_t + a\,u\,w_x = u_x\,w - \nu\,\Lambda^s\,w, \qquad \Lambda = (-\partial_{xx})^{1/2}, \ \ \nu > 0, \ \ s\in(0,2].
\end{equation}

\subsection{The scaling-critical exponent}

In self-similar variables $w = (T-t)^{-1} W(y,\tau)$, $y = x/(T-t)^{c_l}$, $\tau = -\log(T-t)$, the inviscid part is autonomous, $\partial_\tau W = -\mathrm{Res}[W]$ with fixed point $\Omega$. Since $\Lambda^s$ scales as $(\mathrm{length})^{-s} = (T-t)^{-s c_l}$, the dissipation enters the rescaled equation with a time-dependent coefficient,
\begin{equation}
\partial_\tau W = -\mathrm{Res}[W] - \nu\,e^{-\gamma\tau}\,\Lambda^s_y\,W, \qquad \gamma := 1 - s\,c_l.
\end{equation}
Define the scaling-critical dissipation exponent
\begin{equation}
s^*(a) = \frac{1}{c_l(a)},
\end{equation}
the value at which $\gamma = 0$. $s^*$ is a formal scaling (dissipation-relevance) threshold read off the self-similar exponent: for a fixed, sufficiently regular perturbation the rescaled dissipation coefficient $e^{-\gamma\tau}$ vanishes for $s < s^*$, so the dissipation is asymptotically subdominant in the self-similar limit. That the inviscid profile then persists as the attractor is the program of Section 8, not established here (the required weighted dissipation-form bound, nonlinear estimate, and modulation closure are open, Appendix A); and $s^*$ is not the sharp critical dissipation curve separating blow-up from global regularity, which for this family remains unknown.
For $s < s^*(a)$ we have $\gamma > 0$: the dissipation coefficient $e^{-\gamma\tau}\to 0$ and, for fixed sufficiently regular perturbations, the dissipation is asymptotically irrelevant in the self-similar limit. The case $s = s^*$ is marginal ($\gamma=0$) and $s > s^*$ is relevant (the open supercritical regime).

Read off the recomputed $c_l(a)$ of Section 2, $s^*(a)$ rises from $1$ at $a=0$ toward $\infty$ as $a\to a_c$ (reaching about $54$ near the branch endpoint, beyond the tabulated points; Figure 3); for the ordinary Laplacian $s=2$ the sub/supercritical boundary sits at $a\approx 0.39$ (where $c_l = 1/2$; the branch inverts to $a = 0.386$). The map is consistent with the two rigorously known cross-checks, both of them sub/supercritical-side checks rather than confirmations of the value $1/c_l$. At $a=0$, $s^* = 1$: the raw CLM blows up even at the full Laplacian $s=2$, which is supercritical ($s=2 > s^*=1$), so $s^*$ is consistent with Schochet only as a formal scaling-relevance exponent (below which the rescaled dissipation coefficient vanishes as $\tau\to\infty$), not as a proved persistence threshold or a blow-up/regularity threshold. Sakajo's generalized-viscosity results sharpen this reading at $a=0$: blow-up persists at sufficiently small viscosity regardless of the derivative order of the dissipation \cite{Sakajo2003blowup}, with global solutions at large viscosity \cite{Sakajo2003global}, so no dissipation order is by itself regularizing there. Since $a=0$ is the exact anchor of the branch ($c_l(0)=1$), this point only confirms that Schochet is not contradictory once $s^*$ is so read; it does not test the computed branch. The branch value is tested at $a=1/2$, where $s^* = 3$ exactly ($c_l(\tfrac12) = \tfrac13$ by the exact solution of \cite{Chen2020}; also \cite[Thm. 2]{LushnikovSilantyevSiegel2021}) and J. Chen \cite{Chen2020} proved blow-up at $s = 2 < 3$ (subcritical), consistent with persistence. The only quantitative validation of the branch itself is the $a_c$ cross-check of Section 2.

\begin{figure}[htbp]\centering
\includegraphics[width=0.75\textwidth]{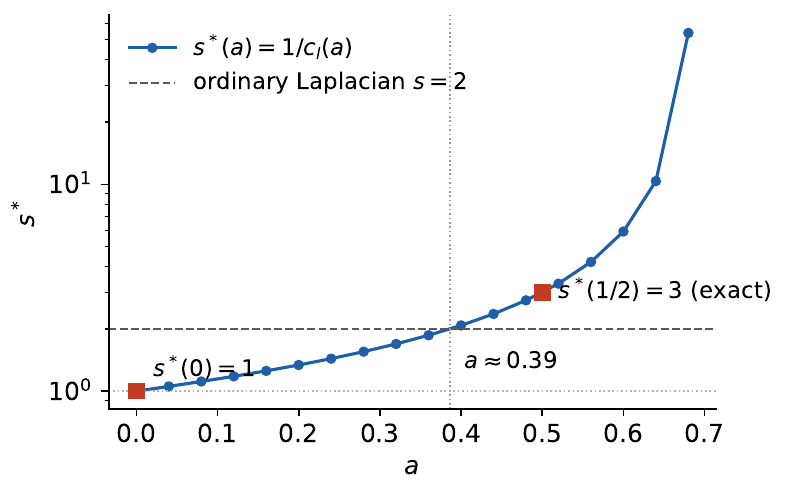}
\caption{The scaling-relevance exponent $s^*(a) = 1/c_l(a)$ read off the recomputed branch of Section 2 (log scale), rising from $s^*(0) = 1$ (Schochet's supercritical full-Laplacian blow-up) through $s^*(1/2) = 3$ (J. Chen's subcritical $s=2$ blow-up) to about $54$ near the branch endpoint (beyond the tabulated points), and crossing the ordinary-Laplacian value $s = 2$ at $a \approx 0.39$ where $c_l = 1/2$. This is a formal scaling diagnostic, not a proved persistence threshold.}
\end{figure}

\section{Numerical methods}

All numerical results in this paper are diagnostic; the closed-form theorems of Sections 3 to 5 and Appendix A do not depend on any of them.

\emph{Profile branch.} The profile solver compactifies the real line by $y = c\tan(\theta/2)$ with the Hilbert transform transplanted exactly onto the circle, and Newton-continues from the exact CLM anchor in steps $\Delta a = 0.04$, reaching Newton residual $10^{-13}$-$10^{-10}$ across the branch (how well the discretized equation is solved, not the solution accuracy). The solution accuracy at $a=0$ is $\sim 3\times 10^{-4}$, set by the first-order transplanted Hilbert transform; the two-grid difference $|c_l(1024) - c_l(2048)|$ is $\lesssim 5\times 10^{-4}$, a self-consistency measure rather than a certified continuum bound. Tabulated $c_l$ off the $\Delta a = 0.04$ grid are degree-6 polynomial interpolations (Section 2). The reference branch is that of \cite{LushnikovSilantyevSiegel2021}.

\emph{Linearized spectrum.} The spectrum is assembled three ways, all diagnostic. (i) A dense finite-difference Jacobian of the rescaled residual on the compactified grid ($N = 1024$-$8192$) at the sampled advections $a = 0.3, 0.4, 0.5, 0.6, 0.65$; strip candidates are classified as resolution-stable or drifting by tracking them across $N$, and a centroid-frequency filter ($< N/16$) removes the near-Nyquist pollution that proliferates as $a\to a_c$ (a heuristic threshold, not a certified separation). (ii) A log-scale (Mellin-adapted) discretization on $x = \ln y$, in which the Hilbert transform becomes an explicit exponential-kernel convolution and the spectral derivative is $F^{-1}\mathrm{diag}(ik)F$ (the code asserts that this annihilates constants and differentiates a Fourier mode to machine accuracy). By the realization dichotomy (Proposition 2) this fixed-lower-cutoff grid renders the maximal-$L^2$ realization: at $a=0.5$ its rightmost mode sits at $\mathrm{Re} = +2.4931$, within $0.3\%$ of the plain-$L^2$ origin indicial line $+5/2$, so it corroborates Proposition 2 rather than certifying the far-field edge. (This value is the output after the principal-value diagonal of the $g$-representation Hilbert kernel was corrected to its finite part $-1/(2\pi)$; the run log ships with the source.) A non-periodic finite-difference variant removes the FFT-periodicity pollution but is not grid-converged below $M_x \approx 4000$. (iii) The exact Hardy/pole-basis computations back Theorem 2 in closed form. A dynamic-rescaling computation at $a=0$ is an exact-profile rate and solver sanity check (it reads the self-consistent scaling rate off the exact CLM profile), not a forward relaxation of the $a>0$ branch.
\emph{Evans-determinant count.} The topological strip count $n_{\mathrm{disc}}(a)$ of Section 3.2 uses a control contour around the pinned symmetry modes and anchors $a=0$ to the proven empty strip of Theorem 2; the per-point phase-increment and integer-defect quantities are recorded as diagnostics but are not enforced as hard gates, and the run outputs are not retained. It returns $0$ for all computed $a\in[0,0.65]$ but is numerical evidence, not proof, with three recorded gaps (a uniform large-imaginary-part bound, trace-ideal membership of the kernel, and quadrature-error bounds in the trace norm), which the code computes as diagnostics rather than enforcing as interval bounds; the code is supplied for rerunning. Upgrading any of these would move this strand toward a computer-assisted proof.

The results above are numerical evidence; at $a=0$ they accompany theorems, and for $a>0$ they stand on their own.

\section{Discussion}

The theorems are at $a=0$. Theorem 1 fixes the essential spectrum of $L_0$ on the origin-$H^2$ realization, in the closed half-plane $\{\mathrm{Re} \ge -\tfrac12\}$, as the single vertical line $\{\mathrm{Re} = -\tfrac12\}$: a Mellin diagonalization of the far-field dilation and a log-widening Weyl sequence place the line in the essential spectrum, and a constructive $X$-resolvent bound, an explicit $\alpha^{-3/2}$ majorant, empties the open strip. Away from $a=0$ the essential spectrum acquires a second, origin line whose position is set by the second-derivative weight of $X$; that both lines lie in the essential spectrum is proven for each admissible smooth profile (Proposition 1, a conditional inclusion), while whether the essential spectrum equals their union for $a>0$ is an open exactness question we do not address here. Emptying the rest of the strip rests on the subordinacy of the nonlocal terms at both endpoints; the Hilbert term is non-compact, so a compact-perturbation argument does not apply. Theorem 2 excludes discrete spectrum in the strip at $a=0$ in closed form, via the Hardy collapse of $L_0$ to a scalar first-order operator with the explicit eigenfunction $u_\lambda = y^{1-\lambda}/(y+\tfrac{i}{2})^2$; its completeness input (Lemma 1) is proven in full, so Theorem 2 is unconditional; and since the family $u_\lambda$ fails to be square-integrable at infinity for $\mathrm{Re}\,\lambda \le -\tfrac12$ (it decays only as $y^{-1-\mathrm{Re}\,\lambda}$), no eigenvalue sits on or to the left of the essential line either, so the full physical point spectrum of $L_0$ over $\mathbb{C}$ is $\{0,1\}$ with no embedded eigenvalues. Combined with Theorem 1, the CLM collapse profile is linearly stable at $a=0$ in the spectral sense, with a gap of $\tfrac12$ on $X$; Appendix A makes the linear semigroup and its exact decay rate explicit on the conjugated weighted space, which is not the $X$ norm. Beyond $a=0$, the discrete exclusion rests on numerical evidence at the sampled advections (Section 3.2). The scaling-relevance exponent $s^*(a) = 1/c_l(a)$ is computed from the recomputed branch and is consistent with its two sub/supercritical cross-checks. The small-$a$ discrete-exclusion program, the dissipative-persistence argument, and the $a>0$ exactness are set out as programs in the next paragraph, and left to future work.

Four directions build on the $a=0$ results above. (i) \emph{The $a>0$ discrete exclusion.} The two symmetry modes are pinned at $0$ and $+1$ for all $a$; a perturbative origin-$H^2$ program excludes eigenvalue emergence on compact strip-subsets for small $a$, and reduces the genuinely open core to a single imaginary-part-independent essential-edge (threshold-resonance) estimate, the closed-form scalar reduction of Theorem 2 being exact only at the single-pole $a=0$ profile. (ii) \emph{The $a>0$ exactness.} Upgrading the two-line inclusion of Proposition 1 to a full essential-spectrum theorem for all $a$ calls for the two-ended, two-weight Fredholm argument sketched in Section 4.5. (iii) \emph{Dissipative persistence.} For $s < s^*(a)$ the inviscid profile is expected to remain the attractor of the rescaled dynamics. What Appendix A supplies is the linear input in its final form: the exact conjugated semigroup $S_{\mathcal Y}(\tau)$ on $\mathcal Y_{3/2} = Y_{3/2}\oplus Y_{3/2}$ with norm $e^{-\tau/2}$, and the bounded transfer $J_X : X\to\mathcal Y_{3/2}$ carrying physical data into it. Three inputs are still missing, and none is supplied here: a lower bound for the dissipation form in the $\mathcal Y_\theta$ norm, a quadratic estimate for $N(\varphi) = \varphi\,H\varphi$ in the same norm, and the modulation closure that removes the two symmetry directions along the flow (plus, for $a>0$, a Gearhart-Prüss semigroup-decay bound for the non-normal $L_a$, since a spectral gap alone does not give one). (iv) \emph{A computer-assisted-proof route.} Combining a certified inviscid gap with the dissipative perturbation bound would give a computer-assisted-proof scaffolding for self-similar blow-up of a dissipative one-dimensional model as a spectral object; the supercritical regime $s > s^*(a)$, where the dissipation is relevant, is outside the present framework.

Two natural approaches do not reach the gap. A weighted-energy/coercivity estimate cannot certify the gap: the far field is already coercive in plain $L^2$ with the sharp edge, but a local positive multiplier $+\varphi\,H\Omega$ near the origin (with $H\Omega(0)>0$) survives every $L^2$-equivalent weight, so no coercivity certificate in an $L^2$-equivalent norm reaches the spectral gap. (The small-$a$ coercivity estimate of Elgindi, Ghoul, and Masmoudi \cite[Prop. 2.1]{ElgindiGhoulMasmoudi2021}, for odd $f$ with $f'(0) = Hf(0) = 0$ in a singular-weighted space, is not a counterexample but an instance of the realization mechanism of Section 3.1: the singular weight and origin constraints define their own realization, in which a gap is certified; it neither contradicts nor supplies the $X$-gap here.) This is expected, since $L_a$ is strongly non-normal (the eigenvector matrix is severely ill-conditioned, its condition number growing with resolution) and the gap lives in the spectrum, not the numerical range. Likewise a naive interval-arithmetic resolvent enclosure is unavailable: a weight is a change of norm, eigenvalues are norm-invariant, and the truncated $L_a$ genuinely carries essential-smear eigenvalues inside the strip (the compactified grid mis-renders the continuous line at the wrong real part), so no weighted enclosure can exclude them. The consequence is that the rigorous discrete exclusion must be \emph{analytical} (the $a=0$ anchor plus continuation) rather than a black-box computer-assisted enclosure; where a computer-assisted proof does enter, it must first be given a spectrally correct rendering of the essential spectrum (a non-periodic log-Mellin operator at high resolution feeding an argument-principle count), not applied to the raw grid. This mirrors the viscous three-dimensional Navier-Stokes nonuniqueness proof of Hou, Wang, and Yang \cite{HouWangYang2025}, where the computer-assisted step certifies invertibility only on a finite-rank image, after the continuous spectrum has been carried by a coercive part and the remainder rendered compact; our spectrally correct essential-line rendering feeding an argument-principle count is the same requirement in the gCLM-collapse setting.

Three bodies of work are the relevant incumbents, and what this paper adds is different in each case. First, the modulation stability of Elgindi, Ghoul, and Masmoudi \cite{ElgindiGhoulMasmoudi2021} proves inviscid stable self-similar blow-up for this family near a known self-similar family; we instead analyze the linearized operator's spectrum directly (essential spectrum and gap) and add fractional dissipation, at the cost that ours is a linear spectral statement at $a=0$, not a completed nonlinear-stability theorem. Second, J. Chen \cite{Chen2020} proves real-line dissipative self-similar blow-up for $a$ near $\tfrac12$ at the Laplacian $s=2$ by nonlinear stability of an approximate profile; this is a stability-plus-persistence result in precisely the regime where we cross-check $s^*$, and our contribution relative to it is the $a=0$ spectral picture together with a continuous-in-$s$ scaling diagnostic across the advection branch, rather than an energy estimate at a single advection value. Third, the pole-dynamics program of Lushnikov, Silantyev, Siegel, and Ambrose \cite{LushnikovSilantyevSiegel2021,AmbroseLushnikovSiegelSilantyev2024,SilantyevLushnikovSiegelAmbrose2025} developed the complex-pole continuation machinery (and the value $a_c$) and exact dissipative gCLM solutions at discrete integer dissipation orders and discrete advection values on the periodic line; the difference is real-line versus periodic, a linear-stability spectral gap versus exact pole trajectories, and a continuous fractional-$s$ scaling-relevance map versus isolated integer orders, built with spectral and weighted-energy machinery rather than pole dynamics. We note that \cite{SilantyevLushnikovSiegelAmbrose2025} already gives exact dissipative gCLM blow-up at $a=0$ and $a=1/2$, the two advection values at which $s^*(a)$ is cross-checked in Section 6; the overlap is at those points, and what is new here is the spectral object and the continuous map, not the blow-up itself. The real-line versus periodic distinction is essential, since on the real line arbitrarily small data can blow up even at $s\ge 1$ \cite{AmbroseLushnikovSiegelSilantyev2024}, making the real-line collapse the nontrivial setting. The inviscid computer-assisted-proof lineage of Chen, Hou, and Huang \cite{ChenHouHuang2021,ChenHou2022partI,ChenHou2023partII} and the smooth-profile constructions of Huang and collaborators \cite{HuangQinWangWei2024} supply the structural fact we exploit, that $a>0$ gives a one-scale smooth profile that is spectrally resolvable; and the machine-learning unstable-singularity thread \cite{WangLaiGomezSerranoBuckmaster2023,DeepMindUnstable2025,WangLegerLaiBuckmaster2025} works on the inviscid incompressible-porous-media, Boussinesq, Euler-with-boundary, and Córdoba-Córdoba-Fontelos equations, deliberately not on gCLM-with-dissipation as a spectral object, so it does not overlap our target.

Dissipative one-dimensional blow-up and inviscid stability for this family are established by other methods, as set out above; what they do not supply, and what the programs above pursue, is the spectral and computer-assisted-proof construction for the dissipative problem across the whole advection branch.

\section*{Acknowledgements}

Claude (Anthropic; models Claude Opus 4.8 and Claude Fable 5) was used in preparing this work, including drafting the manuscript, prior-art and literature search, and adversarial checking of the arguments. All mathematical results were derived and verified by the author, who takes full responsibility for the correctness and integrity of the work.

The author reports no conflicts of interest.

\section*{Data availability statement}

The data that support the findings of this study are available from the author upon reasonable request.

\appendix
\renewcommand{\thesection}{Appendix~\Alph{section}}
\renewcommand{\thesubsection}{\Alph{section}.\arabic{subsection}}
\renewcommand{\theequation}{\Alph{section}.\arabic{equation}}
\section{\texorpdfstring{The exact linear semigroup at $a=0$}{The exact linear semigroup at a=0}}

This appendix records the exactly solvable linear structure that underlies the persistence program of Section 8. With fractional dissipation of order $s < s^*(0) = 1$ added to the Constantin-Lax-Majda (CLM) equation and the collapse rescaled, the linearized generator is the operator $L_0$ of Section 3, whose Hardy diagonalization makes its semigroup and decay rate explicit. We give the non-autonomous rescaled setup (Section A.1), the Hardy-reduced operator (Section A.2), the exact semigroup (Section A.3), and its exact decay rate (Section A.4). These are the proven linear inputs to the persistence argument; the nonlinear closure that would complete persistence into a theorem is summarized as a program in Section 8.

\subsection{The rescaled equation and the perturbation problem}

Add fractional dissipation to CLM:
\begin{equation}
w_t = w\,Hw - \nu\,\Lambda^s w, \qquad \Lambda = (-\partial_{xx})^{1/2}, \quad \nu > 0, \quad s \in (0,1),
\end{equation}
with $H$ the Hilbert transform. In the self-similar variables of Section 6,
\begin{equation}
w(x,t) = (T-t)^{-1}\,W(y,\tau), \qquad y = \frac{x}{(T-t)^{c_l}}, \qquad \tau = -\log(T-t),
\end{equation}
with the exact CLM exponent $c_l = 1$, the rescaled equation is non-autonomous with an exponentially small dissipation coefficient,
\begin{equation}
\partial_\tau W = -\mathrm{Res}[W] - \nu\,e^{-\gamma\tau}\,\Lambda^s_y W, \qquad \gamma = 1 - s\,c_l = 1 - s > 0,
\end{equation}
whose inviscid part has the fixed point $\Omega(y) = -y/(y^2 + \tfrac14)$. Writing $W = \Omega + \varphi$ and separating the linear part $L_0 = -\,d\,\mathrm{Res}/d\Omega$ (the $a=0$ operator of Section 3),
\begin{equation}
\partial_\tau\varphi = L_0\varphi - \nu\,e^{-\gamma\tau}\Lambda^s_y\varphi + N(\varphi) + f(\tau), \qquad f(\tau) = -\nu\,e^{-\gamma\tau}\Lambda^s_y\Omega,
\end{equation}
with quadratic remainder $N(\varphi) = \varphi\,H\varphi$, the CLM nonlinearity. Let $P$ project onto the two exact symmetry modes of Section 3.2 (eigenvalues $0$ and $+1$) and let $Q = I - P$ (its realization as a bounded transfer map $J_X : X \to \mathcal Y_{3/2} = Y_{3/2}\oplus Y_{3/2}$ into the conjugated weighted decay space is constructed in Section A.4). The standard modulation removes the $P$-component by choosing the collapse time and spatial scale, so persistence is the statement that the $Q$-component $Q\varphi(\tau)$ decays and $W(\tau)\to\Omega$.

Three structural facts drive the argument. (i) $L_0$ has a spectral gap: by Theorem 2 and Lemma 1 its spectrum to the right of the essential line $\{\mathrm{Re} = -\tfrac12\}$ consists only of the modes $\{0,+1\}$. (ii) The dissipative term is dissipative in the flat $L^2$ pairing (its Fourier symbol $-\nu e^{-\gamma\tau}|k|^s \le 0$) and carries a coefficient that vanishes as $\tau\to\infty$; it is not a bounded perturbation to be dominated, $\Lambda^s$ being of order $s$. Whether it is also form-nonpositive, or merely harmless, in the weighted norm $Y_\theta$ of the decay bound is exactly the dissipation-form input left open in the program (Section 8). (iii) The forcing $f$ is finite and decays like $e^{-\gamma\tau}$: the exact profile gives
\begin{equation}
\widehat{\Lambda^s\Omega}(k) = i\pi\,\mathrm{sgn}(k)\,|k|^s\,e^{-|k|/2}, \qquad \lVert\Lambda^s\Omega\rVert_{L^2}^2 = \pi\,\Gamma(2s+1) < \infty .
\end{equation}
The forcing reaches the decay estimate through $J_X$, not directly. In the conjugated direct-sum norm used for the decay bound (Section A.4) the unprojected forcing is not even in $\mathcal Y_\theta$ for any $\theta > 0$: its Hardy component $f_+ = P_+ f$ has $f_+(0) = -\nu e^{-\gamma\tau}\tfrac{i}{2}\Gamma(s+1)2^{s+1} \neq 0$ (the Hardy projection does not inherit the odd real function's linear vanishing), so the core integrand $\sim y^{-2\theta-1}$ diverges. The projection removes exactly this: $J_X$ subtracts the zeroth and first Taylor coefficients on each block, and since this forcing is analytic at the origin the projected blocks are $O(y^2)$ there, with core integrand $\sim y^{3-2\theta}$ integrable for $\theta < 2$; the tail is finite for every $\theta > 1$ because the projected forcing is closed-form, $P_+\Lambda^s\Omega = \tfrac{i}{2}\Gamma(s+1)(\tfrac12 - iy)^{-(s+1)}$, decaying like $|y|^{-1-s}$; hence $\lVert J_X f\rVert_{\mathcal Y_\theta} < \infty$ for $\theta\in(1,2)$. The remaining analytic content is a semigroup-decay bound off the two modes, a quadratic estimate for $N$, and a lower bound for the dissipation form in the norm of the decay bound. The first is supplied at $a=0$ by the explicit Hardy diagonalization of $L_0$, developed below; the quadratic estimate and the dissipation-form lower bound are the open inputs of the persistence program (Section 8).

\subsection{The Hardy-reduced operator}

By the Hardy reduction of Theorem 2, the exact profile makes the nonlocal operator scalar. On the upper-half-plane Hardy space $H^2_+$ (where $H = -i$) the single-simple-pole identity
\begin{equation}
H\Omega - i\,\Omega = \frac{i}{y + \tfrac{i}{2}}
\end{equation}
holds, so $L_0$ block-diagonalizes on the Hardy split $\varphi = \varphi_+ + \varphi_-$, $\varphi_\pm \in H^2_\pm$,
\begin{equation}
L_0 = L_0^+ \oplus L_0^-, \qquad L_0^- = \overline{L_0^+}, \qquad L_0^+ u = -u - y\,u' + \frac{i}{y + \tfrac{i}{2}}\,u .
\end{equation}
The nonlocality has been fully absorbed: $L_0^+$ is a local first-order transport operator with a smooth bounded zeroth-order term. On the $X_+$-realization (the origin condition $u'' \in L^2$) its point spectrum is $\{0, +1\}$, realized by
\begin{equation}
u_0(y) = \frac{y}{(y + \tfrac{i}{2})^2} \ \ (\text{eigenvalue } 0), \qquad u_1(y) = \frac{1}{(y + \tfrac{i}{2})^2} \ \ (\text{eigenvalue } +1),
\end{equation}
the Hardy representatives of the scaling mode $y\,\Omega'$ and the time/amplitude mode $\Omega + y\,\Omega'$; the essential spectrum is the line $\{\mathrm{Re} = -\tfrac12\}$ of Theorem 1. Everything below is carried out on $L_0^+$; the bound for the full $L_0$ follows from the direct sum and $L_0^- = \overline{L_0^+}$.

\subsection{The exact semigroup}

\begin{namedthm}{Proposition}{A.1}{explicit semigroup}
The semigroup of $L_0^+$ is, for arbitrary amplitude $u$,
\begin{equation}
\big(e^{L_0^+\tau}u\big)(y) = e^{\tau}\left(\frac{y\,e^{-\tau} + \tfrac{i}{2}}{y + \tfrac{i}{2}}\right)^{2} u\big(y\,e^{-\tau}\big).
\end{equation}
\end{namedthm}

\emph{Proof.} Solve $\partial_\tau u = L_0^+ u = -u - y\,\partial_y u + \tfrac{i}{y + i/2}\,u$ by characteristics. The equation $dy/d\tau = y$ gives $y(\tau) = y_0 e^\tau$, so the characteristic through $(y,\tau)$ has foot $y_0 = y e^{-\tau}$. Along it $du/d\tau = \big(-1 + \tfrac{i}{y(\tau) + i/2}\big)u$, whence
\begin{equation}
\big(e^{L_0^+\tau}u\big)(y) = e^{-\tau}\,u(y e^{-\tau})\,\exp\!\left(\int_0^\tau \frac{i}{y e^{-\sigma} + \tfrac{i}{2}}\,d\sigma\right).
\end{equation}
Substituting $w = y e^{-\sigma}$ (so $d\sigma = -dw/w$, and $\sigma : 0 \to \tau$ maps $w : y \to y e^{-\tau}$) and using the partial fraction $\tfrac{i}{w(w + i/2)} = \tfrac{2}{w} - \tfrac{2}{w + i/2}$,
\begin{equation}
\int_0^\tau \frac{i}{y e^{-\sigma} + \tfrac{i}{2}}\,d\sigma
= \int_{y e^{-\tau}}^{y} \frac{i}{w(w + \tfrac{i}{2})}\,dw
= 2\Big[\ln w - \ln(w + \tfrac{i}{2})\Big]_{y e^{-\tau}}^{y}.
\end{equation}
Exponentiating and simplifying with $y/(y e^{-\tau}) = e^\tau$ yields the multiplier $e^{2\tau}\big((y e^{-\tau} + \tfrac{i}{2})/(y + \tfrac{i}{2})\big)^2$; combined with the $e^{-\tau}$ prefactor this is the stated formula. $\qquad\blacksquare$

The formula solves the evolution identically for any amplitude, and acts on the eigenfunctions $u_\lambda(y) = y^{1-\lambda}/(y + \tfrac{i}{2})^2$ of Theorem 2 as $e^{L_0^+\tau}u_\lambda = e^{\lambda\tau}u_\lambda$ (both verified symbolically, and against an independent finite-difference matrix exponential to relative error $\sim 10^{-5}$).

\subsection{The scale-covariant space and the exact decay rate}

Conjugating away the potential term exposes a pure dilation. Let $M$ denote multiplication by $(y + \tfrac{i}{2})^2$ and set $v = M u = (y + \tfrac{i}{2})^2 u$. Then (verified symbolically)
\begin{equation}
M\,L_0^+\,M^{-1} = A := 1 - y\,\partial_y, \qquad \big(e^{A\tau}v\big)(y) = e^{\tau}\,v(y e^{-\tau}),
\end{equation}
a pure weighted dilation, which one reads directly off Proposition A.1. The eigenmodes of $A$ are $v = y^n$ with eigenvalue $1 - n$; the two physical modes are the lowest two, $v = 1$ ($u_1$, eigenvalue $+1$) and $v = y$ ($u_0$, eigenvalue $0$). The left eigenvectors dual to $y^n$ are the Taylor-coefficient functionals $\ell_n(v) = v^{(n)}(0)/n!$, so the spectral projection onto the two modes and its complement are
\begin{equation}
P v = v(0) + v'(0)\,y, \qquad Q v = v - v(0) - v'(0)\,y .
\end{equation}
Thus $Q$ is the regular part at the collapse point: it removes the zeroth and first Taylor coefficients of $v$ at $y = 0$, matching the origin-regularity mechanism of the realization $X$ of Theorem 1. So $Qv$ has vanishing zeroth and first traces at the origin; for $C^2$ or analytic data $Qv = O(y^2)$ as $y\to 0$, while for transferred $X$-data one has only $Qv = o(y^{3/2})$, with the $Y_{3/2}$ integrability supplied by the double Hardy inequality below.

\begin{namedthm}{Definition}{A.2}{the weighted semigroup space $Y_\theta$}
The decay estimate lives on a weighted space of the conjugated variable $v = b^2 u$, $b = y+\tfrac{i}{2}$, which is distinct from the physical realization $X$. For $\theta > 0$, on the half-line,
\begin{equation}
\lVert v\rVert_{Y_\theta}^2 = \int_0^\infty |v(y)|^2\,y^{-2\theta - 1}\,dy
= \int_0^\infty |u(y)|^2\,(y^2 + \tfrac14)^2\,y^{-2\theta - 1}\,dy .
\end{equation}
Because the weight $y^{-2\theta-1}$ blows up at the origin, elements of $Y_\theta$ carry no pointwise origin traces; the modulation projection is therefore defined on the physical $X$ (which has traces) and mapped into $Y_\theta$, not applied within $Y_\theta$ (the transfer estimate below).
\end{namedthm}

\begin{namedthm}{Proposition}{A.3}{exact operator norm}
For every $\theta > 0$,
\begin{equation}
\lVert e^{A\tau}\rVert_{Y_\theta \to Y_\theta} = e^{(1 - \theta)\tau} \qquad (\text{constant } 1, \text{ every function}).
\end{equation}
\end{namedthm}

\emph{Proof.} For a pure-power weight the weighted-composition norm is exact and function-independent. With $T_\tau = e^{A\tau}$ and the push-forward $z = y e^{-\tau}$,
\begin{equation}
\lVert T_\tau v\rVert_{Y_\theta}^2 = e^{3\tau}\!\int |v(z)|^2 (z e^{\tau})^{-2\theta - 1}\,dz = e^{(2 - 2\theta)\tau}\lVert v\rVert_{Y_\theta}^2 ,
\end{equation}
the weight ratio $W(z e^\tau)/W(z) = e^{-(2\theta + 1)\tau}$ being constant in $z$. Taking square roots gives the claim. $\qquad\blacksquare$

The substitution $y = e^t$ makes the structure behind Proposition A.3 explicit and supplies the generator. The map
\begin{equation}
(Wv)(t) := e^{-\theta t}\,v(e^t)
\end{equation}
is unitary from $Y_\theta$ onto $L^2(\mathbb{R},dt)$, since $\int_0^\infty \lvert v\rvert^2 y^{-2\theta-1}\,dy = \int_{\mathbb{R}} \lvert v(e^t)\rvert^2 e^{-2\theta t}\,dt$. Under $W$ the weighted dilation becomes a translation times a scalar,
\begin{equation}
W\,e^{A\tau}\,W^{-1} = e^{(1-\theta)\tau}\,\mathcal T_\tau, \qquad (\mathcal T_\tau g)(t) = g(t-\tau),
\end{equation}
because $W(e^{A\tau}v)(t) = e^{\tau}e^{-\theta t}v(e^{t-\tau}) = e^{(1-\theta)\tau}(Wv)(t-\tau)$. Since $(\mathcal T_\tau)_{\tau\in\mathbb{R}}$ is a strongly continuous unitary group on $L^2(\mathbb{R})$, $(e^{A\tau})_{\tau\in\mathbb{R}}$ is a strongly continuous group on $Y_\theta$ with $\lVert e^{A\tau}\rVert_{Y_\theta\to Y_\theta} = e^{(1-\theta)\tau}$ exactly, which is Proposition A.3 again. Its generator is $A = 1 - y\partial_y$, carried by $W$ to $(1-\theta) - \partial_t$, with domain
\begin{equation}
D(A_\theta) = \{\,v\in Y_\theta : y\,v'\in Y_\theta\,\} = W^{-1}H^1(\mathbb{R}),
\end{equation}
the identification following from $(Wv)'(t) = W\big(y v' - \theta v\big)(t)$; the smooth compactly supported functions on $(0,\infty)$ form a core. In particular $A_\theta$ is normal on $Y_\theta$, equal to $(1-\theta)$ plus a skew-adjoint part, with spectrum the single vertical line $\{\mathrm{Re} = 1-\theta\}$ and $e^{A\tau}$ equal to $e^{(1-\theta)\tau}$ times a unitary group; this normality is the structural reason the operator norm is exact and function-independent. Choosing $\theta = \tfrac32$ places the rate at the essential edge $1 - c_l/2 = \tfrac12$ of Theorem 1:
\begin{equation}
\lVert e^{A\tau}\rVert_{Y_{3/2}} = e^{-\tau/2} .
\end{equation}
In $Y_{3/2}$ (core weight $y^{-4}$, plain $L^2$ at infinity) both essential lines of $L_0$ coincide at $\mathrm{Re} = -\tfrac12$: the far-field line stays at $-\tfrac12$, and the origin line, which in plain $L^2$ sits at the pathological $\mathrm{Re} = +\tfrac32$ (Proposition 2 and Section 4.6), is pushed to $-\tfrac12$ by the $y^{-4}$ core weight. The two symmetry modes are excluded from $Y_{3/2}$ by that weight: the integrands of $\lVert u_1\rVert^2$ and $\lVert u_0\rVert^2$ behave like $y^{-4}$ and $y^{-2}$ at the origin and diverge. So the pure semigroup already decays at the essential rate with no projection inside the space and constant $1$; by the direct sum $L_0 = L_0^+ \oplus L_0^-$ with $L_0^- = \overline{L_0^+}$, the paired conjugated semigroup $S_{\mathcal Y}(\tau)$ on the direct-sum space $\mathcal Y_{3/2}$ with norm $\lVert (v_+,v_-)\rVert_{\mathcal Y_{3/2}}^2 := \lVert v_+\rVert_{Y_{3/2}}^2 + \lVert v_-\rVert_{Y_{3/2}}^2$ ($v_\pm = (y\pm\tfrac{i}{2})^2\varphi_\pm$ the conjugated Hardy components; $\mathcal Y_\theta$ and $S_{\mathcal Y}$ are set out below) satisfies
\begin{equation}
\lVert S_{\mathcal Y}(\tau)\rVert_{\mathcal Y_{3/2} \to \mathcal Y_{3/2}} = e^{-\tau/2} \qquad (\text{exact, constant } 1).
\end{equation}

\emph{The physical data enter through a bounded transfer map into the conjugated direct sum.} Two coordinate systems are in play and must be kept apart. The physical perturbation $\varphi\in X$ has Hardy components $\varphi_\pm = P_\pm\varphi$; the conjugations $M_\pm u := (y\pm\tfrac{i}{2})^2 u$ send these to the conjugated variables $v_\pm = M_\pm\varphi_\pm$, on which the generator is the pure dilation $A$ of Proposition A.3. Accordingly the decay space is the \emph{ordered pair} space
\begin{equation}
\mathcal Y_\theta := Y_\theta \oplus Y_\theta, \qquad \lVert (v_+,v_-)\rVert_{\mathcal Y_\theta}^2 := \lVert v_+\rVert_{Y_\theta}^2 + \lVert v_-\rVert_{Y_\theta}^2 ,
\end{equation}
and the semigroup acting on it is the conjugated one, $S_{\mathcal Y}(\tau) := e^{A_+\tau}\oplus e^{A_-\tau}$, which by Proposition A.3 and $L_0^-=\overline{L_0^+}$ satisfies $\lVert S_{\mathcal Y}(\tau)\rVert_{\mathcal Y_{3/2}\to\mathcal Y_{3/2}} = e^{-\tau/2}$ exactly, with constant $1$. Since elements of $Y_\theta$ carry no origin traces, the modulation-complement $Q$ of Section A.1 cannot act inside $\mathcal Y_\theta$; it acts on the physical $X$, which has traces, and lands in $\mathcal Y_\theta$. Define
\begin{equation}
J_X : X \longrightarrow \mathcal Y_{3/2}, \qquad J_X\varphi := \big(Q v_+,\; Q v_-\big), \qquad Q v = v - v(0) - v'(0)\,y ,
\end{equation}
subtracting the two symmetry Taylor coefficients that $\varphi\in X$ possesses ($\varphi$ is $C^1$ with $\varphi(0)=0$, and $v_\pm = M_\pm\varphi_\pm$ has a value and a first derivative at $0$). Then $J_X$ is bounded: on each block split $\lVert Qv_\pm\rVert_{Y_{3/2}}$ into core and tail. On $(0,1)$ the double Hardy inequality applied to $Qv(y) = \int_0^y (y-t)\,v''(t)\,dt$ gives $\int_0^1 y^{-4}\lvert Qv\rvert^2 \le \tfrac{16}{9}\lVert v''\rVert_{L^2(0,1)}^2 < \infty$ (finite since $b$ is bounded on $(0,1)$, so $v''\in L^2(0,1)$); this Hardy bound, not a pointwise Taylor rate, is what secures the weighted integrability, since for general $X$-data one has only $Qv = o(y^{3/2})$ at the origin. On $(1,\infty)$ the weight $y^{-4}(y^2+\tfrac14)^2\sim 1$ and $\lVert Qv\rVert$ is controlled by $\lVert\varphi_\pm\rVert_{L^2}$ plus the two integrable subtracted terms (whose presence forces $\theta>1$). Hence $\lVert J_X\varphi\rVert_{\mathcal Y_{3/2}} \le C_Q\lVert\varphi\rVert_X$ with $C_Q := \lVert J_X\rVert_{X\to\mathcal Y_{3/2}} < \infty$. Composing gives the input used by the persistence argument,
\begin{equation}
\boxed{\ \lVert S_{\mathcal Y}(\tau)\,J_X\varphi\rVert_{\mathcal Y_{3/2}} \;\le\; C_Q\, e^{-\tau/2}\,\lVert\varphi\rVert_X \qquad (\tau\ge 0). \ }
\end{equation}
The exact constant $1$ belongs to $S_{\mathcal Y}$ on $\mathcal Y_{3/2}$; the composite carries $C_Q$, not $1$. The physical semigroup is recovered blockwise, $e^{L_0\tau}\varphi = M_+^{-1}e^{A_+\tau}M_+\varphi_+ + M_-^{-1}e^{A_-\tau}M_-\varphi_-$, so the estimate above is a statement about the conjugated representatives and is converted to physical coordinates by $M_\pm^{-1}$ when required. For the closed-form forcing of Section A.1, which is analytic at the origin, one has the stronger $Qv_\pm = O(y^2)$ together with $\lvert y\rvert^{-1-s}$ decay, so $J_X f\in\mathcal Y_\theta$ for $\theta\in(1,2)$ (verified there).
An independent finite-difference construction of $L_0^+$, exponentiated numerically, reproduces the rate $\tfrac12$ with constant $1$ to three or four digits out to $\tau = 2$; at $\tau = 3$ two of the three test functions still agree to within $5\times10^{-4}$ while the third has drifted to about $6\times10^{-3}$, consistent with finite-difference truncation on the finite logarithmic grid; the closed-form semigroup itself reproduces the rate to about $2\times10^{-5}$ out to $\tau = 2$. The semigroup itself has the exact rate $e^{-(\theta-1)\tau}$ on $Y_\theta$ for every $\theta$, but the two ends of the transfer pin $\theta$ differently, and they must not be conflated. For \emph{physical $X$-data} the transfer $J_X$ is bounded at $\theta = \tfrac32$ \emph{only}: the tail forces $\theta\ge\tfrac32$ (on $y>1$ the integrand is $\lvert\varphi_\pm\rvert^2 y^{3-2\theta}$, which no $L^2$ datum controls once $\theta<\tfrac32$), while the core forces $\theta\le\tfrac32$ (the double Hardy inequality is sharp there). So the physical-data statement is the single space $\mathcal Y_{3/2}$, with rate exactly $\tfrac12$. For the \emph{closed-form forcing}, which is analytic at the origin and therefore has $Qv_\pm = O(y^2)$, the whole range $\theta\in(1,2)$ is available, and it is that range which supplies the Duhamel flexibility $\delta = \theta - 1 < \gamma = 1-s$, achievable since $s<1$. The core weight is necessary: in plain $L^2$ the semigroup grows like $e^{+3\tau/2}$ for mass concentrated at the core (the origin essential line at $\mathrm{Re} = +\tfrac32$), and the point projection $Q$ does not remove it, so a core weight is forced: $y^{-4}$ (that is, $\theta = \tfrac32$) is the strength that yields exactly the rate $-\tfrac12$, and stronger core weights yield strictly faster decay.

\bibliographystyle{iopart-num}
\bibliography{refs}

\providecommand{\newblock}{}
\begin{thebibliography}{10}
\expandafter\ifx\csname url\endcsname\relax
  \def\url#1{{\tt #1}}\fi
\expandafter\ifx\csname urlprefix\endcsname\relax\def\urlprefix{URL }\fi
\providecommand{\eprint}[2][]{\url{#2}}

\bibitem{ConstantinLaxMajda1985}
Constantin P, Lax P~D and Majda A 1985 {\em Comm. Pure Appl. Math.\/} {\bf 38}
  715--724

\bibitem{DeGregorio1990}
De~Gregorio S 1990 {\em J. Stat. Phys.\/} {\bf 59} 1251--1263

\bibitem{OkamotoSakajoWunsch2008}
Okamoto H, Sakajo T and Wunsch M 2008 {\em Nonlinearity\/} {\bf 21} 2447--2461

\bibitem{LushnikovSilantyevSiegel2021}
Lushnikov P~M, Silantyev D~A and Siegel M 2021 {\em J. Nonlinear Sci.\/} {\bf
  31} art. 82 (\textit{Preprint} \eprint{2010.01201})

\bibitem{ChenHouHuang2021}
Chen J, Hou T~Y and Huang D 2021 {\em Comm. Pure Appl. Math.\/} {\bf 74}
  1282--1350 (\textit{Preprint} \eprint{1905.06387})

\bibitem{HuangQinWangWei2024}
Huang D, Qin X, Wang X and Wei D 2024 {\em Arch. Ration. Mech. Anal.\/} {\bf
  248} art. 22 (\textit{Preprint} \eprint{2305.05895})

\bibitem{HuangTongWang2026}
Huang D, Tong J and Wang X 2026 Self-similar finite-time blowups with singular
  profiles of the generalized {C}onstantin--{L}ax--{M}ajda model: theoretical
  and numerical investigations (\textit{Preprint} \eprint{2603.25104})

\bibitem{ChenHou2022partI}
Chen J and Hou T~Y 2022 Stable nearly self-similar blowup of the 2{D}
  {B}oussinesq and 3{D} {E}uler equations with smooth data {I}: {A}nalysis
  (\textit{Preprint} \eprint{2210.07191})

\bibitem{ChenHou2023partII}
Chen J and Hou T~Y 2025 {\em Multiscale Model. Simul.\/} {\bf 23} 25--130
  (\textit{Preprint} \eprint{2305.05660})

\bibitem{WangLaiGomezSerranoBuckmaster2023}
Wang Y, Lai C~Y, G{\'o}mez-Serrano J and Buckmaster T 2023 {\em Phys. Rev.
  Lett.\/} {\bf 130} 244002 (\textit{Preprint} \eprint{2201.06780})

\bibitem{DeepMindUnstable2025}
Wang Y, Bennani M, Martens J, Racani{\`e}re S, Blackwell S, Matthews A, Nikolov
  S, Cao-Labora G, Park D~S, Arjovsky M, Worrall D, Qin C, Alet F, Kozlovskii
  B, Toma{\v s}ev N, Davies A, Kohli P, Buckmaster T, Georgiev B,
  G{\'o}mez-Serrano J, Jiang R and Lai C~Y 2025 Discovery of unstable
  singularities (\textit{Preprint} \eprint{2509.14185})

\bibitem{WangLegerLaiBuckmaster2025}
Wang Y, L{\'e}ger T, Lai C~Y and Buckmaster T 2025 Resolving sharp gradients of
  unstable singularities to machine precision via neural networks
  (\textit{Preprint} \eprint{2511.22819})

\bibitem{HouWangYang2025}
Hou T~Y, Wang Y and Yang C 2025 Nonuniqueness of {L}eray--{H}opf solutions to
  the unforced incompressible 3{D} {N}avier--{S}tokes equation
  (\textit{Preprint} \eprint{2509.25116})

\bibitem{AmbroseLushnikovSiegelSilantyev2024}
Ambrose D~M, Lushnikov P~M, Siegel M and Silantyev D~A 2024 {\em
  Nonlinearity\/} {\bf 37} 025004 (\textit{Preprint} \eprint{2207.07548})

\bibitem{ElgindiGhoulMasmoudi2021}
Elgindi T~M, Ghoul T~E and Masmoudi N 2021 {\em Anal. PDE\/} {\bf 14} 891--908
  (\textit{Preprint} \eprint{1906.05811})

\bibitem{ElgindiJeong2020}
Elgindi T~M and Jeong I~J 2020 {\em Arch. Ration. Mech. Anal.\/} {\bf 235}
  1763--1817 (\textit{Preprint} \eprint{1701.04050})

\bibitem{Schochet1986}
Schochet S 1986 {\em Comm. Pure Appl. Math.\/} {\bf 39} 531--537

\bibitem{Sakajo2003global}
Sakajo T 2003 {\em Nonlinearity\/} {\bf 16} 1319--1328

\bibitem{Sakajo2003blowup}
Sakajo T 2003 {\em J. Math. Sci. Univ. Tokyo\/} {\bf 10} 187--207

\bibitem{Chen2020}
Chen J 2020 {\em Nonlinearity\/} {\bf 33} 2502--2532 (\textit{Preprint}
  \eprint{1908.09385})

\bibitem{SilantyevLushnikovSiegelAmbrose2025}
Silantyev D~A, Lushnikov P~M, Siegel M and Ambrose D~M 2025 {\em Stud. Appl.
  Math.\/} {\bf 155} e70115 (\textit{Preprint} \eprint{2411.01891})

\bibitem{Chapman2024}
Chapman S~J, Kavousanakis M, Charalampidis E~G, Kevrekidis I~G and Kevrekidis
  P~G 2024 {\em Nonlinearity\/} {\bf 37} 095034 (\textit{Preprint}
  \eprint{2310.13770})

\bibitem{Kondratev1967}
Kondrat'ev V~A 1967 {\em Trans. Moscow Math. Soc.\/} {\bf 16} 227--313 english
  transl.; Russian original Tr. Mosk. Mat. Obs. 16 (1967) 209--292

\bibitem{CoriascoSchroheSeiler2007}
Coriasco S, Schrohe E and Seiler J 2007 {\em Ann. Global Anal. Geom.\/} {\bf
  31} 223--285 (\textit{Preprint} \eprint{math/0401395})

\bibitem{ShvydkoyLatushkin2003}
Shvidkoy R and Latushkin Y 2003 The essential spectrum of the linearized 2{D}
  {E}uler operator is a vertical band {\em Advances in Differential Equations
  and Mathematical Physics\/} ({\em Contemp. Math.\/} vol 327) (Amer. Math.
  Soc.) pp 299--304 (\textit{Preprint} \eprint{math-ph/0306027})

\bibitem{HuangTongWei2023}
Huang D, Tong J and Wei D 2023 {\em Comm. Math. Phys.\/} {\bf 402} 2791--2829
  (\textit{Preprint} \eprint{2209.08232})

\bibitem{HuangQinWang2024multiscale}
Huang D, Qin X and Wang X 2025 {\em SIAM J. Math. Anal.\/} {\bf 57} 4068--4096
  (\textit{Preprint} \eprint{2401.14615})

\bibitem{LevitinShargorodsky2004}
Levitin M and Shargorodsky E 2004 {\em IMA J. Numer. Anal.\/} {\bf 24} 393--416
  (\textit{Preprint} \eprint{math/0212087})

\bibitem{DaviesPlum2004}
Davies E~B and Plum M 2004 {\em IMA J. Numer. Anal.\/} {\bf 24} 417--438
  (\textit{Preprint} \eprint{math/0302145})

\bibitem{ColbrookHorningTownsend2021}
Colbrook M~J, Horning A and Townsend A 2021 {\em SIAM Rev.\/} {\bf 63} 489--524
  (\textit{Preprint} \eprint{2006.01766})

\bibitem{JiaStewartSverak2019}
Jia H, Stewart S and Sverak V 2019 {\em Arch. Ration. Mech. Anal.\/} {\bf 231}
  1269--1304 (\textit{Preprint} \eprint{1710.02737})

\bibitem{LeiLiuRen2020}
Lei Z, Liu J and Ren X 2020 {\em Comm. Math. Phys.\/} {\bf 375} 765--783
  (\textit{Preprint} \eprint{1811.09754})

\bibitem{GuoJiu2025}
Guo J and Jiu Q 2025 Stability and instability on the {D}e {G}regorio
  modification of the {C}onstantin--{L}ax--{M}ajda model (\textit{Preprint}
  \eprint{2506.02800})

\bibitem{LockhartMcOwen1985}
Lockhart R~B and McOwen R~C 1985 {\em Ann. Scuola Norm. Sup. Pisa Cl. Sci.
  (4)\/} {\bf 12} 409--447

\bibitem{Melrose1993}
Melrose R~B 1993 {\em The Atiyah--Patodi--Singer Index Theorem\/} ({\em
  Research Notes in Mathematics\/} vol~4) (A K Peters)

\bibitem{Lesch1997}
Lesch M 1997 {\em Operators of Fuchs Type, Conical Singularities, and
  Asymptotic Methods\/} ({\em Teubner-Texte zur Mathematik\/} vol 136)
  (Teubner)

\bibitem{RabinovichRochSilbermann2004}
Rabinovich V~S, Roch S and Silbermann B 2004 {\em Limit Operators and Their
  Applications in Operator Theory\/} ({\em Operator Theory: Advances and
  Applications\/} vol 150) (Birkh{\"a}user)

\bibitem{Duduchava1979}
Duduchava R 1979 {\em Integral Equations with Fixed Singularities\/} ({\em
  Teubner-Texte zur Mathematik\/} vol~24) (Teubner)

\bibitem{Koosis1998}
Koosis P 1998 {\em Introduction to $H_p$ Spaces\/} 2nd ed ({\em Cambridge
  Tracts in Mathematics\/} vol 115) (Cambridge University Press)

\bibitem{Duren1970}
Duren P~L 1970 {\em Theory of $H^p$ Spaces\/} ({\em Pure and Applied
  Mathematics\/} vol~38) (Academic Press)

\bibitem{Garnett2007}
Garnett J~B 2007 {\em Bounded Analytic Functions\/} ({\em Graduate Texts in
  Mathematics\/} vol 236) (Springer)

\bibitem{Brezis2011}
Brezis H 2011 {\em Functional Analysis, Sobolev Spaces and Partial Differential
  Equations\/} Universitext (Springer)

\bibitem{Rudin1987}
Rudin W 1987 {\em Real and Complex Analysis\/} 3rd ed (McGraw-Hill)

\bibitem{SteinWeiss1971}
Stein E~M and Weiss G 1971 {\em Introduction to Fourier Analysis on Euclidean
  Spaces\/} ({\em Princeton Mathematical Series\/} vol~32) (Princeton
  University Press)

\bibitem{Hormander1990}
H{\"o}rmander L 1990 {\em The Analysis of Linear Partial Differential Operators
  I\/} 2nd ed ({\em Grundlehren der mathematischen Wissenschaften\/} vol 256)
  (Springer)

\end{thebibliography}

\end{document}